\documentclass[twoside]{article}
\usepackage{fleqn,graphics,epsfig}
\begin{document}
\date{8 July 2002}
\newcounter{problem}
\newcommand{\bgm}[1]{\mbox{\boldmath $#1$}}
\def \X   {{ X}}
\def \tX   {{\tilde X}}
\def \tx   {{\tilde x}}
\def \Y  {{ Y}}
\def \tY   {{\tilde Y}}
\def \F  {{\cal F}}
\def \b1  {{T}}
\def \l  {{\lambda(x)}}
\title{Thinking about the brain\\
{\small Based on lectures at Les Houches Session LXXV,  July 2001}\footnote{
To be published in {\em Physics of Biomolecules and Cells,} H.
Flyvbjerg, F. J\"ulicher, P. Ormos, \& F. David, eds. (EDP Sciences, Les
Ulis; Springer-Verlag, Berlin, 2002).}}
\author{William Bialek\\
\\
\small NEC Research Institute,
4 Independence Way,
\small Princeton, New Jersey 08540 USA\\
\small Department of Physics,
Princeton University,
\small Princeton, New Jersey 08544 USA\footnote{Present address.}}

\maketitle

\begin{abstract}
We all are fascinated by the phenomena of intelligent behavior, as
generated both by our own brains and by the brains of other animals.
As physicists we would like to understand if there are some general
principles that govern the structure and dynamics of the neural
circuits that underlie these phenomena.  At the molecular level there
is an extraordinary universality, but these mechanisms are
surprisingly complex.  This raises the question of how the brain selects from these
diverse mechanisms and adapts to compute   ``the right thing'' in each context.
One approach is to ask what problems the brain really solves.  There are several
examples---from the ability of the visual system to count photons on a dark night to our
gestalt recognition of statistical tendencies toward symmetry in random
patterns---where the performance of the system in fact approaches some
fundamental physical or statistical limits.  This suggests that some sort
of optimization principles may be at work, and there are  examples where
these principles have been formulated clearly and generated predictions
which are confirmed in new experiments; a central theme in this work is
the matching of the coding and computational strategies of the brain to
the statistical structure of the world around us.  Extension of these
principles to the problem of learning leads us into interesting
theoretical questions about how to measure the complexity of the data
from which we learn and the complexity of the models that we use in
learning, as well as opening some new opportunities for experiment.   
This combination of theoretical and experimental work gives us some new
(if still speculative) perspectives on classical problems and
controversies in cognition.
\end{abstract}

\vfill\newpage
\vfill

\tableofcontents
\vfill\newpage

\section{Introduction}

Here in Les Houches we are surrounded by many beautiful and dramatic
phenomena of nature.   In the last century we have come to understand
the powerful physical forces that shaped the landscape, creating the
peaks that reach thousands of meters into the sky.    As we stand and
appreciate the view, other powerful forces also are at work:  we are
conscious of our surroundings, we parse a rich scene into natural and
manmade objects that have meaningful relationships to one another and to
us, and we learn about our environment so that we can navigate even in the
dark after long hours of discussion in the bar.  These aspects of
intelligent behavior---awareness, perception, learning---surely are among
the most dramatic natural phenomena that we experience directly.  As
physicists our efforts to  provide a predictive, mathematical description
of nature  are animated by the belief that qualitatively striking
phenomena should have deep theoretical explanations.  The challenge, then,
is to tame the evident complexities of intelligent behavior and to uncover
these deep principles.

Words such as ``intelligent''  perhaps are best viewed as colloquial
rather than technical:  intelligent behavior refers to a class of
phenomena exhibited by humans and by many other organisms, and membership
in this class is by agreement among the participants in
the conversation. There also is a technical meaning of
``intelligence,'' determined by the people who construct intelligence
tests.  This is an area fraught with political and sociological
difficulties, and there also is some force to Barlow's
criticism that intelligence tends to be defined as what the tests
measure \cite{hbb-intell}.  For now let us leave the general
term ``intelligence'' as an informal one, and try to be precise about
some particular aspects of intelligent behavior.  

Our first task, then, is to choose some subset of intelligent behaviors which we can
describe in quantitative terms.  I shall have nothing to say about consciousness, but
for learning and perception we can go some way toward constructing a
theoretical framework within which quantitative experiments can be
designed and analyzed. Indeed, because perception constitutes our personal
experience of the physical world, there is a tradition of physicists
being interested in perceptual phenomena that reaches back (at least) to
Helmholtz, Rayleigh, Maxwell and Ohm, and a correspondingly rich body of
quantitative experimental work.  If we can give a quantitative
description of the phenomena it is natural to hope that some regularities
may emerge, and that these could form the basis of a real theory.

I will argue that there is indeed one very striking regularity that
emerges when we look quantitatively at the phenomena of perception, and
this is a notion of optimal performance.  There are well defined limits
to the reliability of our perceptions set by noise at the sensory input,
and this noise in turn often has fundamental physical origins.  In
several cases the brain approaches these limits to reliability,
suggesting that the circuitry inside the brain is doing something like an
optimal processing of the inputs or an optimal extraction of the
information relevant for its tasks.  It would be very attractive if this
notion of optimization---which grows out of the data!---could be elevated
to a principle, and I will go through one example in detail where we have
tried to carry out this program.  

The difficulty with collecting evidence for optimization is that we might be left only
with a list of unrelated examples:  There is a set of tasks for which performance is near
optimal, and for each task we have a theory of how the brain does the task based on
optimization principles.  But precisely because the brain is not a general purpose
computer, some tasks are done better than others.  What we would like is not a list, but
some principled view of what the brain does well.  Almost since Shannon's
original papers there has been some hope that information theory could
provide such organizing principles, although much of the history is
meandering rather than conclusive.  I believe that in the past several
years there has been substantial progress toward realizing the old dreams.
On the one hand we now have direct experimental demonstrations that the
nervous system can adapt to the statistical structure of the sensory
world in ways that serve to optimize the efficiency of information
transmission  or representation.  On the other hand, we have a new
appreciation of how information theory can be used to assess the
relevance of sensory information and the complexity of data streams. 
These theoretical developments unify ideas that have arisen in fields as
diverse as coding theory, statistics and dynamical systems ... and hold
out some hope for a unified view of many different tasks in neural
computation.  I am very excited by all of this, and I hope to communicate
the reasons for my excitement.

A very different direction is to ask about the microscopic basis for the
essentially macroscopic phenomena of perception and learning.  In the
last decade we have seen an explosion in the experimental tools for
identifying molecular components of biological systems, and as these
tools have been applied to the brain this has created a whole new field
of molecular neurobiology.  Indeed, the volume of data on the molecular
``parts list'' of the brain is so vast that we have to ask carefully what
it is we would like to know, or more generally why we are asking for a
microscopic description. One possibility is that there is no
viable theory at a macroscopic level: if we want to know why we
perceive the world as we do, the answer might be found only in a detailed
and exhaustive investigation of what all the molecules and cells are doing
in the relevant regions of the brain.  This is too horrible to
discuss.

One very good reason for looking at the microscopic basis of neural computation is that
molecular events in the cells of the brain (neurons) provide  prototypes for thinking
about molecular events in all cells, but with the advantage that important parts of the
function of neurons involve electrical signals which are wonderfully accessible to
quantitative measurements.  Fifty years of work has
brought us a nearly complete list of molecular components involved in the
dynamics of neural signalling and computation, quantitative experiments
on the properties of these individual molecules, and accurate mathematical
models of how these individual molecular properties combine to determine
the dynamics of the cell as a whole.  The result is that the best
characterized networks of molecular interactions in cells are the
dynamics of ion channels in neurons.  This firm foundation puts us in a
position to ask questions about the emergent properties of these
networks, their stability and robustness, the role of noise, ...  all in
experimentally accessible systems where we really know the relevant
equations of motion and even most of the relevant parameters.

A very different reason for being interested in the molecular basis of perception and
learning is  because, as  it turns out, the brain is a very peculiar
computational device.  As in all of biology, there is no obvious blueprint or wiring
diagram; everything organizes itself.  More profoundly, perhaps, all the
components are impermanent.  Particularly when we think about storing
what we have learned or hope to remember, the whole point seems to be
a search for permanence, yet almost every component of the relevant
hardware in the brain will be replaced on a time scale of weeks, roughly the
duration of this lecture series.  Nonetheless we expect you to remember
the events here in Les Houches for a time much longer than the
duration of the school.  Not only is there a problem of understanding how
one stores information in such a dynamic system, there is the problem of
understanding how such a system maintains stable function over long
periods of time.  Thus if the computations carried out by a neuron   are
determined by the particular combination of ion channel proteins that the
cell expresses and inserts in the membrane, how does the cell ``know''
and maintain the correct expression levels as proteins are constantly
replaced?   Typical neurons express of order ten different kinds of ion
channels at once, and it is not clear what functions are made possible by
this molecular complexity---what can we do with ten types of channel that
we can't do with nine?  Finally, as in many aspects of life, crucial
aspects of neural computation are carried out by surprisingly small
numbers of molecules, and we shall have to ask how the system achieves
reliability in the presence of the noise associated with these small
numbers.

The plan is to start by examining the evidence for optimal performance in
several systems, and then to explore information theoretic ideas that
might provide some more unified view.  Roughly speaking we will proceed
from things that are very much grounded in data---which is important,
because we have to convince ourselves that working on brains can involve
experiments with the ``look and feel'' of good physics---toward more
abstract problems.  I would hope that some of the abstract ideas will
link back to experiment, but I am still unclear about how to do this.
Near the end of the course I will circle back to   outline some of the issues
in understanding the microscopic basis of neural computation, and
conclude with some speculative thoughts on the ``hard problems'' of
understanding cognition.

Some general references on the optimality of sensory and neural systems
are reviews which Barlow \cite{barlow81} and I
\cite{bialek87,bialek92} have written, as well as
sections of the book {\em Spikes} \cite{spikes}, which may provide
a useful reference for a variety of issues in the lectures. Let me warn the reader that
the level of detail, both in the text and in the references, is a bit uneven (as were
the lectures, I suspect).  I have, however, taken the liberty of scattering some
problems throughout the text.  One last caveat:  I am almost pathologically
un--visual in my thinking, and so I wrote this text without figures.  I think it can
work, in that the essential ideas are summarizable in words and equations, but you
really should look at original papers to see the data that support the theoretical
claims and (more importantly) to get a feeling for how the experiments really were done.

\section{Photon counting}

Sitting quietly in a dark room, we can detect the arrival
of individual photons at our retina.  This observation
has a beautiful history, with its roots
in a suggestion by Lorentz in 1911.\footnote{Reviews on photon counting and
closely related issues include Refs.
\cite{barlow81,bialek87,fred-rmp} and  
Chapter 4 of Ref. \cite{spikes}.}  Tracing
through the steps from photon arrival to perception we see a sampling of
the physics problems posed by biological systems, ranging from the
dynamics of single molecules through amplification and adaptation in
biochemical reaction networks, coding and computation in neural networks,
all the way ``up''  to learning and cognition.  For photon counting some
of these problems are solved, but even in this well studied case many
problems are open and ripe for new theoretical and experimental work.  I
will try to use the photon counting example as a way of motivating some
more general questions.

Prior to Lorentz' suggestion, there was a long history of measuring the
energy of a light flash   that just barely can be seen.  There is, perhaps
surprisingly, a serious problem in relating this  minimum energy of a visible flash
to the number of photons at the retina, largely because of uncertainties about
scattering and absorption in the eye itself.   The compelling alternative is a
statistical argument, as first exploited by Hecht, Shlaer and Pirenne (in New York)
and  van der Velden (in the Netherlands) in the early 1940s
\cite{hsp,velden}:
\begin{itemize}
\item The mean number of photons $\langle n\rangle$ at the retina
is proportional to the intensity $I$ of the flash.
\item With conventional light sources the actual number $n$ of photons
that arrive from any single flash will obey Poisson statistics,
\begin{equation}
P(n| \langle n \rangle ) = \exp(-\langle n \rangle ) {{\langle n
\rangle^n}\over{n!}}
\label{poissoncounts}
\end{equation}
\item Suppose that we can see when at least $K$ photons arrive.  Then the
probability of seeing is
\begin{equation}
P_{\rm see}  = \sum_{n \geq K} P(n| \langle n \rangle ).
\end{equation}
\item We can ask an observer whether he or she sees the flash, and the
first nontrivial observation is that seeing really is probabilistic for
dim flashes, although this could just be fluctuations in attention. 
\item The key point is that however we measure the intensity $I$, we have
$\langle n \rangle = \alpha I$, with $\alpha$ some unknown proportionality
constant, so that
\begin{equation}
P_{\rm see} (I)  = \sum_{n \geq K} P(n| \langle n \rangle = \alpha I ).
\end{equation}
If we plot $P_{\rm see}$ vs. $\log I$, then one can see that the  
{\em shape} of the curve depends crucially on the threshold photon count
$K$, but changing the unknown constant $\alpha$ just translates the curve
along the x-axis.  So we have a chance to measure the threshold $K$ without knowing
$\alpha$ (which is hard to measure).
\end{itemize}
Hecht, Shlaer and Pirenne did exactly this and found a beautiful fit
to $K=5$ or $K=7$; subjects with different age had very different values
for $\alpha$ but similar values of $K$.  This sounds good: maybe the
probabilistic nature of our perceptions just reflects  the physics of
random photon arrivals.

\medskip
\addtocounter{problem}{1}
{\small \noindent {\bf Problem \theproblem:  Poisson
processes.}\footnote{Many of you have seen this before, so this is just a
refresher.  For the rest, you might look at Appendices 4 and 5 in {\em
Spikes} which give a fairly detailed step--by--step discussion of Poisson
processes \cite{spikes}.} To understand what is going on here it would be
a good idea if you review some facts about Poisson processes.  By a
`process' we mean in this case the time series of discrete events
corresponding to photon arrivals or absorption.  If the typical time
between events is long compared to any intrinsic correlation times in the
light source, it is plausible that each photon arrival will be
independent of the others, and this is the definition of a
Poisson process.\footnote{There is also the interesting fact that certain
light sources will generate Poisson photon counting distributions no
matter how frequently the photons arrive:  recall that for a harmonic
oscillator  in a coherent state (as for the field
oscillations in an ideal single mode laser),  measurements of the
number of quanta  yield a Poisson distribution, exactly.}  Thus, if
we look at a very small time interval
$dt$, the probability of counting one event will be $rdt$, where
$r$ is the mean counting rate.  If we count in a time window of size $T$,
the mean count clearly will be
$\langle n
\rangle = rT$. 
\begin{itemize}
\item[a.] Derive the probability density for events at
times
$t_1 , t_2, \cdots , t_n$;  remember to include terms for the probability
of {\em not} observing events at other times.  Also, the events are
indistinguishable, so you need to include a combinatorial factor.  The
usual derivation starts by taking  discrete time bins of size
$dt$, and then at the end of the calculation you let $dt \rightarrow 0$.
You should find that
\begin{equation}
P(t_1 , t_2, \cdots , t_n ) = {1\over{n!}} r^n \exp(-rT) .
\end{equation}
Note that this corresponds to an `ideal gas' of indistinguishable events.
\item[b.] Integrate over all the times $t_1, t_2 , \cdots , t_n$ in the
window $t\in [0,T]$ to find the probability of observing $n$ counts. 
This should agree with Eq. (\ref{poissoncounts}), and you should verify
the normalization.  What is the relation between the mean and variance of
this distribution?
\end{itemize}
\label{poisson}}
\bigskip

An important point is that the $5$ to $7$ photons are distributed across a
broad area on the retina, so that the probability of one receptor (rod)
cell getting more than one photon is very small.  Thus the experiments on
human behavior suggest that individual photoreceptor cells generate
reliable responses to single photons.  This is a  lovely example of using
macroscopic experiments to draw conclusions about single cells.  

It took many years before anyone could measure directly the responses of
photoreceptors to single photons.  It was done first in the 
(invertebrate) horseshoe crab,  and
eventually by Baylor and coworkers in toads
\cite{baylor79} and then in monkeys \cite{baylor84}.  The complication in the
lower vertebrate systems is that the cells are coupled together, so that
the retina can do something like adjusting the size of pixels as a
function of light intensity.  This means that the nice big current
generated by one cell is spread as a small voltage in many cells, so the
usual method of measuring the voltage across the membrane of one cell
won't work; you have to suck the cell into a pipette and collect the
current, which is what Baylor et al. managed to do.  Single photon
responses observed in this way are about a picoamp in amplitude vs. a
continuous background noise of 0.1 pA rms, so these are easily detected.

A slight problem is that van der Velden found $K=2$, far from the $K=
5-7$ found by Hecht, Shlaer and Pirenne.  Barlow explained this
discrepancy by noting that even when counting single photons we may have
to discriminate (as in photomultipliers) against a background of dark
noise
\cite{barlow56}.  Hecht, Shlaer and Pirenne inserted blanks in their
experiments to be sure that you never say ``I saw it'' when nothing is
there [that is, $P_{\rm see}(I=0) = 0$], which means you have to set a
high threshold to discriminate against the noise.  On the other hand, van
der Velden was willing to allow for some false positive responses, so his
subjects could afford to set a lower threshold.  Qualitatively this makes
sense, but to be a quantitative explanation the noise has to be at the
right level.  Barlow reasoned that one source of noise was if the pigment
molecule rhodopsin spontaneously (as a result of thermal fluctuations)
makes the transitions that normally are triggered by a photon; of course
these random events would be indistinguishable from photon arrivals. He
found that everything works if this spontaneous event rate is
equivalent to roughly 1 event per 1000 years per molecule: there are a
billion molecules in one rod cell, which gets us to one event per minute
per cell (roughly) and when we integrate over hundreds of cells for
hundreds of milliseconds we find a mean event count of $\sim 10$, which
means that to be sure we see something we will have to count many more
than $\sqrt{10}$ extra events, corresponding to what Hecht, Shlaer and Pirenne
found in their highly reliable observers.

One of the key
points here is that Barlow's explanation only works if people actually
can adjust the ``threshold'' $K$ in response to different situations. 
The realization that this is possible was part of the more general
recognition that detecting a sensory signal does not involve a true
threshold  between (for example) seeing and not
seeing \cite{green+swets}.  Instead we should imagine that---as 
when we try to measure something in a physics experiment---all sensory tasks involve a
discrimination between signal and noise, and hence there are different strategies
which provide different ways of trading off among the different kinds of errors. 

Suppose, for example, that you get to observe $x$ which could be drawn
either from the probability distribution $P_+(x)$ or from the
distribution $P_-(x)$; your job is to tell me whether it was $+$ or $-$.  Note that the
distribution could be controlled completely by the experimenter (if you play loud but
random noise sounds, for example) or the distribution could be a model of noise
generated in the receptor elements or even deep in the brain.  At least
for simple forms of the distributions  $P_\pm (x)$, we can make a decision
about how to assign a particular value of $x$ by simply setting a
threshold $\theta$; if $x > \theta$ we say that $x$ came from the $+$
distribution, and conversely.  How should we set $\theta$?  Let's try to
maximize the probability that we get the right answer.  If $x$ is chosen
from $+$ with probability $P(+)$, and similarly for $-$, then the
probability that our threshold rule gets the correct answer is 
\begin{equation}
P_{\rm correct} (\theta) = P(+) \int_\theta^\infty dx P_+ (x)
+P(-) \int_{-\infty}^\theta dx P_- (x) .
\label{pc}
\end{equation}
To maximize $P_{\rm correct} (\theta)$ we differentiate with respect to
$\theta$ and set this equal to zero:
\begin{eqnarray}
{{dP_{\rm correct} (\theta)}\over {d\theta}}
&=&  0\\
\Rightarrow P(+)P_+ (\theta ) &=& P(-)P_- (\theta ) .
\end{eqnarray}
In particular if $P(+) = P(-) = 1/2$, we set the threshold at the point
where $P_+(\theta ) = P_-(\theta )$; another
way of saying this is that we assign each $x$ to the probability
distribution that has the larger density at $x$---``maximum likelihood.''
Notice that if the probabilities of the different signals $+$ and $-$
change, then the optimal setting of the threshold changes.  

\medskip
\addtocounter{problem}{1}
{\small \noindent {\bf Problem \theproblem: More careful discrimination.}
Assume as before that $x$ is chosen either from $P_+(x)$ or from
$P_-(x)$.  Rather than just setting a threshold, consider the possibility
that when you see $x$ you  assign it to the $+$ distribution with  a {\em
probability} $f(x)$.  
\begin{itemize}
\item[a.] Express the probability of a correct answer in terms of $f(x)$,
generalizing Eq. (\ref{pc}); this is a functional $P_{\rm correct}
[f(x)]$.
\item[b.] Solve the optimization problem for the function $f(x)$; that
is, solve the equation
\begin{equation}
{{\delta P_{\rm correct}
[f(x)]}\over {\delta f(x)}} = 0.
\end{equation}
Show that the solution is deterministic
[$f(x) =1$ or $f(x) =0$], so that if the goal is to be correct as often
as possible you shouldn't hesitate to make a crisp assignment even at values of $x$
where you aren't sure (!).
\item[c.] Consider the case where $P_\pm (x)$ are Gaussian distributions
with the same variance but different means.  Evaluate the minimum
error probability (formally) and give asymptotic results for large and
small differences in mean.  How large do we need to make this `signal'
to be guaranteed only 1\% errors?
\item[d.]  Generalize these results to multidimensional Gaussian
distributions, and give a
geometrical picture of the assignment rule.  This problem is easiest if
the different Gaussian variables are independent and have equal
variances.
What happens in the more general case of arbitrary covariance matrices?
\end{itemize}}
\bigskip

There are classic experiments to show that people will adjust their
thresholds automatically when we change the a priori probabilities, as
expected for optimal performance.  This can be done without any explicit
instructions---you don't have to tell someone that you are changing the
value of $P(+)$.  At least implicitly, then, people learn something about
probabilities and adjust their assignment criteria appropriately. 
As we will discuss later in the course, there are other ways of showing
that people (and other animals) can learn something about probabilities
and use this knowledge to change their behavior in sensible ways.
Threshold adjustments also can be driven by changing the rewards for
correct answers or the penalties for wrong answers.  In this 
view, it is likely that Hecht et al. drove their observers to high
thresholds by having a large effective penalty for false positive
detections.  Although it's not  a huge literature, people have since
manipulated these penalties and rewards in HSP--style experiments, with
the expected results. Perhaps more dramatically, modern quantum
optics techniques have been used to manipulate the statistics of photon arrivals at the
retina, so that the tradeoffs among the different kinds of errors are
changed ... again with the expected results \cite{teich}.

Not only did Baylor and coworkers detect the single photon responses from
toad photoreceptor cells, they also  found that single receptor
cells in the dark show spontaneous photon--like events at just the right
rate to be the source of dark noise identified by Barlow
\cite{baylor80}!  Just to be clear, Barlow identified a {\em maximum}
dark noise level; anything higher and the observed reliable detection is
impossible.  The fact that the real rod cells have essentially this level
of dark noise means that the visual system is operating near the limits
of reliability set by thermal noise in the input. It would be nice,
however, to make a more direct test of this idea.  

In the lab we often lower the noise level of photodetectors by cooling
them.  This isn't so easy in humans, but it does work with cold blooded
animals like frogs and toads. So, Aho et al.  \cite{aho88} convinced
toads to strike with their tongues at small worm--like objects
illuminated by dim flashes of light, and measured how the threshold for
reliable striking varied with temperature.  It's important that the
prediction is for more reliable behavior as you cool down---all the way
down to the temperature where behavior stops---and this is what Aho et al. observed. 
Happily, Baylor et al. also measured the temperature dependence of the noise in the
detector cells.  The match of behavioral and cellular noise levels vs. temperature is
perfect, strong evidence that visual processing in dim lights really is
limited by input noise and not by any inefficiencies of the brain.

\medskip
\addtocounter{problem}{1}
{\small \noindent {\bf Problem \theproblem: Should you absorb all the
photons?} Consider a rod photoreceptor cell of length $\ell$, with
concentration $C$ of rhodopsin; let the absorption cross section of
rhodopsin be $\sigma$.  The probability that a single photon incident on
the rod will be counted is then $p = 1- \exp(-C\sigma\ell)$, suggesting
that we should make $C$ or $\ell$ larger in order to capture more of the
photons.  On the other hand, as we  increase the number of Rhodopsin
molecules ($CA\ell$, with $A$ the area of the cell) we also increase the
rate of dark noise events.  Show that the signal--to--noise ratio for
detecting a small number of incident photons is maximized at a nontrivial
value of
$C$ or $\ell$, and calculate the capture probability $p$ at this optimum. 
Do you find it strange that the best thing to do is to let some of the
photons go by without counting them?  Can you see any way to design an
eye which gets around this argument?  Hint:  Think about what you see
looking into a cat's eyes at night.}
\bigskip

These observations on the ability of the visual system to count single
photons---down to the limit set by thermal noise in rhodopsin
itself---raise questions at several different levels:
\begin{itemize}
\item At the level of single molecules, there are many interesting physics problems in
the dynamics of rhodopsin itself.
\item At the level of single cells, there are challenges in understanding how a network
of biochemical reactions converts individual molecular events into macroscopic
electrical currents across the rod cell membrane.
\item At the level of the retina as a whole, we would like to understand the rules
whereby these signals are encoded into the stereotyped pulses which are the universal
language of the brain.
\item At the level of the whole organism, there are issues about how the brain learns
to make the discriminations which are required for optimal performance.
\end{itemize}
Let's look at these questions in order.   The goal here is more to provide pointers to
interesting and exemplary issues than to provide answers.

At the level of single molecule dynamics, our ability to see in the dark
ultimately is limited by the properties of rhodopsin (because everything
else works so well!).  Rhodopsin consists of a medium--sized organic
pigment, retinal, enveloped by  a large protein, opsin; the
photo--induced reaction is isomerization of the retinal, which ultimately
couples to structural changes in the protein.  One obvious function of the
protein is to tune the absorption spectrum of retinal so that the same
organic pigment can work at the core of the molecules in rods and in all
three different cones, providing the basis for color vision.  Retinal has
a spontaneous isomerization rate of $\sim$ 1/yr, 1000 times that of
rhodopsin, so clearly the protein acts to lower the dark noise level. 
This is not so difficult to understand, since one can imagine how a big
protein literally could get in the way of the smaller molecule's motion
and raise the barrier for thermal isomerization. Although this
sounds plausible, it's probably wrong:  the activation {\em energies}
for thermal isomerization in retinal and in rhodopsin are almost the
same.  Thus one either must believe that the difference is in an entropic
contribution to the barrier height or in dynamical terms which determine
the prefactor of the transition rate.  I don't think the correct answer
is known.

On the other hand, the photo--induced isomerization rate of
retinal is only $\sim 10^9\,{\rm s}^{-1}$, which is slow enough that
fluorescence competes with the structural
change.\footnote{Recall from quantum mechanics that the
spontaneous emission rates from electronic excited states are constrained
by sum rules if they are dipole--allowed.  This means that emission
lifetimes for visible photons are order 1 nanosecond for almost all of the simple
cases ... .}  Now
fluorescence is a disaster for visual pigment---not only don't you get to
count the photon where it was absorbed, but it might get counted
somewhere else, blurring the image.  In fact rhodopsin does not
fluoresce:  the quantum yield or branching ratio for fluorescence is $\sim 10^{-5}$,
which means that the molecule is changing its structure and escaping the immediate
excited state in tens of femtoseconds \cite{doukas84}.  Indeed, for years
every time people built faster pulsed lasers, they went back to rhodopsin
to look at the initial events, culminating in the direct demonstration of
femtosecond isomerization \cite{schoenlein91}, making this one of the
fastest molecular events ever observed.  

The combination of faster photon induced isomerization and slower thermal
isomerization means that the protein opsin acts as an electronic state
selective catalyst:  ground state reactions are inhibited, excited state
reactions accelerated, each by orders of magnitude.  It is fair to say
that if these state dependent changes in reaction rate did not occur
(that is, if the properties of rhodopsin were those of retinal) we simply
could not see in the dark.

Our usual picture of molecules and their transitions comes from chemical
kinetics:  there are reaction rates, which represent the probability per
unit time for the molecule to make transitions among states which are
distinguishable by some large scale rearrangement; these transitions are
cleanly separated from the time scales for molecules to come to
equilibrium in each state, so we describe chemical reactions (especially
in condensed phases) as depending on temperature not on energy.  The
initial isomerization event in rhodopsin is so fast that this
approximation certainly breaks down.  More profoundly, the time scale of the
isomerization is so fast that it competes with the processes that destroy
quantum mechanical coherence among the relevant electronic and
vibrational states \cite{wang94}.  The whole notion of an irreversible
transition from one state to another necessitates the loss of coherence
between these states (recall Schr\"odinger's cat), and so in this sense
the isomerization is proceeding as rapidly as possible.  I don't think we
really understand, even qualitatively, the physics here.\footnote{That's
not to say people aren't trying; the theoretical literature also is huge,
with much of it (understandably) focused on how the protein
influences the absorption spectra of the chromophore. The dynamical
problems are less well studied, although again there is a fairly large
pile of relevant papers in the quantum chemistry literature (which I
personally find very difficult to read). In the late 1970 and early
1980s, physicists got interested in the electronic properties of
conjugated polymers because of the work by Heeger and others showing that
these quasi--1D materials could be doped to make good conductors. Many
people must have realized that the dynamical models being used by
condensed matter physicists for (ideally) infinite chains might also have
something to say about finite chains, but again this was largely the
domain of chemists who had a rather different point of view. Kivelson and
I tried to see if we could make the bridge from the physicists' models to
the dynamics of rhodopsin, which was very ambitious and never quite
finished; there remains a rather inaccessible conference proceeding
outlining some of the ideas \cite{bgk}. Our point of view was rediscovered
and developed by Aalberts and coworkers a decade later
\cite{aalberts1,aalberts2}.}  If rhodopsin were the only example of this
`almost coherent chemistry' that would be good enough, but in fact the
other large class of photon induced events in biological
systems---photosynthesis---also proceed so rapidly as to compete with
loss of coherence, and the crucial events again seem to happen (if you
pardon the partisanship) while everything is still in the domain of
physics and not conventional chemistry
\cite{vos91,vos94}.  Why biology pushes to these extremes is a good
question.  How it manages to do all this with big floppy molecules in
water at roughly room temperature also is  a great question.   

At the level of single cells, the  biochemical circuitry of the rod takes one
molecule of activated rhodopsin and turns this into a macroscopic response.  
Briefly, the
activated rhodopsin is a catalyst that activates many other molecules, which in turn act
as catalysts and so on.  Finally there is a catalyst (enzyme) that eats
cyclic GMP, but cGMP binds to and opens ionic channels in the cell
membrane.  So when the cGMP concentration falls, channels close, and the
electrical current flowing into the cell is reduced.\footnote{Actually we
can go back to the level of single molecules and ask questions about the
`design' of these rather special channels ... .} The gain of this system
must be large---many molecules of cGMP are broken down for each single
activated rhodopsin---but gain isn't the whole story.  First, most models
for such a chemical cascade would predict large fluctuations in the
number of molecules at the output since the lifetime of the active state
of the single active rhodopsin fluctuates wildly (again, in the simplest
models)
\cite{fred-rmp,rieke98}.  Second, as the lights gradually turn on one has
to regulate the gain, or else the cell will be overwhelmed by the
accumulation of a large constant signal; in fact, eventually all the
channels close and the cell can't respond at all.  Third, since various
intermediate chemicals are present in finite concentration, there is a
problem of making sure that signals rise above the fluctuations in these
concentrations---presumably while not expending to much energy too make
vast excesses of anything. To achieve the required combination of gain,
reliability, and adaptation almost certainly requires a network with
feedback.  The quantitative and even qualitative properties of such
networks depend on the concentration of various protein components, yet
the cell probably cannot rely on precise settings for these
concentrations, so this robustness creates yet another
problem.\footnote{If the cell does regulate molecule counts very
accurately, one problem could be solved, but then you have to explain the
mechanism of regulation.}  

Again if photon counting were the only example all of this it might be
interesting enough, but in fact there are many cells which build single
molecule detectors of this sort, facing all the same problems.  The different
systems use molecular components that are sufficiently similar that one can
recognize the homologies at the level of the DNA sequences which code for the
relevant proteins---so much so, in fact, that one can go searching for unknown
molecules by seeking out these homologies.  This rough universality of tasks and
components cries out for a more principled view of how such networks work
(see, for example, Ref. \cite{detwiler00}); photon counting is such an
attractive example because there is an easily measurable electrical
output and because there are many tricks for manipulating the network
components (see, for example, Ref.
\cite{rieke96}).

At the level of the retina as a whole,  the output
which gets transmitted to the brain is not a continuous voltage or current
indicating (for example) the light intensity or photon counting rate;
instead signals are encoded as sequences of discrete identical pulses
called action potentials or spikes.  Signals from the
photodetector cells are sent from the eye to the brain (after some
interesting processing within the retina itself ... ) along 
$\sim 10^6$ cables---nerve cell `axons'---that form the optic nerve.   Roughly
the same number of cables carry signals from the sensors in our skin, for
example, while each ear sends $\sim 40,000$ axons into the central nervous
system.  It is very likely that our vision
in the photon counting regime is triggered by the occurrence of just a few extra
spikes along at most a few optic nerve axons \cite{spikes,bly}.  For the
touch receptors in our fingertips there is direct evidence that our
perceptions can be correlated with the presence or absence of a single
action potential along one out of the million input cables \cite{valbo}. 
To go beyond simple detection we have to understand how the complex,
dynamic signals of the sensory world can be represented in these
seemingly sparse pulse trains \cite{spikes}.

Finally, the problem of photon counting---or any simple detection task---hides a deeper
question:  how does the brain ``know'' what it needs to do in any given task?  Even
in our simple example of setting a threshold to maximize the probability of a
correct answer, the optimal observer must at least implicitly acquire knowledge of
the relevant probability distributions.  Along these lines, there is more to
the `toad cooling' experiment than a test of photon counting and dark noise.  The
retina has adaptive mechanisms that allow the response to speed up at higher levels
of background light, in effect integrating for shorter times when we can be sure
that the signal to noise ratio will be high.  The flip side of this mechanism
is that the retinal response slows down dramatically in the dark.  In moderate
darkness (dusk or bright moonlight) Aho  et al. found that the slowing of
the retinal response is reflected directly in a slowing of the animal's behavior
\cite{aho93}:   it is as if the toad experiences an illusion because
images of its target are delayed, and it strikes at the delayed
image.\footnote{We see this illusion too.  Watch a pendulum swinging
while wearing glasses that have a neutral density filter over one eye, so
the mean light intensity in the two eyes is different.  The dimmer light
results in a  slower retina, so the signals from the two eyes are not
synchronous.  As we try to interpret these signals in terms of motion, we
find that even if the pendulum is swinging in a plane parallel to the
line between our eyes, what we see is motion in 3D.  The magnitude of the
apparent depth of oscillation is related to the neutral density and hence
to the slowing of signals in the `darker' retina.}  But if this continued
down to the light levels in the darkest night, it would be a disaster,
since the delay would mean that the toad inevitably strikes behind the
target! In fact, the toad does not strike at all in the first few trials
of the experiment in dim light, and then strikes well within the target.  It is hard to
escape the conclusion that the animal is learning about the typical
velocity of the target and then using this knowledge to extrapolate and
thereby correct for retinal delays.\footnote{As far as I know there are
no further experiments that probe this learning more directly, e.g. by
having the target move at variable velocities.}  Thus, performance in the
limit where we count photons involves not only efficient processing of
these small signals but also learning as much as possible about the world
so that these small signals become interpretable.

We take for granted that life operates within boundaries set by physics---there
are no vital forces.\footnote{Casual acceptance of this statement of course
reflects a hard fought battle that stretched from debates about
conservation of energy in $\sim$1850 to the discovery of the DNA structure in
$\sim$1950.  If you listen carefully, some people who talk about the mysteries of
the brain and mind still come dangerously close to a vitalist position, and the
fact that we can't really explain how such complex structures evolved leaves room
for wriggling, some of which makes it into the popular press.  Note also that, as
late as 1965, Watson was compelled to have a section heading in {\em
Molecular Biology of the Gene} which reads ``Cells obey the laws of physics and
chemistry.''  Interestingly, this continues to appear in later editions.}  What is
striking about the example of photon counting is that in this case life operates
{\em at} the limit:  you can't count half a photon, your counting can't be any
more reliable than allowed by thermal noise, chemical reactions can't happen faster
than loss of quantum coherence, and so on.  Could this be a general principle?  Is
it possible that, at least for crucial functions which have been subjected to eons
of evolutionary pressure, all biological systems have found solutions which are optimal
or extremal in this physicist's sense?    If so, we have the start of a real program to
describe these systems using variational principles to pick out optimal functions,
and then sharp questions about how realistic dynamical models can implement this
optimization.   Even if the real systems aren't optimal, the exercise of
understanding what the optimal system might look like will help guide us in
searching for new phenomena and maybe in understanding some puzzling old
phenomena.  We'll start on this project in the next lecture.

\section{Optimal performance at more complex tasks}

Photon counting is pretty simple, so it might be a good idea to look at
more complex tasks and see if any notion of optimal performance still
makes sense. The most dramatic example is from bat echolocation, in a series of
experiments by Simmons and colleagues culminating in the demonstration that bats can
discriminate reliably  among echoes that differ by just $\sim 10 - 50$ {\em
nano}seconds in delay \cite{simmons-nano}.  In these experiments, bats stand
at the base of a Y with loudspeakers on the two arms.  Their ultrasonic calls are
monitored by microphones and returned through the loudspeakers with programmable
delays.  In a typical experiment, the `artificial echoes' produced by one side of the Y
are at a fixed delay $\tau$, while the other side alternately produces delays of
$\tau\pm\delta\tau$.  The bat is trained to take a step toward the side which
alternates, and the question is how small we can make $\delta\tau$ and still have the
bat make reliable decisions. 

Early experiments from Simmons and
coworkers suggested that delays differences of
$\delta\tau\sim 1\,\mu{\rm sec}$ were detectable, and perhaps more surprisingly that
delays of $\sim 35\,\mu{\rm sec}$ were less detectable.  The latter result might make
sense if the bat were trying to measure delays by matching the detailed waveforms of
the call and echo, since these sounds have most of their power at frequencies near
$f\sim 1/(35\, \mu{\rm sec})$---the bat can be confused by delay differences
which correspond to an integer number of periods in the acoustic waveform, and one
can even see the $n=2$ `confusion resonance' if one is careful.

The problem with these results on delay discrimination in the $1-50\,\mu{\rm sec}$
range is not that they are too precise but that they are not precise enough.  One can
measure the background acoustic noise level (or add noise so that the level is
controlled) and given this noise level a detector which looks at the detailed
acoustic waveform and integrates over the whole call should be able to estimate
arrival times much more accurately than $\sim 1\,\mu{\rm sec}$,.  Detailed calculations
show that the smallest detectable delay differences should be tens of nanoseconds.
I think this was viewed as so obviously absurd that it was grounds for throwing out the
whole idea that the bat  uses detailed waveform
information.\footnote{The alternative is that the bat bases delay estimates on the
envelope of the returning echo, so that one is dealing with structures on the
millisecond time scale, seemingly much better matched to the intrinsic time scales of
neurons.}  In an absolutely stunning development, however, Simmons and company went back
to their experiments, produced delays in the appropriate range---convincing yourself that
you have control of acoustic and electronic delays with nanosecond precision is not so
simple---and found that the bats could do what they should be able to do as ideal
detectors. Further,
they added noise in the background of the echoes and showed that performance of the
bats tracked the ideal performance over a range of noise levels.

\medskip
\addtocounter{problem}{1} 
{\small \noindent {\bf Problem \theproblem: Head movements and delay accuracy.} 
Just to be sure you understand the scale of things ... .  When bats are asked to make
``ordinary'' discriminations in the Y apparatus, they move their head from arm to
arm with each call.  How accurately would they have to reposition their head to be
sure that the second echo from one arm is not shifted by more than $\sim 1\,\mu{\rm
sec}$?  By more than $10\,{\rm nsec}$?  Explain the behavioral strategy that bats
could use to avoid this problem.  Can you position (for example) your hand with
anything like the precision that the bat needs for its head in these experiments?}
\bigskip

Returning to vision, part of the problem with photon counting is that
it almost seems inappropriate to dignify such a simple task as detecting
a  flash of light with the name ``perception.''  Barlow and colleagues
have studied a variety of problems that seem richer, in some cases
reaching into the psychology literature for examples of gestalt
phenomena---where our perception is of the whole rather than its parts
\cite{barlow80}.  One such example is the recognition of symmetry in
otherwise random patterns. Suppose that we want to make a random texture pattern.  One
way to do this is to draw the contrast $C({\bf \vec x})$ at each point ${\bf \vec x}$ in
the image from some simple probability distribution that we can write down.  An example
is to make a Gaussian random texture, which corresponds to
\begin{equation}
P[C({\bf \vec x})] \propto \exp\left[ -{1\over 2}\int d^2x \int d^2x' C({\bf \vec x})
K({\bf \vec x } - {\bf \vec x'}) C({\bf \vec x'})\right] ,
\end{equation}
where $K({\bf \vec x } - {\bf \vec x'})$ is the kernel or propagator that describe the
texture.  By writing $K$ as a function of the difference between coordinates we
guarantee that the texture is homogeneous; if we want the texture to be isotropic we
take $K({\bf \vec x } - {\bf \vec x'}) = K(|{\bf \vec x } - {\bf \vec x'}|)$.
Using this scheme, how do we make a texture with symmetry, say with respect to
reflection across an axis?

The statement that texture has symmetry across an an axis is that for each point ${\bf
\vec x}$ we can find the corresponding reflected point ${\bf R\cdot \vec x}$, and that
the contrasts at these two points are very similar; this should be true for every
point.  This can be accomplished by choosing
\begin{eqnarray}
P_\gamma [ C({\bf \vec x}) ] &\propto& 
\exp\left[ 
-{1\over 2}
\int d^2x \int d^2x'
C({\bf \vec x}) K({\bf \vec x } - {\bf \vec x'}) 
C({\bf \vec x'}) \right.
\nonumber\\ 
&& \,\,\,\,\,\,\,\,\,\,\,\,\,\,\,\,\,\,\,\, + {\gamma\over 2}  
\left. \int d^2 x | C({\bf \vec x}) - C({\bf R\cdot \vec x})|^2 \right] ,
\end{eqnarray}
where $\gamma$ measures the strength of the tendency toward symmetry.  Clearly as
$\gamma \rightarrow \infty$ we have an exactly symmetric pattern,  quenching
half of the degrees of freedom in the original random texture.  On the other hand, as
$\gamma \rightarrow 0$, the weakly symmetric textures drawn from $P_\gamma$ become
almost indistinguishable from a pure random texture ($\gamma = 0$).  Given images of a
certain size, and a known kernel $K$, there is a limit to the smallest value of $\gamma$
that can be distinguished reliably from zero, and we can compare this statistical
limit to the performance of human observers.  This is more or less what Barlow did,
although he used blurred random dots rather than the Gaussian textures considered here;
the idea is the same (and must be formally the same in the limit of many dots).  The
result is that human observers come within a factor of two of the statistical limit
for detecting $\gamma$ or its analog in the random dot patterns.

One can use similar sorts of visual stimuli to think about motion, where rather than
having to recognize a match between two halves of a possibly symmetric image we have
to match successive frames of a movie. Here again human observers can
approach the statistical limits \cite{barlow+tripathy}, as long as we stay in the
right regime:  we seem not to make use of fine dot positioning (as would be generated
if the kernel $K$ only contained low order derivatives) nor can we integrate
efficiently over many frames. These results are interesting because they show the
potentialities and limitations of optimal visual computation, but also because the
discrimination of motion in random movies is one of the places where people have tried
to make close links between perception and neural activity in the (monkey) cortex
\cite{newsome-review}.    In addition to symmetry and motion, other examples of
optimal or near optimal performance include other visual texture discriminations and
auditory identification of complex pitches in the auditory system; even  bacteria
can approach the limits set by physical noise sources as they detect and react to
chemical gradients, and there is a species of French cave beetle that can sense
milliKelvin temperature changes, almost at the limit set by thermodynamic temperature
fluctuations in their sensors.  

I would like to discuss one case in detail, because it shows how
much we can learn by stepping back and looking for a simple example (in
proper physics tradition).  Indeed, I believe that one of the crucial
things one must do in working at the interface of physics and biology is
to take some particular biological system and dive into the details. 
However much we believe in general principles, we have to confront
particular cases.  In thinking about brains it would be nice to have some
``simple system'' that we can explore, although one must admit that the
notion of a simple brain seems almost a non--sequitur. Humans tend to be
interested in the brains of other humans, but as physicists we know
that we are not at the center of the universe, and we might worry
that excessive attention to our own brains reflects a sort of
preCopernican prejudice.  It behooves us, then, to look around the
animal kingdom for  accessible examples of what we hope are
general phenomena.  For a variety of reasons, our attention is
drawn to invertebrates---animals without backbones---and to
insects in particular.

First, most of the animals on earth are insects, or, more precisely,
arthropods. Second, the nervous system of a typical invertebrate has far
fewer neurons than found in a mammal or even a `lower vertebrate' such as
a fish or frog.  The fly's entire visual brain has roughly $5\times 10^5$
cells, while just the primary visual cortex of a monkey has $\sim 10^9$.
Third, many of the cells in the invertebrate nervous system are {\it
identified}:  cells of essentially the same structure
occur in every individual, and that if one records the response of these
cells to sensory stimuli (for example) these responses are reproducible
from individual to individual.\footnote{This should not be taken to mean
that the properties of neurons in the fly's brain are fixed by genetics,
or that all individuals in the species are identical.  Indeed, we will
come to the question of individuality vs. universality in what follows. 
What is important here is that neural responses are sufficiently
reproducible that one can speak meaningfully about the properties of
corresponding cells in different individuals.} Thus the cells can be
named and numbered based on their structure or function in the neural
circuit.\footnote{If you want to know more about the structure of a fly's
brain, there is a beautiful book by Strausfeld \cite{atlas}, but this is
very hard to find. An alternative is the online flybrain project that
Strausfeld and others have been building \cite{flybrain}.
Another good general reference is the collection of articles edited by 
Stavenga and Hardie \cite{facets}.} Finally, the
overall physiology of invertebrates allows for very long, stable
recordings of the electrical activity of their neurons. In short, experiments on
invertebrate nervous systems look and feel like the physics experiments
with which we all grew up---stable samples with quantitatively
reproducible behavior.

Of course what I give here is meant to sound like a rational account of
why one might choose the fly as a model system.  In fact my own choice
was driven by the good fortune of finding myself as a postdoc in
Groningen some years ago, where in the next office Rob de Ruyter van
Steveninck was working on his thesis.  When Rob showed me the kinds of
experiments he could do---recording from a single neuron for a week, or
from photoreceptor cells all afternoon---we both realized that this was a
golden opportunity to bring theory and experiment together in studying a
variety of problems in neural coding and computation.  Since this is a
school, and hence the lectures have in  part the flavor of advice to
students, I should point out that (1) the whole process of
theory/experiment collaboration was made easier by the fact that Rob
himself is a physicist, and indeed the Dutch were then far ahead of the
rest of the world in bringing biophysics into physics departments, but
(2) despite all the positive factors, including the fact that as postdoc
and student we had little else to do,
it still took months for us to formulate a reasonable first step in our
collaboration.  I admit that it is only after having some success 
 with Rob that I have have had the courage to venture into
collaborations with real biologists.

What do flies actually do with their visual brains?  If you  watch a fly
flying around in a room or outdoors, you will notice that flight paths
tend to consist of rather straight segments interrupted by sharp turns. 
These observations can be quantified through the measurement of
trajectories during free  \cite{land+collett74,wagner} or lightly
tethered \cite{hans-flight1,hans-flight2} flight and in experiments where
the fly is suspended from a torsion balance \cite{reichardt+poggio}.
Given the aerodynamics for an object of the fly's dimensions, even flying
straight is tricky. In the torsion balance one can demonstrate directly
that motion across the visual field drives the generation of torque, and
the sign is such as to stabilize flight against rigid body rotation of the
fly. Indeed one can close the feedback loop by measuring  the torque
which the fly produces and using this torque to (counter)rotate the
visual stimulus, creating an imperfect `flight simulator' for the fly in
which the only cues to guide the flight are visual; under natural
conditions the fly's mechanical sensors play a crucial role.  Despite the
imperfections of the flight simulator, the tethered fly will fixate small
objects, thereby stabilizing the appearance of straight flight. 
Similarly, Land and Collett  showed that aspects of flight behavior under
free flight conditions can be understood if flies generate torques in
response to motion across the visual field, and that this response is
remarkably fast, with a latency of just $\sim 30$ msec
\cite{land+collett74}.   The combination of free flight and torsion
balance experiments strongly suggests that flies can estimate  their
angular velocity from visual input alone, and then produce motor outputs
based on this estimate \cite{reichardt+poggio}.

When you look down on the head of a fly, you see---almost to the
exclusion of anything else---the large compound eyes.  
Each little hexagon  that you see on the fly's head is a separate lens, and in
large flies  there are
$\sim 5,000$ lenses in each eye, with approximately 1 receptor cell behind each
lens,\footnote{This is the sort of sloppy physics speak which annoys biologists.  The
precise statement is different in different insects.  For flies there are
eight receptors behind each lens.  Two provide sensitivity to
polarization and some color vision, but these are not used for motion
sensing.  The other six receptors look out through the same lens in different
directions, but as one moves to neighboring lenses one finds that there is one
cell under each of six neighboring lenses which looks in the same direction. 
Thus these six cells are equivalent to one cell with six times larger photon
capture cross section, and the signals from these cells are collected and
summed in the first processing stage (the lamina). One can even see the
expected six fold improvement in
signal to noise ratio \cite{deruyter96}.} and roughly 100 brain cells per
lens devoted to the processing of visual information. The lens focuses
light on the receptor, which is small enough to act as an optical
waveguide. Each receptor sees only a small portion of the world,
just as in our eyes; one difference between flies and us is that diffraction
is much more significant for organisms with compound eyes---because
the lenses are so small, flies have
an angular resolution of about $1^\circ$, while we do about 100$\times$
better.  There is a beautiful literature on optimization principles for the
design of the compound eye; the topic even makes an appearance in the Feynman
lectures.

Voltage signals from the receptor cells
are processed by several layers of the brain, each layer having cells
organized on a lattice which parallels the lattice of lenses visible
from the outside of the fly.
After passing through the
lamina, the medulla, and the lobula, signals arrive at the lobula plate.
Here there is a stack of about
50 cells which are are sensitive to motion \cite{hausen84,hausen89}.
The cells all are identified in the sense defined above, and are specialized to detect
different kinds of motion.
If one kills individual cells in the lobula plate then the
simple experiment of moving a stimulus and recording
the flight torque no longer works \cite{hausen83}, strongly
suggesting that these cells are an obligatory link in the pathway from the
retina to the flight motor.  If one lets the fly watch a randomly moving
pattern, then it is possible to ``decode'' the responses of the movement
sensitive neurons and to reconstruct the time dependent angular
velocity signal \cite{bialek91}, as will be discussed below.
Taken together, these observations support a picture in which the fly's
brain uses photoreceptor signals to estimate angular velocity,
and encodes this estimate in the activity of a few neurons.\footnote{Let
me emphasize that you should be skeptical of any specific claim about
what the brain computes.  The fact that flies can stabilize their flight
using visual cues, for example, does {\em not} mean that they compute
motion in any precise sense---they could use a form of `bang--bang'
control that needs knowledge only of the algebraic sign of the velocity,
although I think that the torsion balance experiments argue against such
a model. It also is a bit mysterious why we find neurons with such
understandable properties:  one could imagine   connecting photoreceptors
to flight muscles via a network of neurons in which there is nothing
that we could  recognize as a motion--sensitive cell.  Thus it
is not obvious either that the fly must compute motion or that there must
be motion--sensitive neurons (one might make the same argument about
whether there needs to be a whole area of motion--sensitive neurons in
primate cortex, as observed).  As you will see, when the dust settles I
will claim that flies in fact compute motion {\em optimally.} The direct
evidence for this claim comes from careful examination of the
responses of single neurons.  We don't know why the fly goes to the
trouble of doing this, and in particular it is hard to point to a
behavior in which this precision has been demonstrated experimentally (or
is plausibly necessary for survival). This is a first example of the
laundry list problem:  if the brain makes optimal estimates of $x,y$ and
$z,$ then we have an opening to a principled theory of $x-,$ $y-,$ and
$z-$perception and the corresponding neural responses, but we don't know
why the system chooses to estimate $x,y,z$ as opposed to $x',y',$ and
$z'$.  Hang in there ... we'll try to address this too!}
Further, we can study the photoreceptor signals (and noise) as well as
the responses of motion--sensitive neurons with a precision almost
unmatched in any other set of neurons:  thus we have a good idea of what
the system is ``trying'' to do, and we have tremendous access to both
inputs and outputs.   I'll try to make several points:
\begin{itemize}
\item Sequences of a few action spikes from the H1 neuron allow for discrimination among
different motions with a precision close to the limit set by noise in the photodetector
array.
\item With continuous motion, the spike train of H1 can be decoded to recover a running
estimate of the motion signal, and the precision of this estimate is again within a
factor of two of the theoretical limit.
\item Analogies between the formal problem of optimal estimation and
statistical physics help us to develop a theory of optimal estimation which predicts the
structure of the computation that the fly {\em should} do to achieve the most reliable
motion estimates.
\item In several cases we can find independent evidence that the fly's computation has
the form predicted by optimal estimation theory, including some features of the neural
response that were found only after the theoretical predictions.
\end{itemize}
We start by getting an order--of--magnitude feel for the theoretical limits.

Suppose that we look at a pattern of typical contrast $C$ and it moves by an angle
$\delta \theta$.   A single photodetector element will see a change in contrast of
roughly $\delta C \sim C\cdot (\delta\theta/\phi_0)$, where $\phi_0$ is the angular
scale of blurring due to diffraction.  If we can measure for a time $\tau$, we will
count an average number of photons $R\tau$, with $R$ the counting rate per detector,
and hence the noise can be expressed a fractional precision in intensity of $\sim
1/\sqrt{R\tau}$.  But fractional intensity is what we mean by contrast,
so $1/\sqrt{R\tau}$ is really the contrast noise in one photodetector.  To get the
signal to noise ratio we should compare the signal and noise in each of the
$N_{\rm cells}$ detectors, then add the squares if we assume (as for photon shot noise)
that noise is independent in each detector while the signal is coherent.  The result is
\begin{equation}
SNR \sim N_{\rm cells} \cdot \left({{\delta\theta}\over{\phi_0}}\right)^2 C^2 R\tau  .
\label{thetaSNR}
\end{equation}
This calculation is  rough,  and we can do a little better \cite{bialek92,santafe}, but
it contains the right ideas.  Motion discrimination is hard for flies because they have
small lenses and hence  blurry images ($\phi_0$ is large) and because they have to
respond quickly ($\tau$ is small).  Under reasonable laboratory conditions the optimal
estimator would reach $SNR =1$ at an angular displacement of $\delta\theta
\sim 0.05^\circ$.

We can test the precision of motion estimation in two very different ways.  One is
similar to the experiments described for photon counting or for bat echolocation:  we
create two alternatives and ask if we can discriminate reliably.  For the motion
sensitive neurons in the fly visual system Rob pursued this line by recording the
responses of a single neuron (H1, which is sensitive, not surprisingly, to horizontal
motions) to steps of motion that have either an amplitude $\theta_+$ or an amplitude
$\theta_-$ \cite{step-disc}.  The cell responds with a brief volley of
action potentials which we can label as occurring at times $t_1, t_2,
\cdots $.  We as observers of the neuron can look at these times and try
to decide whether the motion had amplitude $\theta_+$ or $\theta_-$; the
idea is exactly the same as in Problem 2, but here we have to measure the distributions
$P_\pm (t_1, t_2, \cdots )$ rather than making assumptions about their form.  Doing the
integrals, one finds that looking at spikes generated in the first
$\sim 30\,{\rm msec}$ after the step (as in the fly's behavior) we can reach
the reliability expected for $SNR = 1$ at a displacement $\delta\theta =
|\theta_+ -
\theta_-| \sim 0.12^\circ$, within a factor of two of the theoretical limit set by
noise in the photodetectors.  These data are quite rich, and it is worth noting a few
more points that emerged from the analysis:
\begin{itemize}
\item On the $\sim 30\,{\rm msec}$ time scale of relevance to behavior, there are only
a handful of spikes.  This is partly what makes it possible to do the analysis so
completely, but it also is a lesson for how we think about the neural representation of
information in general.
\item Dissecting the contributions of individual spikes, one finds that each successive
spike makes a nearly independent contribution to the signal to noise ratio for
discrimination, so there is essentially no redundancy.
\item Even one or two spikes are enough to allow discrimination of motions much smaller
than the lattice spacing on the retina or the nominal ``diffraction limit'' of angular
resolution. Analogous phenomena have been known in human vision for more than a
century and are called hyperacuity; see Section 4.2 in {\em Spikes} for a discussion
\cite{spikes}.
\end{itemize}

The step discrimination experiment gives us a very clear view of reliability in the
neural response, but as with the other discrimination experiments discussed above it's
not a very natural task.  An alternative is to ask what happens when the motion
signal (angular velocity $\dot\theta(t)$) is a complex function of time.  Then we can
think of the signal to noise ratio in Eq. (\ref{thetaSNR}) as being equivalent to a
spectral density of displacement noise $N_\theta^{\rm eff}
\sim \phi_0^2/(N_{\rm cells} C^2 R)$, or a  generalization in which the photon
counting rate is replaced by an effective (frequency dependent) rate related
to the noise characteristics of the photoreceptors \cite{santafe}.  It seems likely,
as discussed above, that the fly's visual system really does make a continuous or
running estimate of the angular velocity, and that this estimate is encoded in the
sequence of discrete spikes produced by neurons like H1.  It is not clear that any
piece of the brain ever ``decodes'' this signal in an explicit way, but if {\em we}
could do such a decoding we could test directly whether the accuracy of our decoding
reaches the limiting noise level set by noise in the photodetectors.  

The idea of using spike trains to recover continuous time dependent signals started
with this analysis of the fly visual system \cite{bialek91,88,bialek+zee}, and has since
expanded to many different systems \cite{spikes}.  Generalizations of these ideas to
decoding from populations of neurons \cite{warland-pop} even have application to
future prosthetic devices which might be able to decode the commands given in motor
cortex to control robot arms \cite{miguel}.  Here our interest is not so much in the
structure of the code, or in the usefulness of the decoding; rather our goal is to use
the decoding as a tool to characterize the precision of the underlying computations.

To understand how we can decode a continuous signal from discrete sequences of spike it
is helpful to have an example, following Ref. \cite{bialek+zee}.  Suppose 
that the signal of interest is
$s(t)$ and the neuron we are looking at generates action potentials according to a
Poisson process with a rate $r(s)$.  Then the probability that we observe spikes at
times $t_1, t_2, \cdots, t_N \equiv \{t_i\}$ given the signal $s(t)$ is (generalizing
from Problem 1) 
\begin{equation}
P[\{t_i\} | s(t)] = {1\over {N!}} \exp\left[ -\int dt \,r(s(t)) \right]
\prod_{i=1}^N r(s(t_i)) .
\end{equation}
For simplicity let us imagine further that the signal $s$ itself comes from a Gaussian
distribution with zero mean, unit variance and a correlation time $\tau_c$, so that
\begin{equation}
P[s(t)] \propto
\exp\left[ -{1\over {4\tau_c}} \int dt \left( \tau_c^2 {\dot s}^2 + s^2\right)\right] .
\end{equation}
Our problem is not to predict the spikes from the signal, but rather given the
spikes to estimate the signal, which means that we are interested in conditional
distribution $P[s(t) | \{t_i\}]$.  From Bayes' rule,
\begin{eqnarray}
P[s(t) | \{ t_i \}] 
&=&
{ {P[\{ t_i \} | s(t)]P[s(t)]} \over {P[\{t_i\}]} } \\
& \propto &
\exp\left[ -{{\tau_c}\over {4}} \int dt \,{\dot s}^2 -\int dt V_{\rm eff}(s(t))
\right] \prod_{i=1}^N r(s(t_i)) ,\nonumber\\
&&\end{eqnarray}
where 
\begin{equation}
V_{\rm eff}(s) = {1\over {4\tau_c}} s^2 + r(s) .
\end{equation}
I write the probability distribution in this form to remind you of the (Euclidean) path
integral description of quantum mechanics, where the amplitude for a particle of mass
$m$ to move along a trajectory $x(t)$ in a potential $V(x)$ is given by
\begin{equation}
A[x(t)]\propto \exp\left[ -{m\over{2}}\int dt \,{\dot x}^2 - \int dt V(x(t))\right] ,
\end{equation}
in units where $\hbar =1$.  If we want to estimate $s(t)$ from the probability
distribution $P[s(t) | \{t_i\}]$, then we can compute the conditional mean, which will
give us the best estimator in the sense that $\chi^2$ between our estimate and the true
signal will be minimized (see Problem 5 below). Thus the estimate at some
particular time
$t_0$ is given by
\begin{eqnarray}
s_{\rm est}(t_0) &=& \int Ds(t)\, s(t_0) P[s(t) | \{t_i\}]\\
&\propto&
\Bigg\langle s(t_0) \prod_{i=1}^N r(s(t_i)) \Bigg\rangle ,
\label{est-exp}
\end{eqnarray}
where $\langle \cdots \rangle$ stands for an expectation value over trajectories drawn
from the distribution
\begin{equation}
P_{\rm eff}[s(t)]\propto \exp\left[ -{{\tau_c}\over {4}} \int dt \,{\dot s}^2 -\int dt
V_{\rm eff}(s(t))
\right] .
\end{equation}
Thus estimating the trajectory involves computing an $N+1$--point function in the
quantum mechanics problem defined by the potential $V_{\rm eff}$.

\medskip
\addtocounter{problem}{1} 
{\small \noindent {\bf Problem \theproblem: Optimal estimators.}  
Imagine that you observe $y$, which is related to another variable $x$ that
you actually would like to know.  This relationship can be described by the joint
probability distribution $P(x,y)$, which you know.  Any estimation strategy can be
described as computing some function $x_{\rm est} = F(y)$.  For any estimate we can
compute the expected value of $\chi^2$,
\begin{equation}
\chi^2 = \int dx dy P(x,y) | x - F(y)|^2 .
\end{equation} 
Show that the estimation strategy which minimizes $\chi^2$ is the computation of the
conditional mean,
\begin{equation}
F_{\rm opt} (y) = \int dx \, x P(x|y) .
\end{equation}
\label{condmean}}
\bigskip

If we took any particular model seriously we could in fact try to compute the relevant
expectation values, but here (and in applications of these ideas to analysis
of real neurons) I don't really want to trust these details; rather I want to focus on
some general features.  First one should note that trying to do the calculations in a
naive perturbation theory will work only in some very limited domain.  Simple forms of
perturbation theory are equivalent to the statement that interesting values of the
signal $s(t)$ do not sample strongly nonlinear regions of the input/output relation
$r(s)$, and this is unlikely to be true in general.  On the other hand, there is a
chance to do something simple, and this is a cluster expansion:
\begin{eqnarray}
\Bigg\langle s(t_0) \prod_{i=1}^N r(s(t_i)) \Bigg\rangle
&\approx&
\langle s(t_0) \rangle \prod_{i=1}^N \langle r(s(t_i)) \rangle
\nonumber\\
&&\,\,\,\,\,\,\,\,\,\, 
+ A \sum_{i=1}^N \langle \delta s(t_0) \delta r(s(t_i))\rangle
+ \cdots ,
\end{eqnarray}
where $\delta s$ refers to fluctuations around around the mean $\langle s\rangle$, and
similarly for $r$, while $A$ is a constant.  As with the usual cluster expansions in
statistical physics, this does not require perturbations to be weak; in particular the
relation
$r(s)$ can have arbitrarily strong (even discontinuous) nonlinearities. What is needed
for a cluster expansion to make sense is that the times which appear in the $N$--point
functions be far apart when measured in units of the correlation times for the underlying
trajectories.  In the present case, this will happen (certainly) if the times between
spikes are long compared with the correlation times of the signal.  Interestingly, as
the mean spike rate gets higher, it is the correlation times computed in the full
$V_{\rm eff}(s)$ which are important, and these are  smaller than the bare
correlation time in many cases. At least in this class of models, then, there is a
regime of low spike rate where we can use a cluster expansion, and this regime extends
beyond the naive crossover determined by the number of spikes per correlation time
of $s$.  As explained in {\em Spikes,} there are good reasons to think
that many neurons actually operate in this limit of spike trains which are ``sparse in
the time domain'' \cite{spikes}.

What are the consequences of the cluster expansion?  If we can get away with keeping the
leading term, and if we don't worry about constant offsets in our estimates (which can't
be relevant ... ), then we have
\begin{equation}
s_{\rm est}(t) = \sum_{i=1}^N f(t-t_i) + \cdots , 
\label{linest}
\end{equation}
where $f(t)$ again is something we could calculate if we trusted the details of the
model.  But this is very simple:  we can estimate a continuous time dependent signal
just by adding up contributions from individual spikes, or equivalently by filtering the
sequence of spikes.  If we don't want to trust a calculation of $f(t)$ we can just use
experiments to find $f(t)$ by asking for that function which gives us the smallest value
of $\chi^2$ between our estimate and the real signal, and this optimization problem is
also simple since $\chi^2$ is quadratic in $f$ \cite{spikes,bialek91}.  So the path is
clear---do a long experiment on the response of a neuron to signals drawn from some
distribution $P[s(t)]$, use the first part of the experiment to find the filter $f(t)$
such that the estimate in Eq. (\ref{linest}) minimizes $\chi^2$,
and then test the accuracy of our estimates on the remainder of the data.  This is
exactly what we did with H1 \cite{bialek91}, and we found that over a broad range of
frequencies the spectral density of errors in our estimates was within a factor of two
of the limit set by noise in the photoreceptor cells.  Further, we could change, for
example, the image contrast and show that the resulting error spectrum scaled as
expected from the theoretical limit \cite{spikes}.

To the extent that the fly's brain can estimate motion with a precision close to the
theoretical limit, one thing we know is that the act of processing itself does not add
too much noise.  But being quiet is not enough:  to make maximally reliable estimates of
nontrivial stimulus features like motion one must be sure to do the correct computation.
To understand how this works let's look at a simple example, and then I'll outline what
happens when we use the same formalism to look at motion estimation. My discussion here
follows joint work with Marc Potters \cite{marc}.

Suppose that someone draws a random number $x$ from a probability distribution $P(x)$. 
Rather than seeing $x$ itself, you get to see only a noisy version, $y = x +\eta$, where
$\eta$ is drawn from a Gaussian distribution with variance $\sigma^2$.  Having seen $y$,
your job is to estimate $x$, and for simplicity let's say that the ``best estimate'' is
best in the least squares sense, as above.  Then from Problem 5 we know
that the optimal estimator is the conditional mean,
\begin{equation}
x_{\rm est}(y)  = \int dx \, x P(x|y).
\end{equation}
Now we use Bayes' rule and push things through:
\begin{eqnarray}
x_{\rm est}(y)  &=& \int dx \, x P(y|x)P(x) {1\over{P(y)}}\\
&=& {1\over{P(y)}} {1\over{\sqrt{2 \pi \sigma^2}}} \int dx \, x P(x) \exp\left[
-{1\over{2\sigma^2}} ( y - x)^2 \right]\\
&=&
{1\over {Z(y)}} \int dx \, x \exp\left[ - {{V_{\rm eff}(x)}\over{k_B T_{\rm eff}}}
+ {{F_{\rm eff} x }\over{k_B T_{\rm eff}}} \right] ,
\end{eqnarray}
where we can draw the analogy with statistical mechanics by noting the correspondences:
\begin{eqnarray}
{{V_{\rm eff}(x)}\over{k_B T_{\rm eff}}}  &=& -\ln P(x) + {{x^2}\over{2\sigma^2}}\\
k_B T_{\rm eff} &=& \sigma^2\\
F_{\rm eff} &=& y .
\end{eqnarray}
Thus, making optimal estimates involves computing expectation values of position, the
potential is (almost) the prior distribution, the noise level is the temperature, and
the data act as an external force.
The connection with statistical mechanics is more than a convenient analogy; it helps us
in finding approximation schemes.  Thus at large noise levels we are at high
temperatures, and all other things being equal the force is effectively small so we can
compute the response to this force in perturbation theory.  On the other hand, at low
noise levels the computation of expectation values must be dominated by ground state or
saddle point configurations.  

\begin{table}
\begin{tabular}{||l|l||}\hline
{\bf Statistical mechanics} & {\bf Optimal Estimation} \\\hline
Temperature & Noise level \\\hline
Potential & (log) Prior distribution\\\hline
Average position & Estimate \\\hline
External force & Input data \\\hline
\end{tabular}
\end{table}

In the case of visual motion estimation, the data are the voltages produced by
the photoreceptors $\{ V_n (t)\}$ and the thing we are trying to estimate is the angular
velocity $\dot\theta (t)$. These are functions rather than
numbers, but this isn't such a big deal if we are used to doing functional integrals. 
The really new point is that $\dot\theta(t)$ does not directly determine the voltages. 
What happens instead is that even if the fly flies along a perfectly straight path, so
that $\dot\theta =0$, the world effectively projects a movie $C({\bf x},t)$ onto the
retina and it is this movie which determines the voltages (up to noise).  The motion
$\theta(t)$ transforms this movie, and so enters in a strongly nonlinear way; if we take
a one dimensional approximation we might write $C(x,t) \rightarrow C(x-\theta(t),t)$.
Each photodetector responds to the contrast as seen through an aperture function
$M(x-x_n)$ centered on a lattice point $x_n$ in the retina, and for simplicity let's
take the noise to be dominated by photon shot noise. Since we have independent
knowledge of the $C(x,t)$, to describe the relationship between $\dot \theta (t)$
and $\{ V_n(t)\}$ we have to integrate over all possible movies, weighted by
their probability of occurrence $P[C]$.

Putting all of these things together with our general
scheme for finding optimal estimates, we have
\begin{eqnarray}
\dot\theta_{\rm est}(t_0) &=& \int D\theta \,\dot\theta (t_0) P[\theta(t) | \{ V_n
(t)\}] 
\\
P[\theta(t) | \{ V_n (t)\}] &=& P[\{ V_n (t)\} | \theta(t)] P[\theta(t)] {1\over
{P[\{ V_n (t)\}]}}\\P[\{ V_n (t)\} | \theta(t)] &\propto& \int DC \, P[C] \exp\left[ 
-{R\over 2} \sum_n \int dt \,
|V_n(t) - {\bar V}_n(t)|^2
\right]\nonumber\\
&&\\
{\bar V}_n(t) &=& \int dx M(x-x_n) C(x-\theta(t),t) .
\end{eqnarray}
Of course we can't do these integrals exactly, but we can look for approximations, and
again we can do perturbation theory at low signal to noise levels and search for saddle
points at high signal to noise ratios.
When the dust settles we find \cite{marc}:
\begin{itemize}
\item At low signal to noise ratios the optimal estimator is quadratic in the receptor
voltages,
\begin{equation}
\dot\theta_{\rm est}(t) \approx \sum_{nm} \int d\tau \int d\tau' V_n(t-\tau) K_{nm}
(\tau,
\tau') V_m (t-\tau')  .
\end{equation}
\item  At moderate signal to noise ratios,   terms with higher powers of the
voltage become relevant and `self energy' terms provide corrections to the  kernel
$K_{nm}(\tau, \tau ')$.
\item At high signal to noise ratios averaging over time becomes less important and the
optimal estimator crosses over to
\begin{equation}
\dot\theta_{\rm est}(t) \approx 
{{\sum_n {\dot V}_n(t) [V_n(t) - V_{n-1}(t)]}\over{{\rm constant} + \sum_n [V_n(t) -
V_{n-1}(t)]^2}} ,
\end{equation}
where ${\rm constant}$ depends on the typical contrast and dynamics in the movies chosen
from $P[C(x,t)]$ and on the typical scale of angular velocities in $P[\theta(t)]$.
\end{itemize}
Before looking at the two limits in detail, note that the whole form of the motion
computation depends on the statistical properties of the visual environment.  Although
the limits look very different, one can show that there is no phase transition and hence
that increasing signal to noise ratio takes us smoothly from one limit to the other;
although this is sort of a side point, it was a really nice piece of work by Marc.  An
obvious question is whether the fly is capable of doing these different computations
under different conditions.

We can understand the low signal to noise ratio limit by realizing that when something
moves there are correlations between what we see at the two space--time points $(x,t)$
and  $(x+v\tau, t+\tau)$.  These correlations extend to very high
orders, but as the background noise level increases  the higher order correlations
are corrupted first, until finally the only reliable thing left is the two--point
function, and closer examination shows that near neighbor correlations are the most
significant:  we can be sure something is moving because signals in neighboring
photodetectors are correlated with a slight delay.   This form of ``correlation
based'' motion computation was suggested long ago by Reichardt and Hassenstein based on
behavioral experiments with beetles \cite{reichardt61}; later work from Reichardt and
coworkers explored the applicability of this model to fly behavior
\cite{reichardt+poggio}.  Once the motion sensitive neurons were discovered it was
natural to check if their responses could be understood in these terms. 

There are two clear signatures of the correlation model. 
First, since the receptor voltage is linear in response to image contrast, the
correlation model confounds contrast with velocity:  all things being equal, doubling
the image contrast causes our estimate of the velocity to increase by a factor of four
(!).  This is an observed property of the flight torque that flies generate in response
to visual motion, at least at low contrasts, and the same quadratic behavior can be seen
in the rate at which motion sensitive neurons generate spikes and even in human
perception (at very low contrast).   Although this might seem strange, it's been known
for decades.  What is interesting here is that this seemingly disastrous confounding of
signals occurs even in the optimal estimator: optimal estimation involves a
tradeoff between systematic and random errors, and at low signal to noise ratio this
tradeoff pushes us toward making larger systematic errors, apparently of a form made by
real brains.

The second signature of correlation computation is that we can   produce movies
which have the right spatiotemporal correlations to generate a nonzero estimate
$\dot\theta_{\rm est}$ but don't really have anything in them that we would describe as
``moving'' objects or features.  Rob de Ruyter has  a simple recipe for doing this
\cite{rob-neclects}, which is quite compelling (I recommend you try it yourself):  Make a
spatiotemporal white noise movie $\psi({\bf
\vec x},t)$,
\begin{equation}
\langle \psi({\bf \vec x},t) \psi({\bf \vec x'},t')\rangle
= \delta ({\bf \vec x} - {\bf \vec x'}) \delta (t - t'),
\end{equation}
and then add the movie to itself with a weight and an offset:
\begin{equation}
C({\bf\vec x},t) = \psi({\bf \vec x},t) + a \psi({\bf \vec x + \Delta\vec x},t +\Delta
t) .
\end{equation}
Composed of pure noise, there is nothing really moving here. If you watch the movie,
however, there is no question that you think it's moving, and the fly's neurons respond
too (just like yours, presumably).  Even more impressive is that if you change the {\em
sign} of the weight $a$ ... the direction of motion reverses, as predicted from the
correlation model.

Because the correlation model has a long history, it is hard to view evidence for this
model as a success of optimal estimation theory. The theory of optimal estimation also
predicts, however, that the kernel $K_{nm}(\tau, \tau')$ should adapt to the statistics
of the visual environment, and it does.  Most dramatically one can just show random
movies with different correlation times and then probe the transient response of H1 to
step motions; the time course of transients (presumably reflecting the details of $K$)
can vary over nearly two orders of magnitude from 30--300 msec in response to different
correlation times \cite{rob-neclects}.  All of this makes sense in light of optimal
estimation theory but again perhaps is not a smoking gun.  Closer to a real test is
Rob's argument that the absolute values of the time constants seen in such adaptation
experiments match the frequencies at which natural movies would fall below $SNR=1$ in
each photoreceptor, so that filtering in the visual system is set (adaptively)  to fit
the statistics of signals and noise in the inputs \cite{rob-neclects,rob-filts}.

What about the high SNR limit? If we remember that voltages are linear
in contrast, and let ourselves ignore the lattice in favor of a continuum limit, then the
high SNR limit has a simple structure,
\begin{equation}
\dot\theta_{\rm est}(t) \approx {{\int dx\, (\partial_t C)(\partial_x C)}\over{
B + \int dx\, (\partial_x C)^2 }}
\rightarrow {{\partial_t C}\over{\partial_x C}} ,
\end{equation}
where the last limit is at high contrasts.  As promised by the lack of a phase
transition, this starts as a quadratic function of contrast just like the correlator,
but saturates as the ratio of temporal and spatial derivatives.  Note that if $C(x,t) =
C(x+vt)$, then this ratio recovers the velocity $v$ exactly.  This
simple ratio computation is not optimal in general because real movies have dynamics
other than rigid motion and real detectors have noise, but there is a limit in which it
must be the right answer.  Interestingly, the correlation model (with multiplicative
nonlinearity) and the ratio of derivatives model (with a divisive nonlinearity) have
been viewed as mutually exclusive models of how brains might compute motion.  One of the
really nice results of optimal estimation theory is to see these seemingly very
different models emerge as different limits of a more general strategy.  But, do flies
know about this more general strategy?

The high SNR limit of optimal estimation predicts that the motion estimate (and hence,
for example, the rate at which motion sensitive neurons generate spikes) should
saturate as a function of contrast, but this contrast--saturated level should vary with
velocity.  Further, if one traces through the calculations in more detail, the
the constant $B$ and hence the contrast level required for (e.~g.) half--saturation
should depend on the typical contrast and light intensity in the environment.  Finally,
this dependence on the environment really is a response to the statistics of that
environment, and hence the system must use some time and a little `memory' to keep track
of these statistics---the contrast response function should reflect the statistics of
movies in the recent past.  All of these things are observed
\cite{rob-neclects,rob-filts,rob-smc}.

So, where are we?  The fly's visual system makes nearly optimal estimates of motion under
at least some conditions that we can probe in the lab.  The theory of optimal estimation
predicts that the structure of the motion computation ought to have some surprising
properties, and many of these are observed---some were observed only when theory said to
go look for them, which always is better.  I would like to get a clearer demonstration
that the system really takes a ratio, and I think we're close to doing that
\cite{features}.   Meanwhile, one might worry that theoretical predictions depend too
much on assumptions about the structure of the relevant probability distributions $P[C]$
and $P[\theta ]$, so Rob is doing experiments where he walks through the woods with both
a camera and gyroscope mounted on his head (!), sampling the joint
distribution of movies and motion trajectories.  Armed with these samples one can do all
of the relevant functional integrals by Monte Carlo, which really is lovely since now we
are in the real natural distribution of signals.  I am optimistic that all of this will
soon converge on a complete picture.  I also believe that the success so far is
sufficient to motivate a more general look at the problem of optimization as a design
principle in neural computation.

\section{Toward a general principle?}

One attempt to formulate a general principle for neural computation goes
back to the work of Attneave \cite{attneave54} and Barlow
\cite{barlow59,barlow61} in the 1950s.  Focusing on the processing of
sensory information, they suggested that an important goal of neural
computation is to provide an efficient representation of the incoming
data, where the intuitive notion of efficiency could be made precise
using the ideas of information theory \cite{shannon48}.  

Imagine describing an image by giving the light intensity
in each pixel.  Alternatively, we could give a description in terms of
objects and their locations.  The latter description almost certainly is
more efficient in a sense that can be formalized using information
theory.   The idea of Barlow and Attneave  was to turn this
around---perhaps by searching for maximally efficient representations of
natural scenes we would be forced to discover and recognize the objects
out of which our perceptions are constructed.  Efficient representation
would have the added advantage of allowing the communication of
information from one brain region to another (or from eye to brain along
the optic nerve) with a minimal number of nerve fibers or  even a minimal
number of action potentials.   How could we test these ideas?
\begin{itemize}
\item If we make a model for the class of computations that neurons can
do, then we could try to find within this class the one computation that
optimizes some information theoretic measure of performance.  This should
lead to predictions for what real neurons are doing at various stages of
sensory processing.
\item We could try to make a direct measurement of the efficiency with
which neurons represent sensory information.
\item Because efficient representations depend on the statistical
structure of the signals we are trying to represent, a truly efficient brain
would adapt its strategies to track changes in these statistics, and we
could search for this ``statistical adaptation.'' Even better would be if
we could show that the adaptation has the right form to optimize
efficiency.
\end{itemize}

Before getting started on this program we need a little review of
information theory itself.\footnote{At Les Houches this review was
accomplished largely by handing out notes based on courses given at
MIT, Princeton and ICTP in 1998--99. In principle they will
become part of a book to be published by Princeton University Press,
tentatively titled {\em Entropy, Information and the Brain}.  I include
this here so that the presentation is self--contained, and apologize for
the eventual self--plagiarism that may occur.} Almost all statistical
mechanics textbooks note that the entropy of a gas measures our lack of
information about the microscopic state of the molecules, but often this
connection is left a bit vague or qualitative. Shannon proved a theorem
that makes the connection precise
\cite{shannon48}:  entropy is the unique measure of available information
consistent with certain simple and plausible requirements. Further,
entropy also answers the practical question of how much space we need to
use in writing down a description of the signals or states that we
observe. 

Two friends, Max and Allan, are having  a conversation.
In the course of the conversation, Max asks Allan what he thinks of the
headline story in this morning's newspaper.   We have the clear intuitive
notion that Max will `gain information' by hearing the answer to his
question, and we would like to quantify this intuition.  Following
Shannon's reasoning, we begin by assuming that Max knows Allan very 
well.  Allan speaks very proper English, being careful to follow the 
grammatical rules even in casual conversation. Since they have had many
political discussions Max has a rather good idea about how Allan will
react to the latest news. Thus Max can make a list of Allan's possible
responses to his question, and he can assign probabilities to each of the
answers. From this list of possibilities and probabilities we can compute
an entropy, and this is done in exactly the same way as we compute  the
entropy of a gas in statistical mechanics or thermodynamics: If the
probability of the ${\rm n}^{\rm th}$ possible response is $p_{\rm n}$, 
then the entropy is 
\begin{equation}
S= - \sum_{\rm n} p_{\rm n} \log_2 p_{\rm n}\,{\rm 
bits.}
\end{equation}

The entropy $S$ measures Max's
uncertainty about what Allan will say in response to his question.  Once Allan
gives his answer, all this uncertainty is removed---one of the responses
occurred, corresponding to $p=1$, and all the others did
not,  corresponding to $p=0$---so the entropy is reduced to zero.  It is  appealing
to equate this reduction in our  uncertainty with the information we gain
by hearing Allan's answer. Shannon proved that this is not just an
interesting analogy; it is the {\em only} definition of  information that
conforms to some simple constraints.

To start, Shannon assumes that the information gained on hearing the
answer can be written as a function of the  probabilities $p_{\rm
n}$.\footnote{In particular, this `zeroth' assumption means that we must
take seriously the notion of enumerating the possible answers.  In this
framework we cannot quantify the information that would be gained upon
hearing a previously unimaginable answer to our question.} Then  if all
$N$ possible answers are
 equally likely  the information gained 
should be a  monotonically increasing function of $N$.   The next
constraint is that if our question consists of two parts, and if these
two parts are entirely independent of one another, then we should be able
to write the total information gained as the sum of the information
gained in response to each of the two subquestions. Finally, more general
multipart questions can be thought of as branching trees, where the
answer to each successive part of the question provides some further
refinement of the probabilities; in this case we should be able to write
the total information gained as the weighted sum of the information
gained at each branch point.  Shannon proved that the only function of
the $\{ p_{\rm n}\}$ consistent with these three
postulates---monotonicity, independence, and branching---is the entropy
$S$, up to a multiplicative constant.

If we phrase the problem of gaining information  from hearing the answer
to a question, then it is natural to think about a discrete set of
possible answers.  On the other hand, if we think about gaining
information from the acoustic waveform that reaches our ears, then there
is a continuum of possibilities.  Naively, we are tempted to write
\begin{equation}
S_{\rm continuum} = - \int dx P(x) \log_2 P(x) ,
\label{ent-continuum}
\end{equation}
or some multidimensional generalization.   The difficulty, of course, is
that probability distributions for continuous variables [like $P(x)$ in
this equation] have units---the distribution of $x$ has units inverse to
the units of $x$---and we should be worried about taking logs of objects
that have dimensions.  Notice that if we wanted to compute a difference
in entropy between two distributions, this problem would go away.  This
is a hint that only entropy differences are going to be
important.\footnote{The problem of defining the entropy for continuous
variables is familiar in statistical mechanics.  In the simple
example of an ideal gas in a finite box, we know that the quantum
version of the problem has a discrete set of states, so that we can
compute the entropy of the gas as a sum over these states.  In the limit
that the box is large, sums can be approximated as integrals, and if the
temperature is high we expect that quantum effects are negligible and one
might naively suppose that Planck's constant should disappear from the
results;  we recall that this is not quite the case.  Planck's constant
has units of momentum times position, and so is an elementary area for
each pair of conjugate position and momentum variables
in the classical phase space;  in the classical limit
the entropy becomes (roughly) the logarithm of the occupied volume  in
phase space, but this volume is measured in units of Planck's constant. 
If we had tried to start with a classical formulation (as did Boltzmann
and Gibbs, of course) then we would find ourselves with the problems of
Eq. (\ref{ent-continuum}), namely that we are trying to take the
logarithm of a quantity with dimensions.  If we measure phase space
volumes in units of Planck's constant, then all is well.  The important
point is that the problems with defining a purely classical entropy do
{\em not} stop us from calculating entropy differences, which are
observable directly as heat flows, and we shall find a similar situation
for the information content of continuous (``classical'') variables.}

Returning to the conversation between Max and Allan, we assumed that Max
would receive a complete answer to his question, and hence that all his
uncertainty would be removed.  This is an idealization, of course.  The
more natural description is that, for example, the world can take on many
states $W$, and by observing data $D$ we learn something but not
everything about $W$.  Before we make our observations, we know
only that states of the world  are chosen from some distribution $P(W)$,
and  this distribution has an entropy $S(W)$.  Once we observe some
particular datum
$D$, our (hopefully improved) knowledge of $W$ is described by the
conditional distribution
$P(W|D)$, and this has an entropy $S(W|D)$ that is smaller than $S(W)$ if we
have reduced our uncertainty about the state of the world by virtue of our
observations.  We identify this reduction in entropy as the information that
we  have gained about $W$. 

\medskip
\addtocounter{problem}{1} 
{\small \noindent {\bf Problem \theproblem: Maximally informative
experiments.}  Imagine that we are trying to gain information about the
correct theory
$T$ describing some set of phenomena.  At some point, our relative
confidence in one particular theory is very high; that is, $P(T = T_*) > 
F \cdot P(T \neq T_* )$ for some large $F$. On the other hand, there are
many possible theories, so our absolute confidence in the theory $T_*$
might nonetheless be quite low, $P(T = T_*) << 1$.  Suppose we follow the
`scientific method' and design an experiment that has a yes or no answer,
and this answer is perfectly correlated with the correctness of theory
$T_*$, but uncorrelated with the correctness of any other possible
theory---our experiment is designed specifically to test or falsify the
currently most likely theory.  What can you say about how much
information you expect to gain from such a measurement?  Suppose instead
that you are completely irrational and design an experiment that is
irrelevant to testing $T_*$ but has the potential to eliminate many
(perhaps half) of the alternatives.  Which experiment is expected to be
more informative?  Although this is a gross cartoon of the scientific
process, it is not such a terrible model of a game like ``twenty
questions.'' It is interesting to ask whether people play such question
games following strategies that might seem irrational but nonetheless
serve to maximize information gain \cite{iris}.  Related but distinct criteria for
optimal experimental design have been developed in the statistical literature
\cite{fedorov}.}
\bigskip

Perhaps this is the point to note that a single observation $D$ is not, in fact,
guaranteed to provide positive information, as emphasized by  DeWeese and Meister
\cite{deweese+meister}. Consider, for instance, data which tell us that all of our
previous measurements have larger error bars than we thought:  clearly such data, at an
intuitive level, reduce our knowledge about the world and should be associated with a
negative information.   Another way to say this is that some data points
$D$ will increase our uncertainty about state $W$ of the world, and hence
for these particular data the conditional distribution $P(W|D)$ has a
larger entropy than the prior distribution $P(D)$.  If we identify
information with the reduction in entropy, $I_D = S(W) - S(W|D)$, then
such data points are associated unambiguously with negative information. 
On the other hand, we might hope that, on average, gathering data
corresponds to gaining information:  although single data points can
increase our uncertainty, the average over all data points does not.

If we average over all possible data---weighted, of course, 
by their probability of occurrence
$P(D)$, we obtain the average information that $D$ provides about $W$,
\begin{eqnarray}
I(D\rightarrow W) &=&  S(W) - \sum_D P(D) S(W|D) \\
&=&
\sum_W \sum_D P(W,D) \log_2\left[{{P(W,D)}\over{P(W)P(D)}}\right] .
\label{defmut}
\end{eqnarray}
Note that the information which $D$ provides about $W$
is symmetric in $D$ and $W$.  This means that we can 
also view the state of the world as providing information about the data we will
observe, and this information is, on average, the same as the data will provide
about the state of the world.  This `information provided' is therefore often
called the mutual information, and this symmetry will be very important in subsequent
discussions; to remind ourselves of this symmetry we write $I(D;W)$ rather than
$I(D\rightarrow W)$.

\medskip
\addtocounter{problem}{1}
{\small
\noindent {\bf Problem \theproblem: Positivity of information.} 
Prove that the mutual information $I(D\rightarrow W)$, defined in Eq.
(\ref{defmut}), is positive.}
\bigskip

One consequence of the symmetry or mutuality of
information is that we can write
\begin{eqnarray}
I(D; W) &=&  S(W) - \sum_D P(D) S(W|D) \\
&=&
S(D) - \sum_W P(W) S(D|W) .
\label{inf=ent-cond}
\end{eqnarray}
If we consider only discrete sets of possibilities then entropies are
positive (or zero), so that these equations imply
\begin{eqnarray}
I(D; W) &\leq&  S(W)\\
I(D; W) &\leq&
S(D) .
\label{capacity}
\end{eqnarray}
The first equation tells us that by observing $D$ we cannot learn more
about the world then there is entropy in the world itself.  This makes sense: 
entropy measures the number of possible states that the world can be in, and we
cannot learn more than we would learn by reducing this set of possibilities down to
one unique state.  Although sensible (and, of course, true), this is not a terribly
powerful statement: seldom are we in the position that our ability to gain knowledge
is limited by the lack of possibilities in the world around us.\footnote{This is not
quite true.  There is a tradition of studying the nervous system as it responds to
highly simplified signals, and under these conditions the lack of possibilities in
the world can be a significant limitation, substantially confounding the
interpretation of experiments.}  The second equation, however, is much more powerful. 
It says that, whatever may be happening in the world, we can never learn more than the
entropy of the distribution that characterizes our data.  Thus, if we ask how much we
can learn about the world by taking readings from a wind detector on top  of the roof,
we can place a bound on the amount we learn just by taking a very long stream of data,
using these data to estimate the distribution 
$P(D)$, and then computing the entropy of this distribution.  

The entropy of our observations\footnote{In the same way that  we speak
about the entropy of a gas I will often speak about the entropy of a
variable or the entropy of a response.  In the gas, we understand from
statistical mechanics that the entropy is defined not as a property of
the gas but as a property of the distribution or ensemble from which the
microscopic states of the gas are chosen; similarly we should really speak
here about ``the entropy of the distribution of observations,'' but this
is a bit cumbersome.  I hope that the slightly sloppy but more compact
phrasing does not cause confusion.}  thus limits how much we can learn no
matter what question we were hoping to answer, and so we can think of the
entropy as setting (in a slight abuse of terminology) the capacity of the
data $D$ to provide or to convey information.  As an example, the entropy
of neural responses sets a limit to how much information a neuron can
provide about the world, and we can estimate this limit even if we don't
yet understand what it is that the neuron is telling us (or the rest of
the brain).  Similarly, our bound on the information conveyed by the wind
detector does not require us to understand how these data might be used
to make predictions about tomorrow's weather. 

Since the information we can gain is limited by the
entropy, it is natural to ask if we can put limits on the entropy using some low order
statistical properties of the data:  the mean, the variance, perhaps higher moments or 
correlation functions, ... .   In particular,
if we can say that the entropy has a maximum value consistent with the observed
statistics, then we have placed a firm upper bound on the information that these data
can convey.  

The problem of finding the maximum entropy given some constraint again is
familiar from statistical mechanics:  the Boltzmann distribution  is the
distribution that has the largest possible entropy given the mean energy.
More generally, let us imagine that we have knowledge not of the whole
probability distribution $P(D)$ but only of some expectation values,
\begin{equation}
\langle f_{\rm i} \rangle = \sum_D P(D) f_{\rm i} (D) ,
\end{equation}
where we allow that there may be several expectation values  known 
(${\rm i} = 1, 2, ... , K$).  Actually there is one more expectation value
that we always know, and this is that the average value of one is one; the
distribution is normalized:
\begin{equation}
\langle f_0 \rangle = \sum_D P(D) = 1 .
\end{equation}
Given the set of numbers $\{\langle f_0\rangle , \langle f_1 \rangle, \cdots ,
\langle f_K \rangle \}$  as constraints on the probability distribution $P(D)$, we would
like to know the largest possible value for the entropy, and we would like to find
explicitly the distribution that provides this maximum.  

The problem of maximizing a quantity subject to constraints  is
formulated using Lagrange multipliers.  In this case, we want to maximize
$S = - \sum P(D) \log_2 P(D)$, so we introduce a function $\tilde S$,
with one Lagrange multiplier $\lambda_{\rm i}$ for each constraint:
\begin{eqnarray}
{\tilde S}[P(D)] &=& -\sum_D P(D) \log_2 P(D) - \sum_{{\rm i}=0}^K \lambda_{\rm i}
\langle f_{\rm i} \rangle\\
&=& -{1\over {\ln 2}} \sum_D P(D) \ln P(D)
- \sum_{{\rm i}=0}^K \lambda_{\rm i} \sum_D P(D) f_{\rm i}(D).
\end{eqnarray}
Our problem is then to find the maximum of the function $\tilde S$, but this is easy
because the probability for each value of $D$ appears independently.  
The result is that
\begin{equation}
P(D) = {1\over Z} \exp\left[ - \sum_{{\rm i}=1}^K \lambda_{\rm i} f_{\rm i}(D) \right] ,
\label{maxentform}
\end{equation}
where $Z = \exp(1+\lambda_0)$ is a normalization constant.

\medskip
\addtocounter{problem}{1}
{\small
\noindent {\bf Problem \theproblem: Details.}   Derive
Eq. (\ref{maxentform}).  In particular, show
that Eq. (\ref{maxentform}) provides a probability distribution which
genuinely  {\em maximizes} the entropy, rather than being just  an
extremum.}
\bigskip

These ideas are enough to get started on ``designing'' some simple neural processes.
Imagine, following Laughlin \cite{laughlin81}, that a neuron is responsible for
representing a single number such as the light intensity $\cal I$ averaged over small
patch of the retina (don't worry about time dependence).  Assume that
this signal will be represented by a continuous voltage $V$, which is true for the
first stages of processing in vision.   This voltage is encoded  into discrete
spikes  only as a second or third step.  The information that
the voltage provides about the intensity is 
\begin{eqnarray}
I(V \rightarrow {\cal I}) &=& \int d{\cal I} \int dV \, P(V,{\cal I})
\log_2\left[{{P(V,{\cal I})}\over{P(V)P({\cal I})}}\right]\\
&=&
\int d{\cal I} \int dV \, P(V,{\cal I})
\log_2\left[{{P(V|{\cal I})}\over{P(V)}}\right] .
\end{eqnarray}
The conditional distribution $P(V|{\cal I})$ describes the process by which the neuron
responds to its input, and so this is what we  should try to ``design.'' 

Let us suppose that the voltage is on average a nonlinear function
of the intensity, and that the dominant source of noise is additive (to the voltage),
independent of light intensity, and small compared with the overall dynamic range of
the cell:
\begin{equation}
V = g({\cal I}) + \xi,
\end{equation}
with some distribution $P_{\rm noise}(\xi)$ for the noise.  Then the conditional
distribution
\begin{equation}
P(V|{\cal I }) = P_{\rm noise} (V- g({\cal I})) ,
\end{equation}
and the entropy of this conditional distribution can be written as
\begin{eqnarray}
S_{\rm cond}  &=& 
- \int dV \, P(V|{\cal I}) \log_2 P(V|{\cal I}) \\
&=& - \int d\xi \, P_{\rm noise}(\xi)\log_2 P_{\rm noise}(\xi ).
\end{eqnarray}
Note that this is a constant, independent both of  the light intensity and of
the nonlinear input/output relation $g({\cal I})$.
This is useful because we can write the information as a difference between the total
entropy of the output variable $V$ and this conditional or noise entropy, as in Eq.
(\ref{inf=ent-cond}):
\begin{equation}
I(V \rightarrow {\cal I}) = -\int dV P(V) \log_2 P(V) - S_{\rm cond} .
\end{equation}
With $S_{\rm cond}$ constant independent of our `design,' maximizing information is
the same as maximizing the entropy of the distribution of output voltages.  Assuming
that there are maximum and minimum values for this voltage, but no other constraints,
then the maximum entropy distribution is just the uniform distribution within the
allowed dynamic range.  But if the noise is small it doesn't contribute much to
broadening $P(V)$ and we calculate this distribution as if there were no noise, so that
\begin{eqnarray}
P(V)dV &=& P({\cal I}) d{\cal I} ,\\
{{dV}\over{d\cal I}} &=& {1\over{P(V)}} \cdot P({\cal I}) .
\end{eqnarray}
Since we want to have $V = g({\cal I})$ and $P(V) = 1/(V_{\rm max} - V_{\rm min})$,
we find
\begin{eqnarray}
{{dg({\cal I})}\over{d{\cal I}}} &=& (V_{\rm max} - V_{\rm min}) P({\cal I}),\\
g({\cal I}) &=& (V_{\rm max} - V_{\rm min}) \int_{\cal I_{\rm min}}^{\cal I}
d{\cal I'} P({\cal I'}) .
\label{sbl-pred}
\end{eqnarray}
Thus, the optimal input/output relation is proportional to the cumulative probability
distribution of the input signals.

The predictions of Eq. (\ref{sbl-pred}) are quite interesting.  First of all it makes
clear that any theory of the nervous system which involves optimizing information
transmission or  efficiency of representation inevitably predicts that the computations
done by the nervous system must be matched to the statistics of sensory inputs (and,
presumably, to the statistics of motor outputs as well).  Here the matching is
simple:  in the right units we could just read off the distribution of inputs by
looking at the (differentiated) input/output relation of the neuron.   Second, this
simple model automatically carries some predictions about adaptation to overall light
levels.  If we live in a world with diffuse light sources that are not directly
visible, then the intensity which reaches us at a point is the product of the
effective brightness of the source and some local reflectances.  As is it gets dark
outside the reflectances don't change---these are material properties---and so we
expect that the distribution $P({\cal I})$ will look the same except for  scaling.
Equivalently, if we view the input as the log of the intensity, then
to a good approximation  $P(\log{\cal I})$ just shifts linearly
along the $\log {\cal I}$ axis as mean light intensity goes up and down.  But then the
optimal input/output relation $g({\cal I})$ would exhibit a similar invariant shape
with shifts along the input axis when expressed as a function of $\log{\cal I}$, and
this is in rough agreement with experiments on light/dark adaptation in  a wide variety
of visual neurons.  Finally, although obviously a simplified version of the real problem
facing even the first stages of visual processing, this calculation does make a
quantitative prediction that would be tested if we  measure both the input/output
relations of early visual neurons and the distribution of light intensities that the
animal encounters in nature.  

Laughlin \cite{laughlin81} made this comparison (20 years ago!) for the fly visual
system.  He built an electronic photodetector with aperture and spectral
sensitivity matched to those of the fly retina and used his photodetector to scan
natural scenes, measuring
$P({\cal I})$ as it would appear at the input to these neurons.  In parallel he
characterized the second order neurons of the fly visual system---the large monopolar
cells which receive direct synaptic input from the photoreceptors---by measuring the
peak voltage response to flashes of light.  The agreement with
Eq. (\ref{sbl-pred}) was remarkable, especially when we remember that there are no free
parameters.    While there are obvious open questions (what
happened to time dependence?), this is a really beautiful result.

Laughlin's analysis focused on the nonlinear input/output properties of neurons but
ignored dynamics. An alternative which has been pursued by several groups is to treat
dynamics but to ignore nonlinearity \cite{atick,hans-opt}, and we tried to review some
of these ideas in section 5.3 of {\em Spikes} \cite{spikes}.  As far as I know there is
not much work which really brings together dynamics and nonlinearity, although there are
striking results about filtering and nonlinearity in the color domain \cite{dan-color}.
While these model problems capture something about real neurons, it would be nice to
confront the real systems more directly.  In particular, most neurons in the brain
represent signals through trains of action
potentials.  As noted in the introduction to this section, we'd like to make
a direct measurement of the information carried by these spikes or of the
efficiency with which the spikes represent the sensory world.

The first question we might
ask is how much information we gain about the sensory inputs by observing the occurrence
of just one spike at some time
$t_0$  \cite{brenner00a}.
For simplicity let us imagine that the inputs are described just by one function
of time $s(t)$, although this is not crucial; what will be crucial is that we can repeat
exactly the same time dependence many times, which for the visual system means showing
the same movie over and over again, so that we can characterize the variability and
reproducibility of the neural response. In general, the information gained about $s(t)$
by observing a set of neural responses is
\begin{eqnarray}
I
&=&
\sum_{\rm responses}\int Ds(\tau) P[s(\tau) \,,\, {\rm resp}]
\log_2\left(
{{P[s(\tau) \, ,\, {\rm resp}]}
\over{P[s(\tau)]P({\rm resp})}}
\right),
\nonumber\\&&
\end{eqnarray}
where information is measured in bits.
In the present case, the response is a single spike, so summing over the full range
of responses is equivalent to integrating over the possible spike arrival times
$t_0$:
\begin{eqnarray}
I_{\rm 1\,spike} &=&
\int dt_0\int Ds(\tau ) P[s(\tau ) \,,\, t_0]
\log_2 \left(
{{P[s(\tau )\,,\,t_0 ]}\over{P[s(\tau)]P[t_0]}}
\right)\\
&=&
\int dt_0 P[t_0]  \int Ds(\tau ) P[s(\tau )| t_0] 
\log_2\left(
{{P[s(\tau )|t_0 ]}\over{P[s(\tau)]}}
\right),\nonumber\\
&&
\end{eqnarray}
where by $P[s(\tau ) | t_0]$ we mean the distribution of stimuli given  that we have 
observed a spike  at time $t_0$.   In the absence of knowledge about the stimulus, all
spike arrival times are equally likely, and hence $P[t_0]$ is uniform.
Furthermore, we expect that the coding of stimuli is stationary in time, so
that the conditional distribution $P[s(\tau)|t_0]$ is of the same shape for
all $t_0$, provided that we measure the time dependence of the
stimulus $s(\tau )$ relative to the spike time
$t_0$:  $P[s(\tau ) | t_0] =P[s(\tau - \Delta t ) | t_0 - \Delta
t]$.  With these simplifications we can write  the information
conveyed by observing a single spike at time
$t_0$ as \cite{88}
\begin{eqnarray}
I_{\rm 1\,spike}
&=&
\int Ds(\tau) P[s(\tau) |  t_0]
\log_2\left(
{{P[s(\tau) |  t_0]}
\over{P[s(\tau)]}}
\right).
\label{info1spikea}
\end{eqnarray}
In this formulation we think of the spike as `pointing to' certain regions in the
space of possible stimuli, and of course the information conveyed is quantified by an
integral that relates to the volume of these regions.  The difficulty is that if we want
to use Eq. (\ref{info1spikea}) in the analysis of real experiments we will need a model
of the distribution $P[s(\tau) |  t_0]$, and this could be hard to come by:  the stimuli
are drawn from a  space of very high dimensionality (a function space, in principle) and
so we cannot sample this distribution thoroughly in any reasonable experiment.
Thus computing information transmission by mapping spikes back
into the space of stimuli involves some model of how the code
works, and then this model is used to simplify the structure of
the relevant distributions, in this case $P[s(\tau) |  t_0]$.  We
would like an alternative approach that does not depend on such
models.\footnote{I hope that it is clear where this could lead:  If
we can estimate information using  a model of what spikes stand
for, and also estimate information without such a model, then by
comparing the two estimates we should be able to test our model of
the code in the most fundamental sense---does our model of what the
neural response represents capture all of the information that
this response provides?}

From Bayes' rule we can
relate the conditional probability of stimuli given spikes to the conditional
probability of spikes given stimuli:
\begin{equation}
{{P[s(\tau) |  t_0]}
\over{P[s(\tau)]}}
=
{{P[ t_0|s(\tau)]}
\over {P[ t_0]}}.
\end{equation}
But we can measure the probability of a spike at $t_0$ given that we know the stimulus
$s(\tau)$ by repeating the stimulus many times and looking in a small bin around $t_0$
to measure the fraction of trials on which a spike occurred.  If we normalize by the size
of the bin in which we look then we are measuring the probability per unit time that a
spike will be generated, which is a standard way of averaging the response over many
trials; the probability per unit time is also called   the time dependent firing
rate $r[t_0;s(\tau)]$, where the notation reminds us that the probability of a spike at
one time depends on the whole history of inputs leading up to that time. 

If we don't know the stimulus then the probability of a spike at
$t_0$ can only be given by the average firing rate over the whole experiment, $\bar r =
\langle r[t;s(\tau)]\rangle$, where the expectation value
$\langle {\bf \cdots}\rangle$ denotes an average over the distribution of stimuli
$P[s(\tau)]$.  Thus we can write
\begin{equation}
{{P[s(\tau) |  t_0]}
\over{P[s(\tau)]}}
=
{{r[t_0; s(\tau )]}
\over{\bar r}} .
\end{equation}
Furthermore, we can substitute this relation into Eq. (\ref{info1spikea}) for
the information carried by a single spike, and then we obtain
\begin{equation}
I_{\rm 1\,spike}
=
\Bigg\langle
\left({{r[t_0; s(\tau )]}
\over{\bar r}}\right)
\log_2
\left(
{{r[t_0; s(\tau )]}
\over{\bar r}}
\right)
\Bigg\rangle ,
\label{ens-avg-I1}
\end{equation}
We can compute the average in Eq. (\ref{ens-avg-I1}) by integrating over time, provided
that the stimulus we use runs for a sufficiently long time that it provides a fair
(ergodic)
sampling of the true distribution $P[s(\tau)]$ from which stimuli are drawn.

Explicitly, then, if we sample the ensemble of possible stimuli by choosing a
single time dependent stimulus $s(t)$ that runs for a long duration $T$, and
then we repeat this stimulus many times to accumulate the time dependent firing
rate $r[t; s(\tau)]$, the information conveyed by a single spike is given
exactly by an average over this firing rate:
\begin{equation}
I_{\rm 1\,spike}
=
{1\over T}
\int_0^T dt\, \left({{r[t; s(\tau )]}
\over{\bar r}}\right)
\log_2
\left(
{{r[t; s(\tau )]}
\over{\bar r}}
\right) .
\label{final1spikeinfo}
\end{equation}
This is an exact formula,independent of any model for the
structure of the neural code.   It makes sense that the information carried
by one spike should be related to the firing rate, since the the rate vs. time gives a
complete description of the `one body' or one spike statistics of the spike train, in
the same way that the single particle density describes the one body statistics of a
gas or liquid.  

\medskip
\addtocounter{problem}{1}
{\small
\noindent {\bf Problem \theproblem: Poisson model and lower bounds.} 
Prove that Eq. (\ref{final1spikeinfo}) provides a lower bound to the information
per spike transmitted if the entire spike train is a modulated Poisson
process \cite{santafe}.}
\bigskip

Another view of the result in Eq. (\ref{final1spikeinfo}) is in terms of the
distribution of times at which a single spike might occur.  First we note that the information
a single spike provides about the stimulus must be the same as the information that
knowledge of the stimulus provides about the occurrence time of a single
spike---information is mutual. Before we know the precise trajectory of the
stimulus
$s(t)$, all we can say is that if we are looking for one spike, it can occur
anywhere in our experimental window of size $T$, so that the probability is
uniform,
$p_0(t) = 1/T$ and the entropy of this distribution is just $\log_2 T$.
Once we know the stimulus, we can expect that spikes will occur preferentially
at times where the firing rate is large, so the probability distribution
should be proportional to $r[t; s(\tau)]$; with proper  normalization  we 
have $p_1(t) = r[t; s(\tau)]/(T\bar r)$.  Then the conditional entropy is
\begin{eqnarray}
S_1 &=& -\int_0^T dt p_1 (t) \log_2 p_1(t) \\
&=& - {1\over T}\int_0^T dt{{r[t; s(\tau)]}\over{\bar r}}
\log_2 \left({{r[t; s(\tau)]}\over{{\bar r}T}}\right) .
\end{eqnarray}
The reduction in entropy is the gain in information, so
\begin{eqnarray}
I_{\rm 1\,spike} &=& S_0 - S_1
\\
&=& 
{1\over T}
\int_0^T dt\, \left({{r[t; s(\tau )]}
\over{\bar r}}\right)
\log_2
\left(
{{r[t; s(\tau )]}
\over{\bar r}}
\right) ,
\end{eqnarray}
as before.

A crucial point about Eq. (\ref{final1spikeinfo})  is that when we
derive it we do not make use of the fact that $t_0$ is the time of
a single spike:  it could be any event that occurs at a well
defined time.  There is considerable interest in the
question of whether `synchronous' spikes from two neurons provide
special information in the neural code.  If we define synchronous
spikes as two spikes that occur within some fixed (small) window
of time then this compound event can also be given an arrival time
(e.g., the time of the later spike), and marking these arrival
times across repeated presentations of the same stimulus we can
build up the rate $r_E [t; s(\tau )]$ for these events of class
$E$ in exactly the same way that we build up an estimate of the
spike rate.  But if we have compound events constructed  from two
spikes, it makes sense to compare the information carried by a
single event $I_E$ with the information that would be carried
independently by two spikes,  $2I_{\rm 1\,spike}$.  If the
compound event conveys more information than the sum of its parts,
then  this compound event  indeed
is a special symbol in the code.  
The same arguments apply to compound events constructed from temporal patterns of spikes
in one neuron.

If a compound event provides an amount of information exactly equal to what we expect by
adding up contributions from the components, then we say that the components or
elementary events convey information  independently.  If there is less than independent
information we say that the elementary events are redundant, and if the compound event
provides more than the sum of its parts we say that there is {\em synergy} among the
elementary events.

When we use these ideas to analyze Rob's experiments on the fly's H1 neuron
\cite{brenner00a}, we find that the occurrence of a single spike can provide from 1 to 2
bits of information, depending on the details of the stimulus ensemble.  More robustly
we find that pairs of spikes separated by less than 10 msec can provide more---and
sometimes vastly more---information than expected just by adding up the contributions of
the individual spikes.  There is a small amount of redundancy among spikes with larger
separations, and if stimuli have a short correlation time then spikes carry independent
information once they are separated by more than 30 msec or so.   It is interesting that
this time scale for independent information is close to the time scales of behavioral
decisions, as if the fly waited long enough to see all the spikes that have a chance of
conveying information synergistically.

We'd like to understand what happens as all the spikes add up to give us a fuller
representation of the sensory signal: rather than
thinking about the information carried by particular events, we want to
estimate the information carried by long stretches of the neural response.
Again the idea is straightforward \cite{deruyter97,strong98}:  use Shannon's
definitions to write the mutual information between stimuli and spikes in terms of
difference between two entropies, and then use a long experiment to sample the relevant
distributions and thus estimate these entropies.  The difficulty is that when we talk
not about single events but about ``long stretches of the neural response,'' the
number of possible responses is (exponentially) larger, and sampling is
more difficult.  Much of the effort in the original papers
thus is in convincing ourselves that we have control over these sampling problems.

Let us look at segments of the spike train with length $T$, and within this time we
record the spike train with time resolution $\Delta\tau$; these
parameters are somewhat arbitrary, and we will need to vary them to be be sure we
understand what is going on.  In this view, however, the response is a ``word'' with $K
= T/\Delta\tau$ letters; for small $\Delta\tau$ there can be only one or zero spikes in a
bin and so the words are binary words, while for poorer time resolution we have a
larger alphabet.  If we let the fly watch a long movie, many different words will be
produced and with a little luck we can get a good estimate of the probability
distribution of these words, $P(W)$. This distribution has an entropy
\begin{equation}
S_{\rm total}(T,\Delta\tau) = -\sum_W P(W)\log_2 P(W)
\end{equation}
which measures the size of the neuron's vocabulary and hence the capacity of the code
given our parameters $T$ and $\Delta\tau$ [cf. Eq. (\ref{capacity}) and the subsequent
discussion].  While a large vocabulary is a good thing, to convey information I have to
associate words with particular things in a reproducible way.  Here we can show the same
movie many times, and if we look across the many trials at a moment $t$ relative to the
start of the movie we again will see different words (since there is some noise in the
response), and these provide samples of the distribution $P(W|t)$.  This distribution in
turn has an entropy which we call the noise entropy since any variation in response to
the same inputs constitutes noise in the representation of those inputs:\footnote{This
is not to say that such variations might not provide information about something else,
but in the absence of some other signal against which we can correlate this ambiguity
cannot be resolved.  It is good to keep in mind, however, that what we call noise could
be signal, while what we call signal really does constitute information about the input,
independent of any further hypotheses.}
\begin{equation}
S_{\rm noise} = \Bigg\langle -\sum_W P(W|t)\log_2 P(W|t) \Bigg\rangle_t ,
\end{equation}
where $\langle \cdots \rangle$ denotes an average over $t$ and hence (by ergodicity)
over the ensemble of sensory inputs $P[s]$.
Finally, the information that the neural response provides about the sensory input is
the difference between the total entropy and the noise entropy, 
\begin{equation}
I(T,\Delta\tau) = S_{\rm total}(T,\Delta\tau) - S_{\rm noise}(T,\Delta\tau),
\end{equation}
as in Eq. (\ref{inf=ent-cond}).  A few points worth emphasizing:
\begin{itemize}
\item By using time averages in place of ensemble averages we can measure the information
that the response provides about the sensory input without any explicit coordinate
system on the space of inputs and hence without making any assumptions about which
features of the input are most relevant.
\item If we can sample the relevant distributions for sufficiently large times windows
$T$, we expect that entropy and information will become extensive quantities, so that it
makes sense to define entropy rates and information rates.
\item As we improve our time resolution, making $\Delta\tau$ smaller, the capacity of
the code $S_{\rm total}$ must increase, but it is an experimental question whether the
brain has the timing accuracy to make efficient use of this capacity.
\end{itemize}

The fly's H1 neuron provides an ideal place to do all of this because of the extreme
stability of the preparation.  In an effort to kill off any concerns about sampling and
statistics, Rob did a huge experiment with order one thousand replays of the
same long movie \cite{deruyter97,strong98}. With this large data set we were able to see
the onset of extensivity, so we extracted information and entropy rates (although this
really isn't essential) and we were able to explore a wide range of time resolutions, 
$800 > \Delta\tau > 2$ ms.
Note that $\Delta\tau = 800$ ms corresponds to counting spikes in bins that
contain typically thirty spikes, while $\Delta\tau =2$ ms
corresponds to timing each spike to within 5\% of the typical interspike
interval.
Over this range, the entropy of the spike train varies over
a factor of roughly 40, illustrating the increasing capacity of the
system to convey information by making use of spike timing. 
The information that the spike train conveys
about the visual stimulus increases
in approximate  proportion to the entropy, corresponding to 
$\sim 50\%$ efficiency, although we start to see some saturation of information at the
very highest time resolutions. Interestingly, this level of efficiency (and its
approximate constancy as a function of time resolution) confirms an earlier measurement
of efficiency in mechanosensor neurons from frogs and crickets that used the ideas of
decoding discussed in the previous section \cite{rieke93}.

What have we learned from this?  First of all,  the
earliest experiments on neural coding showed that the rate of spiking encodes
stimulus amplitude for static stimuli, but this left open the question of whether
the precise timing of spikes carries additional information.  The first application of
information theory to neurons (as far as I know) was MacKay and McCulloch's
calculation of the capacity of neurons to carry information given
different assumptions about the nature of the code \cite{mackay}, and of
course they drew attention to the fact that the capacity increases as we allow fine
temporal details of the spike sequence to become distinguishable symbols in the code.
Even MacKay and McCulloch were skeptical about whether real neurons could
use a significant fraction of their capacity, however, and other investigators were more
than skeptical.  The debate about whether spike timing is important raged on,
and I think that one of the important contributions of an information theoretic
approach has been to make precise what we might mean by `timing is important.'

There are two senses in which the timing of action potentials could
be important to the neural code.  First there is the simple question of whether
marking spike arrival times to higher resolution really allows us to extract more
information about the sensory inputs.  We know (following MacKay and McCulloch) that
if we use higher time resolution the entropy of the spike train increases,
and hence the capacity to transmit information also increases. The question
is whether this capacity is used, and the answer (for one neuron, under one set of
conditions ... ) is in Ref. \cite{strong98}: yes, unambiguously.

A second notion of spike timing being important is that temporal patterns of spikes
may carry more information than would be expected by adding the contributions of single
spikes.  Again the usual setting for this sort of code is in a population of neurons,
but the question is equally well posed for patterns across time in a single cell. 
Another way of asking the question is whether the high information rates observed for
the spike train as a whole are ``just'' a reflection of rapid, large amplitude
modulations in the spike rate.\footnote{Note that in many cases this formulation
obscures the fact that the a single ``large amplitude modulation'' is a bump in the
firing rate with area of order one, so what might be called a large rate change is
really one spike.} Equation \ref{final1spikeinfo} makes clear that ``information carried
by rate modulations'' is really the information carried by single spikes.  The results
of Ref. \cite{brenner00a} show  that pairs of spikes can carry more than twice the single
spike information, and the analysis of longer windows of responses shows that this
synergy is maintained, so that the spike train as a whole is carrying 30--50\% more
information than  expected by summing the contributions of single spikes.  Thus the
answer to the question of whether temporal patterns are important, or whether there is
``more than just rate'' is again: yes, unambiguously.

These results on coding efficiency and synergy are surprisingly robust
\cite{multifly}:  If we analyze the responses of H1 neurons from many different flies,
all watching the same movie, it is easy to see that the flies are very
different---average spike rates can vary by a factor of three among individuals, and by
looking at the details of how each fly associates stimuli and responses we can
distinguish a typical pair of flies from just $30\,{\rm msec}$ of data.  On the other
hand, if we look at the coding efficiency---the information divided by the total
entropy---this is constant to within 10\% across the population of flies in our
experiments, and this high efficiency always has a significant contribution from synergy
beyond single spikes.

The direct demonstration that the neural code is efficient in this information theoretic
sense clearly depends on using complex, dynamic sensory inputs rather than the
traditional quasistatic signals.  We turned to these dynamic inputs not because
they are challenging to analyze but rather because we thought that
they would provide a better model for the problems encountered by the brain under
natural conditions.  This has become part of a larger effort in the community to 
analyze the way in which the nervous system deals with natural signals, and it probably
is fair to point out that this effort has not been without its share of controversies
\cite{canberra}.  One way to settle the issue is to strive for ever more natural
experimental conditions. I think that Rob's recent experiments hold the record for
`naturalness': rather than showing movies in the lab, he has taken his whole experiment
outside into the woods where the flies are caught and recorded from H1 while rotating
the fly through angular trajectories like those observed for freely flying flies
\cite{natural}.  This is an experimental tour de force (I can say this without
embarrassment since I'm a theorist) because you have to maintain stable electrical 
recordings of neural activity while the sample is spinning at thousands of degrees per
second and and accelerating to reverse direction within 10 msec.  The reward is that
spike timing is even more reproducible in the ``wild'' than in the lab, coding
efficiencies and information rates are higher and are maintained to even smaller values
$\Delta\tau$.

All of the results above point to the idea that spike trains really do provide an
efficient representation of the sensory world, at least in one precise information
theoretic sense.  Many experimental groups are  exploring whether similar results can be
obtained in other systems, in effect asking if these theoretically appealing features of
the code are universal.  Here I want to look at a different question, namely whether
this efficient code is fixed in the nervous system or whether it adapts and develops in
response to the surroundings.  The ideas and results that we have on these issues are, I
think, some of the clearest evidence available for optimization in the neural code. 
Once again all the experiments are done in Rob de Ruyter's lab using H1 as the test
case, and most of what we currently know is from work by Naama Brenner and Adrienne
Fairhall \cite{brenner00b,fairhall-nature}.

There are two reasons to suspect that the efficiency we have measured is achieved
through adaptation rather than through hard wiring.  First, one might guess that a fixed
coding scheme could be so efficient and informative only if we choose the right ensemble
of inputs, and you have to trust me that we didn't search around among many ensembles to
find the results that I quoted.  Second, under natural conditions the signals we
want to encode are intermittent---the fly may fly straight, so that typical angular
velocities are $\sim 50^\circ/{\rm sec}$, or may launch into acrobatics where typical
velocities are $\sim 2000^\circ/{\rm sec}$, and there are possibilities in between.  At
a much simpler level, if we look across a natural scene, we find regions where the
variance of light intensity or contrast is small, and nearby regions in which it is large
\cite{scaling-prl}.  Observations on contrast variance in natural
images led to the suggestion that neurons in the retina might adapt in real time to
this variance, and this was confirmed \cite{stelios}.  Here we would like to look at the
parallel issue for velocity signals in the fly's motions sensitive neurons.

In a way what we are looking for is really contained in Laughlin's model problem
discussed above.  Suppose that we could measure the strategy that the fly actually uses
for converting continuous signals into spikes; we could characterize this by giving the
probability that a spike will be generated at $t$ given the signal $s(\tau < t)$, which
is what we have called the firing rate $r[t;s(\tau )]$.  We are hoping that the neuron
sets this coding strategy using some sort of optimization principle, although it is
perhaps not so clear what constraints are relevant once we move from the model problem
to the real neurons.  On the other hand, we {\em can} say something about the nature of
such optimization problems if we think about scaling:  when we plot $r[s]$, what sets the
scale along the $s$ axis?  

We know from the previous section that there is a limit to
the smallest motion signals that can be reliably estimated, and of course there is a
limit to the highest velocities that the system can deal with (if you move sufficiently
fast everything blurs and vision is impossible).  Happily, most of the signals we (and
the fly) deal with are well away from these limits, which are themselves rather far
apart.  But this means that there is nothing intrinsic to the system which sets the
scale for measuring angular velocity and encoding it in spikes, so if $r[s]$ is to
emerge as the solution to an optimization problem then the scale along the $s$ axis must
be determined by outside world, that is by the distribution $P[s]$ from which the
signals are drawn.  Further, if we scale the distribution of inputs, $P[s] \rightarrow
\lambda P[\lambda s]$ then the optimal coding strategy also should scale, $r[s]
\rightarrow r[\lambda s]$.\footnote{I'm being a little sloppy here:  $s$ is really a
function of time, not just a single number, but I hope the idea is clear.}  The
prediction, then, is that if the system can optimize its coding strategy in relation to
the statistical structure of the sensory world, then we should be able to draw signals
from a family of scaled distributions and see the input/output relations of the neurons
scale in proportion to the input dynamic range.  To make a long story short, this is
exactly what we saw in H1 \cite{brenner00b}.\footnote{There is one technical issue in the
analysis of Ref. \cite{brenner00b} that I think is of broader theoretical interest.  In
trying to characterize the input/output relation
$r[t;s(\tau)]$ we face the problem that the inputs $s(\tau)$ really live in a function
space, or in more down to earth terms a space of very large dimensionality.  Clearly we
can't just plot the function $r[t;s(\tau)]$ in this space.  Further, our intuition is
that neurons are not equally sensitive to all of the dimensions or features of their
inputs.  To make progress (and in fact to get the scaling results) we have to make this
intuition precise and find the relevant dimensions in stimulus space; along the way it
would be nice to provide some direct evidence that the number of dimensions is actually
small (!).  If we are willing to consider Gaussian $P[s(\tau )]$, then we can show that
by computing the right correlation functions between $s(\tau )$ and the stream of
spikes $\rho(t) = \sum_i \delta (t- t_i)$ we can both count the number of relevant
dimensions and provide an explicit coordinate system on the relevant subspace
\cite{features,brenner00b}.   These techniques are now being used to analyze other
systems as well, and we are trying to understand if we can make explicit use of
information theoretic tools to move beyond the analysis of Gaussian inputs and low order
correlation functions.  I find the geometrical picture of neurons as selective for a
small number of dimensions rather attractive as well being useful, but it is a bit off
the point of this discussion.}

The observation of scaling behavior is such a complex system certainly warms the hearts
of physicists who grew up in a certain era.  But Naama realized that one could do 
more.  The observation of scaling tells us (experimentally!) that the system has a choice
among a one parameter family of input/output relations, and so we can ask why the system
chooses the one that it does.  The answer is striking:  the exact scale factor chosen by
the system is the one that maximizes information transmission.

If the neural code is adapting in response to changes of the input distribution, and
further if this adaptation serves to maximize information transmission, then we should
be able to make sudden a change between two very different input distributions and
``catch'' the system using the wrong code and hence transmitting less than the maximum
possible information.  As the system collects enough data to be sure that the
distribution has changed, the code should adapt and information transmission should
recover.  As with the simpler measurements of information transmission in steady state,
the idea here is simple enough but finding an experimental design that actually gives
enough data to avoid all statistical problems is a real challenge, and this is what
Adrienne did in Ref. \cite{fairhall-nature}. The result is clear:  when we switch from
one $P[s]$ to another we can detect the drop in efficiency of information transmission
associated with the use of the old code in the new distribution, and we can measure the
time course of information recovery as the code adapts.  What surprised us (although it
shouldn't have) was the speed of this recovery, which can be complete in much less than
100 msec.  In fact, for the conditions of these experiments, we can actually calculate
how rapidly an optimal processor could make a reliable decision about the change in
distribution, and when the dust settles the answer is that the dynamics of the
adaptation that we see in the fly are running within a factor of two of the maximum
speed set by these limits to statistical inference.\footnote{Again there is more to
these experiments than what I have emphasized here.  The process of adaptation in fact
has multiple time scales, ranging from tens of milliseconds out to many minutes.  These
rich dynamics offer possibilities for longer term statistical properties of the spike
train to resolve the ambiguities (how does the fly know the absolute scale of velocity
if it is scaled away?) that arise in any adaptive coding scheme.  The result is that
while information about the scaled stimulus recovers quickly during the process of
adaptation, some information about the scale itself is preserved and can be read out by
simple algorithms.}

I have the feeling that my presentation of these ideas mirrors their development.  It
took us a long time to build up the tools that bring information theory and optimization
principles into contact with real experiments on the neural coding of complex, dynamic
signals.  Once we have the tools, however, the results are clear, at least for this one
system where we (or, more precisely, Rob) can do almost any experiment we want:
\begin{itemize}
\item Spike trains convey information about natural input signals with $\sim 50\%$
efficiency down to millisecond time resolution.
\item This efficiency is enhanced significantly by synergistic coding in which
temporal patterns of spikes stand for more than the sum of their parts.
\item Although the detailed structure of the neural code is highly individualized,
these basic features are strikingly constant across individuals.
\item Coding efficiency and information rates are higher under more natural conditions.
\item The observed information transmission rates are the result of an adaptive coding
scheme which takes a simple scaling form in response to changes in the dynamic range of
the inputs.
\item The precise choice of scale by the real code is the one which maximizes
information transmission.
\item The dynamics of this adaptation process are almost as fast as possible given the
need to collect statistical evidence for changes in the input distribution.
\end{itemize}
I think it is fair to say that this body of work provides very strong evidence in
support of information theoretic optimization as a ``design principle'' within which we
can understand the phenomenology of the neural code.

\section{Learning and complexity}

The world around us, thankfully, is a rather structured place.
Whether we are doing a careful experiment in the laboratory or are gathering sense data
on a walk through the woods, the signals that arrive at our brains are far from
random noise;  there appear to be some underlying regularities or
rules.  Surely one task  our brain must face is the learning or extraction of
these rules and regularities.  Perhaps the simplest example of learning a rule
is fitting a function to data---we believe in advance that the rule belongs
to a class of possible rules that can be parameterized, and as we collect data
we learn the values of the parameters.   This simple example introduces us to many deep
issues:
\begin{itemize}
\item If there is noise in the data then really we are
trying to learn a probability distribution, not just a functional relation.
\item We
would like to compare models (classes of possible rules) that have different
numbers of parameters, and incorporate the intuition that `simpler' models are
better.
\item We might like to step outside the restrictions of finite
parameterization and consider the possibility that the data are described by
functions that are merely `smooth' to some degree.
\item We would like to quantify how much we are learning
(or how much {\em can} be learned) about the underlying rules.
\end{itemize}
In the last
decade or so, a rich literature has emerged, tackling these problems with a
sophistication far beyond the curve fitting exercises that we all performed in
our student physics laboratories.  I will try to take a path through these
developments, emphasizing the connections of these learning problems
to problems in statistical mechanics and the implications of this statistical
approach for an information theoretic characterization of how much we learn.
Most of what I have to say on this subject is drawn from collaborations with Ilya
Nemenman and Tali Tishby \cite{bnt,bnt-israel,tpb}; in particular the first of these
papers is long and has lots of references to more standard things which I will outline
here without attribution.

Let's just plunge in with the classic example: We
observe two streams of data $x$ and $y$, or equivalently a stream of
pairs $(x_{\rm 1} , y_{\rm 1})$, $(x_{\rm 2} , y_{\rm 2})$, $\cdots$ ,
$(x_{\rm N} , y_{\rm N})$.  Assume that we know in advance that the
$x$'s are drawn independently and at random from a distribution
$P(x)$, while the $y$'s are noisy versions of some function acting on
$x$,
\begin{eqnarray}
y_{\rm n} = f(x_{\rm n} ; {\bgm\alpha} ) + \eta_{\rm n} ,
\end{eqnarray}
where $f(x; {\bgm\alpha})$ is one function from a class of functions parameterized by
$\bgm\alpha \equiv \{\alpha_1, \alpha_2, \cdots , \alpha_K\}$ and $\eta_{\rm n}$ is
noise, which for simplicity we will assume is Gaussian with known standard deviation
$\sigma$.  We can even start with a {\em very} simple case, where the function class
is just a linear combination of basis functions, so that
\begin{eqnarray}
f(x; {\bgm\alpha}) = \sum_{\rm \mu =1}^K \alpha_\mu \phi_\mu (x) .
\label{basis}
\end{eqnarray}
The usual problem is to estimate, from $N$ pairs $\{x_{\rm i} , y_{\rm i}\}$, the values
of the parameters $\bgm\alpha$; in favorable cases such as this we might even be able to
find an effective regression formula.  Probably you were taught that the way to do this
is to compute $\chi^2$,
\begin{equation}
\chi^2 = \sum_{\rm n} [y_{\rm n} - f(x_{\rm n} ; {\bgm\alpha} )]^2 ,
\end{equation}
and then minimize to find the correct parameters ${\bgm\alpha}$.  You may or may not
have been taught {\em why} this is the right thing to do.

With the model described above, the probability that we will observe the pairs $\{x_{\rm
i} , y_{\rm i}\}$ can be written as
\begin{equation}
P(\{x_{\rm i} , y_{\rm i}\}|{\bgm \alpha}) = 
\exp\left[ - {N\over 2}\ln (2\pi\sigma^2) - {\chi^2\over{2\sigma^2}}\right]
\prod_{\rm n} P(x_{\rm n}) ,
\end{equation}
assuming that we know the parameters.  Thus finding parameters which minimize $\chi^2$
also serves to maximize the probability that our model could have given rise to the
data.  But why is this a good idea?

We recall that the entropy is the expectation value of $-\log P$, and that it is possible
to encode signals so that the amount of ``space'' required to specify each signal
uniquely is on average equal to the entropy.\footnote{This is obvious for uniform
probability distributions with $2^n$ alternatives, since then the binary number
representing each alternative is this code we want.  For nonuniform
distributions we need to think about writing things down many times and taking an
average of the space we use each time, and the fact that the answer comes out the same
(as the entropy) hinges on the ``typicality'' of such data streams, which is the
information theorist's way of talking about the equivalence of canonical and
microcanonical ensembles.}  With a little more work one can show that each possible
signal $s$ drawn from $P(s)$ can be encoded in a space of $-\log_2 P(s)$ bits. 
Now any model probability distribution implicitly defines a scheme for coding signals that
are drawn from that distribution, so if we make sure that our data have high probability
in the distribution (small values of $-\log P$) then we also are making sure that our
code or representation of these data is compact.  What this means is that good old
fashioned curve fitting really is all about finding efficient representations of data,
precisely the principle enunciated by Barlow for the operation of the nervous system
(!).

If we follow this notion of efficient representation a little further we can do better
than just maximizing $\chi^2$.  The claim that a model provides a code for the data is
not complete, because at some point I have to represent my knowledge of the model
itself.  One idea is to do this explicitly---estimate how accurately you know each of the
parameters, and then count how many bits you'll need to write down the parameters to
that accuracy and add this to the length of your code; this is the point of view taken
by Risannen and others in a set of ideas called ``minimum description length'' or MDL.
Another idea is more implicit---the truth is that I don't really know the parameters,
all I do is estimate them from the data, so it's not so obvious that I should separate
coding the data from coding the parameters (although this might emerge as an
approximation).  In this view what we should do is to integrate over all possible values
of the parameters, weighted by some prior knowledge (maybe just that the parameters are
bounded), and thus compute the probability that our data could have arisen from the
{\em class} of models we are considering.   

To carry out this program of computing the total probability of the data given the model
class we need to do the integral
\begin{eqnarray}
P(\{x_{\rm i} , y_{\rm i}\}|{\rm class}) &= &
\int d^K\alpha\,P({\bgm\alpha}) P[\{x_{\rm i} , y_{\rm i}\} | {\bgm\alpha}]\\
&=&
\left[\prod_{\rm n} P(x_{\rm n})\right]\nonumber\\
&&\times
\int d^K\alpha\,P({\bgm\alpha}) \exp\left[ - {N\over 2}\ln
(2\pi\sigma^2) - {\chi^2\over{2\sigma^2}}\right].
\end{eqnarray}
But remember that $\chi^2$ as we have defined it is a sum over data points, which means
we expect it (typically) will be proportional to $N$.  This means that at large $N$ we
are doing an integral in  which the exponential has terms proportional to $N$---and so
we should use a saddle point approximation.  The saddle point of course is close to the
place where $\chi^2$ is minimized, and then we do the usual Gaussian (one loop) integral
around this point; actually if we stick with the simplest case of Eq. (\ref{basis})
then this Gaussian approximation becomes exact. When the dust settles we find
\begin{equation}
-\ln P(\{x_{\rm i} , y_{\rm i}\}|{\rm class}) = -\sum_{\rm n}\ln P(x_{\rm n})
+ {{\chi_{\rm min}^2}\over{2\sigma^2}} + {K\over 2}\ln N + \cdots ,
\end{equation}
and we recall that this measures the length of the shortest code for $\{x_{\rm i},y_{\rm
i}\}$ that can be generated given the class of models. 
The first term averages to $N$ times the entropy of the distribution
$P(x)$, which makes sense since by hypothesis the $x$'s are being chosen at random.  The
second term is as before, essentially the length of the code required to describe the
deviations of the data from the predictions of the best fit model; this also grows in
proportion to
$N$.  The third term must be related to coding our knowledge of the model itself, since
it is proportional to the number of parameters.  We can understand the ${(1/2)}\ln
N$ because each parameter is determined to an accuracy of $\sim 1/\sqrt N$, so if we
start with a parameter space of size $\sim 1$ there is a reduction in volume by a factor
of $\sqrt N$ and hence a decrease in entropy (gain in information)  by
${(1/2)}\ln N$.  Finally, the terms $\cdots$ don't grow with $N$.

What is crucial about the term $(K/2)\ln N$ is that it depends explicitly on the number
of parameters.  In general we expect that by considering models with more parameters we
can get a better fit to the data, which means that $\chi^2$ can be reduced by
considering more complex model classes.  But we know intuitively that this has to
stop---we don't want to use arbitrarily complex models, even if they do provide a good
fit to what we have seen.  It is attractive, then, that if we look for the shortest code
which can be generated by a class of models, there is an implicit penalty or coding cost
for increased complexity.  It is interesting from a physicist's point of view that this
term emerges essentially from consideration of phase space or volumes in
model space.  It thus is  an entropy--like quantity in its own right, and the selection
of the best model class could be thought of as a tradeoff between this entropy and the
``energy'' measured by $\chi^2$.  If we keep going down this line of thought we can
imagine a thermodynamic limit with large numbers of parameters and data points, and
there can be ``aha!'' types of phase transitions from poor fits to good fits as we
increase the ratio $N/K$ \cite{aha}.

The reason we need to control the complexity of our models is because the real problem
of learning is neither the estimation of parameters nor the compact representation of
the data we have already seen.  The real problem of learning is {\em generalization}:  we
want to extract the rules underlying what we have seen because we believe that these
rules will continue to be true and hence will describe the relationships among data that
we will observe in the future.  Our experience is that overly complex models might
provide a more accurate description of what we have seen so far but do a bad job at
predicting what we will see next.  This suggests that there are connections between
predictability and complexity.  

There is in fact a completely different motivation for quantifying complexity, and this
is to make precise an impression that some systems, such as life on
earth or a turbulent fluid flow, evolve toward a state of higher
complexity; one might even like to classify these states.  These problems
traditionally are in the realm of dynamical systems theory and statistical physics.   A
central difficulty in this effort is to distinguish complexity from
randomness---trajectories of dynamical systems can be regular, which we take to mean
``simple'' in the intuitive sense, or chaotic, but what we mean by complex is
somewhere in between.  The field of complexology (as Magnasco likes to call it) is filled
with multiple definitions of complexity and confusing remarks about what they all might
mean.  In this noisy environment, there is a wonderful old paper by Grassberger
\cite{grassberger} which gives a clear signal:  Systems with regular or chaotic/random
dynamics share the property that the entropy of sample trajectories is almost exactly
extensive in the length of the trajectory, while for systems that we identify
intuitively as being complex there are large corrections to extensivity which can even
diverge as we take longer and longer samples.  In the end Grassberger suggested that
these subextensive terms in the entropy really do quantify our intuitive notions of
complexity, although he made this argument by example rather than axiomatically.

We can connect the measures of complexity that arise in learning problems with those that
arise in dynamical systems  by
noticing that the subextensive components of entropy identified by
Grassberger in fact determine the information available for making
predictions.\footnote{The text of the discussion here follows Ref. \cite{bnt-israel}
rather closely, and I thank my colleagues for permission to include it here.}  This also
suggests a connection to the importance or value of information, especially in a
biological or economic context: information is valuable if it can be used to guide our
actions, but actions take time and hence observed data can be useful only to the extent
that those data inform us about the state of the world at later times.  It would be
attractive if what we identify as ``complex'' in a time series were also the ``useful''
or ``meaningful'' components.

While prediction may come in various forms, depending on context,
information theory allows us to treat all of them on the same footing.
For this we only need to recognize that all predictions are
probabilistic, and that, even before we look at the data, we know that
certain futures are more likely than others. This knowledge can be
summarized by a prior probability distribution for the futures.  Our
observations on the past lead us to a new, more tightly concentrated
distribution, the distribution of futures conditional on the past
data. Different kinds of predictions are different slices through or
averages over this conditional distribution, but information theory
quantifies the ``concentration'' of the distribution without making
any commitment as to which averages will be most interesting.

Imagine that we observe a stream of data $x(t)$ over a time interval
$-T < t < 0$; let all of these past data be denoted by the shorthand
$x_{\rm past}$.  We are interested in saying something about the
future, so we want to know about the data $x(t)$ that will be observed
in the time interval $0 < t < T'$; let these future data be called
$x_{\rm future}$.  In the absence of any other knowledge, futures are
drawn from the probability distribution $P(x_{\rm future})$, while
observations of particular past data $x_{\rm past}$ tell us that
futures will be drawn from the conditional distribution $P(x_{\rm
  future} | x_{\rm past})$. The greater concentration of the
conditional distribution can be quantified by the fact that it has
smaller entropy than the prior distribution, and this reduction in
entropy is  the information that the past
provides about the future.  We can write the average of this {\em
  predictive information} as
\begin{eqnarray}
{\cal I}_{\rm pred} (T,T') &=& 
{\Bigg\langle} \log_2 \left[ {{P(x_{\rm future}| x_{\rm past})} 
\over{P(x_{\rm future})}}\right]\Bigg\rangle
  \\ 
&=& -\langle\log_2 P(x_{\rm future})\rangle 
- \langle\log_2 P( x_{\rm past})\rangle
\nonumber\\
&&\,\,\,\,\,\,\,\,\,\, 
-\left[-\langle\log_2 P(x_{\rm future}, x_{\rm past})\rangle\right]\,,
\label{ents}
\end{eqnarray}
where $\langle \cdots \rangle$ denotes an average over the joint
distribution of the past and the future, $P(x_{\rm future} , x_{\rm
  past})$.

Each of the terms in Eq.~(\ref{ents}) is an entropy. Since we are
interested in predictability or generalization, which are associated
with some features of the signal persisting forever, we may assume
stationarity or invariance under time translations. Then the entropy
of the past data depends only on the duration of our observations, so
we can write $ -\langle\log_2 P( x_{\rm past})\rangle = S(T) $, and by
the same argument $-\langle\log_2 P( x_{\rm future})\rangle = S(T')$.
Finally, the entropy of the past and the future taken together is the
entropy of observations on a window of duration $T+T'$, so that $
-\langle\log_2 P(x_{\rm future} , x_{\rm past})\rangle = S(T+T')$.
Putting these equations together, we obtain
\begin{equation}
{\cal I}_{\rm pred}(T,T') = S(T) +S(T') - S(T+T') . \label{IpredandST}
\end{equation}

In the same way that the entropy of a gas at fixed density is
proportional to the volume, the entropy of a time series
(asymptotically) is proportional to its duration, so that
$\lim_{T\rightarrow\infty} {{S(T)}/ T} = {\cal S}_0$; entropy is an
extensive quantity.  But from Eq.~(\ref{IpredandST}) any extensive
component of the entropy cancels in the computation of the predictive
information: {\em predictability is a deviation from extensivity}.  If
we write
\begin{equation}
  S(T) = {\cal S}_0 T +S_1(T)\,,
\end{equation}
then Eq.~(\ref{IpredandST}) tells us that the predictive information
is related {\em only} to the nonextensive term $S_1(T)$.

We know two general facts about the behavior of $S_1(T)$.  First, the
corrections to extensive behavior are positive, $S_1(T) \geq 0$.
Second, the statement that entropy is extensive is the statement that
the limit 
\begin{equation}
\lim_{T\rightarrow\infty} {{S(T)} / T} = {\cal S}_0 
\end{equation}
exists, and for this to be true we must also have $ \lim_{ T
  \rightarrow \infty} {{S_1(T)} / T} = 0.$ Thus the nonextensive terms
in the entropy must be {\em sub}extensive, that is they must grow with
$T$ less rapidly than a linear function.  Taken together, these facts
guarantee that the predictive information is positive and
subextensive.  Further, if we let the future extend forward for a very
long time, $T' \rightarrow \infty$, then we can measure the
information that our sample provides about the entire future,
\begin{equation}
 I_{\rm pred} (T) = 
\lim_{T' \rightarrow \infty} {\cal I}_{\rm pred}(T,T') = S_1 (T)\,,
\end{equation}
and this is precisely equal to the subextensive entropy.

If we have been observing a time series for a (long) time $T$, then
the total amount of data we have collected in is measured by the
entropy $S(T)$, and at large $T$ this is given approximately by ${\cal
  S}_0 T$.  But the predictive information that we have gathered
cannot grow linearly with time, even if we are making predictions
about a future which stretches out to infinity. As a result, of the
total information we have taken in by observing $x_{\rm past}$, only a
vanishing fraction is of relevance to the prediction:
\begin{equation}
\lim_{T\rightarrow\infty} {{\rm Predictive\ Information} \over{\rm
Total\ Information}} = {{I_{\rm pred} (T)} \over {S(T)}}
\rightarrow 0. \label{chuck}
\end{equation}
In this precise sense, most of what we observe is irrelevant to the
problem of predicting the future.  

Consider the case where time is measured in discrete steps, so that we
have seen $N$ time points $x_1, x_2 , \cdots , x_N$. How much is there
to learn about the underlying pattern in these data? In the limit of
large number of observations, $ N \to \infty$ or $T \to \infty$, the
answer to this question is surprisingly universal: predictive
information may either stay finite, or grow to infinity together with
$T$; in the latter case the rate of growth may be slow (logarithmic)
or fast (sublinear power).

The first possibility, $\lim_{T\rightarrow\infty} I_{\rm pred} (T) = $
constant, means that no matter how long we observe we gain only a
finite amount of information about the future. This situation
prevails, in both extreme cases mentioned above. For example, when the
dynamics are very regular, as for a purely periodic system,
complete prediction is possible once we know the phase, and if we
sample the data at discrete times this is a finite amount of
information; longer period orbits intuitively are more complex and
also have larger $I_{\rm pred}$, but this doesn't change the limiting
behavior $\lim_{T\rightarrow\infty} I_{\rm pred} (T) =$ constant.

Similarly, the predictive information can be small when the dynamics
are irregular but the best predictions are controlled only by the
immediate past, so that the correlation times of the observable data
are finite. This happens, for example, in many physical systems far
away from phase transitions. Imagine, for example, that we observe
$x(t)$ at a series of discrete times $\{t_{\rm n}\}$, and that at each
time point we find the value $x_{\rm n}$. Then we always can write the
joint distribution of the $N$ data points as a product,
\begin{eqnarray}
P(x_1 , x_2 , \cdots , x_N ) &=& P(x_1 ) P(x_2 | x_1) 
P(x_3 | x_2 , x_1) \cdots . 
\end{eqnarray}
For Markov processes, what we observe at $t_{\rm n}$ depends only on
events at the previous time step $t_{\rm n-1}$, so that
\begin{eqnarray}
P(x_{\rm n} | \{x_{\rm 1\leq i \leq n-1}\}) &=& 
P(x_{\rm n} | x_{\rm n-1}) ,
\end{eqnarray}
and hence the predictive information reduces to
\begin{eqnarray}
I_{\rm pred} = \Bigg\langle \log_2 \left[ {{P(x_{\rm n}|x_{\rm n-1})}
\over{P(x_{\rm n})}} \right] \Bigg\rangle .
\end{eqnarray}
The maximum possible predictive information in this case is the
entropy of the distribution of states at one time step, which in turn
is bounded by the logarithm of the number of accessible states. To
approach this bound the system must maintain memory for a long time,
since the predictive information is reduced by the entropy of the
transition probabilities. Thus systems with more states and longer
memories have larger values of $I_{\rm pred}$.

\medskip
\addtocounter{problem}{1}
{\small
\noindent {\bf Problem \theproblem: Brownian motion of a spring.} 
Consider the Brownian motion of an overdamped particle bound a spring.
The Langevin equation describing the particle position $x(t)$ is
\begin{equation}
\gamma {{dx(t)}\over{dt}} + \kappa x(t) = F_{\rm ext} (t) + \delta F(t),
\end{equation}
where $\gamma$ is the damping constant, $\kappa$  is the stiffness of
the spring, $F_{\rm ext}(t)$ is an external force that might be applied
to the particle, and $\delta F(t)$ is the Langevin force.  The Langevin
force is random, with Gaussian statistics and white noise correlation
properties,
\begin{equation}
\langle \delta F(t) \delta F(t') \rangle = 2\gamma k_B T \delta (t-t') .
\end{equation}
Show  that the
correlation function has a simple exponential form,
\begin{equation}
\langle x(t) x(t') \rangle = \langle x^2 \rangle \exp(-|t - t'|/\tau_{\rm c}),
\end{equation}
and evaluate the correlation time.
Now take
the original Langevin equation and form a discrete version, introducing a small time step
$\Delta t$; be sure that your discretization preserves exactly the
observable variance
$\langle x^2
\rangle$. You should be able to find a natural discretization in which the evolution of
$x$ is Markovian, and then you can compute the predictive information for the time
series $x(t)$.  How does this result depend on temperature?  Why?  Express the
dependence on $\Delta t$ in units of the correlation time $\tau_{\rm c}$.  Is there a
well defined limit as $\Delta t \rightarrow 0$?  Again, why (or why
not)?} 
\bigskip

More interesting are those cases in which $I_{\rm pred}(T)$ diverges
at large $T$. In physical systems we know that there are critical
points where correlation times become infinite, so that optimal
predictions will be influenced by events in the arbitrarily distant
past. Under these conditions the predictive information can grow
without bound as $T$ becomes large; for many systems the divergence is
logarithmic, $I_{\rm pred} (T\rightarrow\infty) \propto \log T$.

Long range correlation also are important in a time series where we
can learn some underlying rules, as in the discussion of curve fitting that started this
section.  Since we saw that curve fitting with noisy data really involves a
probabilistic model, let us talk explicitly about the more general problem of learning
distributions.   Suppose a series of random vector
variables $\{{\vec x}_{\rm i}\}$ are drawn independently from the same
probability distribution $Q({\vec x} | {\bgm\alpha})$, and this
distribution depends on a (potentially infinite dimensional) vector of
parameters $\bgm{\alpha}$. The parameters are unknown, and before the
series starts they are chosen randomly from a distribution ${\cal
  P}(\bgm{\alpha})$.  In this setting, at least implicitly, our
observations of $\{{\vec x}_{\rm i}\}$ provide data from which we can
learn the parameters $\bgm{\alpha}$.  Here we put aside (for the
moment) the usual problem of learning---which might involve
constructing some estimation or regression scheme that determines a
``best fit'' $\bgm{\alpha}$ from the data $\{{\vec x}_{\rm i}\}$---and
treat the ensemble of data streams $P[\{{\vec x}_{\rm i}\}]$ as we
would any other set of configurations in statistical mechanics or
dynamical systems theory.  In particular, we can compute the entropy
of the distribution $P[\{{\vec x}_{\rm i}\}]$ even if we can't provide
explicit algorithms for solving the learning problem.

As shown in \cite{bnt},  the crucial quantity in such analysis is
the density of models in the vicinity of the target
$\bar{\bgm\alpha}$---the parameters that actually generated the
sequence. For two distributions, a natural distance measure is the
Kullback--Leibler divergence,
\begin{equation}
D_{\rm KL}(\bar{\bgm\alpha} || {\bgm\alpha}) =
\int d{\vec x} Q({\vec x} | \bar{\bgm\alpha})\log \left[{{ Q({\vec x} |
  \bar{\bgm\alpha}) }\over{Q({\vec x} | {\bgm\alpha}) }}\right],
\end{equation} and the
density is
\begin{equation}
\rho (D;{\bar{\bgm\alpha}}) = \int d^K\alpha  {\cal P} ({\bgm\alpha})
\delta [D - D_{\rm KL} ( \bar{\bgm\alpha} || {\bgm\alpha} )].
\end{equation}
If $\rho$ is large as $D\to 0$, then one easily can get close to the
target for many different data; thus they are not very informative. On
the other hand, small density means that only very particular data
lead to $\bar{\bgm\alpha}$, so they carry a lot of predictive
information.  Therefore, it is clear that the density, but not the
number of parameters or any other simplistic measure, characterizes
predictability and the complexity of prediction. If, as often is the
case for $\dim {\bgm\alpha} <\infty$, the density behaves in the way
common to finite dimensional systems of the usual statistical
mechanics,
\begin{equation}
  \rho(D\to 0, \bar{\bgm\alpha}) \approx A D^{(K-2)/2}\,,
\end{equation}
then the predictive information to the leading order is
\begin{equation}
  I_{\rm pred}(N) \approx {K\over 2} \log N\,.
\end{equation}

The modern theory of learning is concerned in large part with
quantifying the complexity of a model class, and in particular with
replacing a simple count of parameters with a more rigorous notion of
dimensionality for the space of models; for a general review of these
ideas see Ref.~\cite{vapnik}, and for a discussion close in spirit to
this one see Ref.~\cite{vijay}.  The important point here is that the
dimensionality of the model class, and hence the complexity of the
class in the sense of learning theory, emerges as the coefficient of
the logarithmic divergence in $I_{\rm pred}$.  Thus a measure of
complexity in learning problems can be derived from a more general
dynamical systems or statistical mechanics point of view, treating the
data in the learning problem as a time series or one dimensional
lattice.  The logarithmic complexity class that we identify as being
associated with finite dimensional models also arises, for example, at
the Feigenbaum accumulation point in the period doubling route to
chaos \cite{grassberger}.

As noted by Grassberger in his original discussion, there are time
series for which the divergence of $I_{\rm pred}$ is stronger than a
logarithm.  We can construct an example by looking at the density
function $\rho$ in our learning problem above: finite dimensional
models are associated with algebraic decay of the density as
$D\rightarrow 0$, and we can imagine that there are model classes in
which this decay is more rapid, for example
\begin{equation}
  \rho (D \rightarrow 0)
  \approx A \exp\left[ -{B /{D^\mu}}\right] ,\,\ \mu>0 \,.
\end{equation}  
In this case it can be shown that the predictive information diverges
very rapidly, as a sublinear power law,
\begin{equation}
  I_{\rm pred} (N) \sim N^{\mu/(\mu+1)} \,.
\end{equation}
One way that this scenario can arise is if the distribution $Q({\vec
  x})$ that we are trying to learn does not belong to any finite
parameter family, but is itself drawn from a distribution that
enforces a degree of smoothness \cite{bcs}.  Understandably, stronger
smoothness constraints have smaller powers (less to predict) than the
weaker ones (more to predict).  For example, a rather simple case of
predicting a one dimensional variable that comes from a continuous
distribution produces $I_{\rm pred}(N) \sim \sqrt{N}$.

As with the logarithmic class, we expect that power--law divergences
in $I_{\rm pred}$ are not restricted to the learning problems that we
have studied in detail.  The general point is that such behavior will
be seen in problems where predictability over long scales, rather then
being controlled by a fixed set of ever more precisely known
parameters, is governed by a progressively more detailed
description---effectively increasing the number of parameters---as we
collect more data.  This seems a plausible description of what happens
in language, where rules of spelling allow us to predict forthcoming
letters of long words, grammar binds the words together, and
compositional unity of the entire text allows  predictions
about the subject of the last page of the book after reading only the
first few.  Indeed, Shannon's classic experiment on the predictability
of English text (by human readers!) shows this behavior
\cite{shannon-lang,hilberg}, and more recently several groups have extracted
power--law subextensive components from the numerical analysis of
large corpora of text.

Interestingly, even without an explicit example, a simple argument
ensures existence of exponential densities and, therefore, power law
predictive information models.  If the number of parameters in a
learning problem is not finite then in principle it is impossible to
predict anything unless there is some appropriate regularization.  If
we let the number of parameters stay finite but become large, then
there is {\em more} to be learned and correspondingly the predictive
information grows in proportion to this number. On the other hand, if
the number of parameters becomes infinite without regularization, then
the predictive information should go to zero since nothing can be
learned.  We should be able to see this happen in a regularized
problem as the regularization weakens: eventually the regularization
would be insufficient and the predictive information would vanish.
The only way this can happen is if the predictive information grows
more and more rapidly with $N$ as we weaken the regularization, until
finally it becomes extensive (equivalently, drops to zero) at the
point where prediction becomes impossible. To realize this scenario we
have to go beyond $I_{\rm pred} \propto \log T$ with $I_{\rm pred}
\propto N^{\mu/(\mu+1)}$; the transition from increasing predictive
information to zero occurs as $\mu \to 1$.

This discussion makes it clear that the predictive information (the
subextensive entropy) distinguishes between problems of intuitively
different complexity and thus, in accord to Grassberger's definitions
\cite{grassberger}, is probably a good choice for a universal
complexity measure. Can this intuition be made more precise?

First we need to decide whether we want to attach measures of
complexity to a particular signal $x(t)$ or whether we are interested
in measures that are defined by an average over the ensemble
$P[x(t)]$. One problem in assigning complexity to single realizations
is that there can be atypical data streams. Second, Grassberger
\cite{grassberger} in particular has argued that our visual intuition
about the complexity of spatial patterns is an ensemble concept, even
if the ensemble is only implicit. The fact that we admit probabilistic
models is crucial: even at a colloquial level, if we allow for
probabilistic models then there is a simple description for a sequence
of truly random bits, but if we insist on a deterministic model then
it may be very complicated to generate precisely the observed string
of bits. Furthermore, in the context of probabilistic models it hardly
makes sense to ask for a dynamics that generates a particular data
stream; we must ask for dynamics that generate the data with
reasonable probability, which is more or less equivalent to asking
that the given string be a typical member of the ensemble generated by
the model.  All of these paths lead us to thinking not about single
strings but about ensembles in the tradition of statistical mechanics,
and so we shall search for measures of complexity that are averages
over the distribution $P[x(t)]$.

Once we focus on average quantities, we can provide an axiomatic proof
(much in the spirit of Shannon's \cite{shannon48} arguments establishing
entropy as a unique information measure) that links $I_{\rm pred}$ to
complexity. We can start by adopting Shannon's postulates as
constraints on a measure of complexity: if there are $N$ equally
likely signals, then the measure should be monotonic in $N$; if the
signal is decomposable into statistically independent parts then the
measure should be additive with respect to this decomposition; and if
the signal can be described as a leaf on a tree of statistically
independent decisions then the measure should be a weighted sum of the
measures at each branching point. We believe that these constraints
are as plausible for complexity measures as for information measures,
and it is well known from Shannon's original work that this set of
constraints leaves the entropy as the only possibility.  Since we are
discussing a time dependent signal, this entropy depends on the
duration of our sample, $S(T)$.  We know of course that this cannot be
the end of the discussion, because we need to distinguish between
randomness (entropy) and complexity.  The path to this distinction is
to introduce other constraints on our measure.

First we notice that if the signal $x$ is continuous, then the entropy
is not invariant under transformations of $x$, even if these reparamterizations  do
not mix points at different times.  It seems reasonable to ask
that complexity be a function of the process we are observing and not
of the coordinate system in which we choose to record our
observations. However, it is not the whole function $S(T)$ which
depends on the coordinate system for $x$; it is only the extensive
component of the entropy that has this noninvariance.  This can be
seen more generally by noting that subextensive terms in the entropy
contribute to the mutual information among different segments of the
data stream (including the predictive information defined here), while
the extensive entropy cannot; mutual information is coordinate
invariant, so all of the noninvariance must reside in the extensive
term.  Thus, any measure complexity that is coordinate invariant must
discard the extensive component of the entropy.

If we continue along these lines, we can think about the asymptotic
expansion of the entropy at large $T$.  The extensive term is the
first term in this series, and we have seen that it must be discarded.
What about the other terms?  In the context of predicting in a
parameterized model, most of the terms in this series depend in detail
on our prior distribution in parameter space, which might seem odd for
a measure of complexity.  More generally, if we consider
transformations of the data stream $x(t)$ that mix points within a
temporal window of size $\tau$, then for $T >> \tau$ the entropy
$S(T)$ may have subextensive terms which are constant, and these are
not invariant under this class of transformations.  On the other hand,
if there are divergent subextensive terms, these {\em are} invariant
under such temporally local transformations. So if we
insist that measures of complexity be invariant not only under
instantaneous coordinate transformations, but also under temporally
local transformations, then we can discard both the extensive and the
finite subextensive terms in the entropy, leaving only the divergent
subextensive terms as a possible measure of complexity.

To illustrate the purpose of these two extra conditions, we may think
of   measuring the velocity of a turbulent fluid flow
at a given point. The condition of invariance under
reparameterizations means that the complexity is independent of the
scale used by the speedometer. On the other hand, the second condition
ensures that the temporal filtering due to the finite  inertia
of the speedometer's needle does not change the estimated complexity
of the flow.

I believe that these arguments (or their slight variation also presented
in \cite{bnt}) settle the question of the unique definition of
complexity. Not only is the divergent subextensive component of the
entropy the unique complexity measure, but it is also a universal one
since it is connected in a straightforward way to many other measures
that have arisen in statistics and in dynamical systems theory. In my mind the really
big open question is whether we can connect {\em any} of these theoretical developments
to experiments on learning by real animals (including humans).

I have emphasized the problem of learning probability distributions or
probabilistic models rather than learning deterministic functions,
associations or rules. In the previous section we have discussed examples where  the
nervous system adapts to the statistics of its inputs; these
experiments can be thought of as  a simple example of the system learning a
parameterized distribution.  When making saccadic eye movements, human subjects alter
their distribution of reaction times in relation to the relative probabilities of
different targets, as if they had learned an estimate of the relevant likelihood ratios
\cite{carpenter-williams}.  Humans also can learn to discriminate
almost optimally between random sequences (fair coin tosses) and
sequences that are correlated or anticorrelated according to a Markov
process; this learning can be accomplished from examples alone, with
no other feedback \cite{lopes+oden}.  Acquisition of language may
require learning the joint distribution of successive phonemes,
syllables, or words, and there is direct evidence for learning of
conditional probabilities from artificial sound sequences, both by
infants and by adults \cite{saffran1,saffran2}.  

Classical examples of learning in animals---such as eye blink conditioning in
rabbits---also may harbor evidence of learning probability distributions.  The usual
experiment is to play a brief sound followed by a puff of air to the eyes, and then the
rabbit learns to blink its eye at the time when the air puff is expected.  But if the
time between a sound and a puff of air to the eyes is chosen from a probability
distribution, then rabbits will perform graded movements of the eyelid that seem to
more or less trace the shape of the distribution, as if trying to have the exposure of
the eye matched to the (inverse) likelihood of the noxious stimulus \cite{mauk}. These
examples, which are not exhaustive, indicate that the nervous system can learn an
appropriate probabilistic model, and this offers the opportunity to analyze the dynamics
of this learning using information theoretic methods: What is the entropy of
$N$ successive reaction times following a switch to a new set of relative probabilities
in the saccade experiment?  How much information does a single reaction time provide
about the relevant probabilities?  

Using information theory to characterize learning is appealing because the predictive
information in the data itself (that is, in the data from which the subject is being
asked to learn) sets a limit on the generalization power that the subject has at his or
her disposal.  In this sense $I_{\rm pred}$ provides an absolute standard against which
to measure learning performance in the same way that spike train entropy provides a
standard against which to measure the performance of the neural code.  I'm not really
sure how to do this yet, but I can imagine that an information theoretic analysis of
learning would thus lead to a measurement of learning efficiency \cite{hbb-intell} that
parallels the measurement of coding efficiency or even detection efficiency in photon
counting.  Given our classification of learning tasks by their complexity, it would be
natural to ask if the efficiency of learning were a critical function
of task complexity: perhaps we can even identify a limit beyond which
efficient learning fails, indicating a limit to the complexity of the
internal model used by the brain during a class of learning tasks.

\section{A little bit about molecules}

It would be irresponsible to spend this many hours (or pages) on the brain without
saying something that touches the explosion in our knowledge of what happens at the
molecular level.  Electrical signals in neurons are carried by
ions, such as potassium or sodium, flowing through
water or through specialized conducting pores. These pores,  or channels,
are large molecules (proteins)  embedded in the cell membrane, and can
thus respond to the electric field or voltage across the membrane.   The
coupled dynamics of channels and voltages makes each neuron into a
nonlinear circuit, and this seems to be the molecular basis for neural
computation. Many cells have the property that these nonlinear dynamics
select stereotyped pulses that can propagate from one cell to another;  these
action potentials  are the dominant form of long distance cell
to cell communication in the brain, and our understanding of how these
pulses occur is one the triumphs of the (now) classical `biophysics.'
Signals also can be carried by small molecules, which trigger various
chemical reactions when they arrive at their targets. In particular,
signal transmission across the synapse, or connection between two
neurons, involves such small molecule messengers called neurotransmitters. Calcium
ions can play both roles, moving in response to voltage gradients and
regulating a number of important biochemical reactions in living cells,
thereby coupling electrical and chemical events.  Chemical events can
reach into the cell nucleus to regulate which protein molecules---which
ion channels and transmitter receptors---the cell produces.  We will try
to get a feeling for this range of phenomena, starting on the back of an
envelope and building our way up to the facts.

Ions and small molecules diffuse freely through water, but cells are surrounded
by a membrane that functions as a barrier to diffusion.  In particular, these
membranes are composed of lipids, which are nonpolar, and therefore cannot
screen the charge of an ion that tries to pass through the membrane.  The
water, of course, is polar and does screen the charge, so pulling an ion out of
the water and pushing it through the membrane would require surmounting a large
electrostatic energy barrier. This barrier means that the membrane provides an
enormous resistance to current flow between the inside and the outside of the cell.  If
this were the whole story there would be no electrical signalling in biology.  In
fact, cells construct specific pores or channels through which ions can pass, and
by regulating the state of these channels the cell can control the flow of electric
current across the membrane.

Ion channels are themselves molecules, but very large ones---they are proteins
composed of several thousand atoms in very complex arrangements.  Let's try,
however, to ask a simple question:  If we open a pore in the cell membrane, how
quickly can ions pass through?  More precisely, since the ions carry current
and will move in response to a voltage difference across the membrane, how
large is the current in response to a given voltage?  We recall that the
ratio of current to voltage is called conductance, so we are really asking for
the conductance of an open channel. Again we only want an order of magnitude
estimate, not a detailed theory.

Imagine that one ion channel serves, in effect, as a hole in the membrane.
Let us pretend that ion flow through this hole is essentially the same as
through water. The electrical current that flows through the channel is
\begin{equation}
J = q_{\rm ion} \cdot [{\rm ionic\ flux}] \cdot [{\rm channel\ area}] ,
\end{equation}
where $q_{\rm ion}$ is the charge of one ion,
and we recall that `flux' measures the
current across a unit area, so that
\begin{eqnarray}
{\rm ionic\ flux}
&=&
{{\rm ions} \over {\rm cm^2  s}}
=
{{\rm ions} \over {\rm cm^3}}
\cdot
{{\rm cm}\over {\rm s}}
\\
&=&
[{\rm ionic\ concentration}]
\cdot
[{\rm velocity\ of\ one\ ion}]\\
&=& cv.
\end{eqnarray}
Major current carriers like sodium and
potassium are at
$c \sim 100$ milliMolar, or $c \sim 6\times 10^{19} \,\, {\rm ions/cm^3} $.
The average velocity is related to the
applied force through the mobility $\mu$, the force on an ion is in turn equal to the
electric field times the ionic charge, and the electric field is (roughly) the voltage
difference $V$ across the membrane divided by the thickness $\ell$ of the
membrane:
\begin{equation}
v = \mu q_{\rm ion} E
\sim \mu q_{\rm ion} {V\over\ell}
\sim
{D\over{k_B T}} q_{\rm ion} {V\over\ell} ,
\end{equation}
where in the last step we recall the Einstein relation between mobility and
diffusion constant.  Putting the various factors together we find the
current
\begin{eqnarray} J &=& q_{\rm ion} \cdot [{\rm ionic\ flux}] \cdot [{\rm
channel\ area}]
\nonumber\\
&=& q_{\rm ion} \cdot [cv] \cdot [\pi d^2 /4]
\\
&\sim&
{\pi \over 4}
q_{\rm ion} \cdot
{{c d^2 D}\over \ell}
\cdot
{{q_{\rm ion} V}\over{k_B T}} ,
\end{eqnarray}
where the channel has a diameter $d$.  If we assume that the ion carries one
electronic charge, as does sodium, potassium, or chloride, then  $q_{\rm ion}
=1.6 \times 10^{-19}$ C and
\begin{equation}
{{q_{\rm ion} V}\over{k_B T}}
= {V\over{25\ {\rm mV}}} .
\end{equation}
Typical values for the channel diameter should be comparable to the diameter of
a single ion, $d \sim 0.3$ nm, and the thickness of the membrane is
$\ell \sim 5$ nm.  Diffusion constants for ions in water are $D\sim\, 2\times
10^{-9} {\rm m^2/sec}$, or $\sim 2\,{\rm (\mu m)^2/sec}$, which is a more natural
unit.  Plugging in the numbers,
\begin{eqnarray}
J &=& gV\\
g &\sim& 2 \times 10^{-11}\ {\rm Amperes/Volt}
= 20\ {\rm picoSiemens}.
\end{eqnarray}
So our order of magnitude argument leads us to predict that the conductance of
an open channel is roughly $20\ {\rm pS}$, which is about right experimentally.

Empirically, cell membranes have resistances of $R_{\rm m} \sim 10^3\,{\rm ohm/cm^2}$,
or conductances of $G_{\rm m} \sim 10^{-3} \,{\rm S/cm^2}$.  If each open channel
contributes roughly $10\,{\rm pS}$, then this membrane conductance corresponds to an
average density of $\sim 10^8$ open channels per ${\rm cm^2}$, or roughly one channel
per square micron.  This is correct but misleading.  First, they are many channels
which are, at one time, not open; indeed the dynamics with which channels open and
close is very important, as we shall see momentarily.  Second, channels are not
distributed uniformly over the cell surface.  At the synapse, or connection between
two neurons, the postsynaptic cell has channels that open in
response to the binding of the transmitter molecules released by the presynaptic
cell.  These `receptor channels' form a nearly close packed crystalline array in the
small patch of cell membrane that forms the closest contact with the presynaptic cell,
and there are other examples of great concentrations of channels in other
regions of the cell.

{\small
\medskip
\addtocounter{problem}{1}
\noindent {\bf Problem \theproblem: Membrane capacitance.} 
From the facts given above, estimate the capacitance of the cell membrane.
You should get $C\sim 1\,{\rm \mu F/cm^2}$.
\bigskip
}

Channels are protein molecules: 
heteropolymers of amino acids. As discussed by other lecturers here, there are twenty
types of amino acid and a protein can be anywhere from 50 to 1000 units in length. 
Channels tend to be rather large, composed of several hundred amino acids; often there
are several subunits, each of this size. 
For physicists, the `protein folding problem' is to understand what it is
about real proteins that allows them to collapse into a unique
structure.  This is, to some approximation, a question about the
equilibrium state of the molecule, since for many proteins we can
`unfold' the molecule either by heating or by chemical treatment and then
recover the original structure by returning to the original
condition.\footnote{Of course there are interesting exceptions to this
rule.} At present, this problem is attracting considerable attention in
the statistical mechanics community.   For a biologist, the protein
folding problem is slightly different:  granting that proteins fold into
unique structures, one would like to understand the mapping from the
linear sequence of amino acids in a particular protein into the three
dimensional structure of the folded state.  Again, this is a very
active---but clearly distinct---field of research.

We actually need a little more than a unique folded state for proteins.  Most proteins
have a few rather similar structures which are stable, and the energy differences
between these structures are several (up to $\sim 10$) $k_B T$, which means that the
molecule can be pushed from one state to another by interesting perturbations, such as
the binding of a small molecule.  For channels, there is a more remarkable fact,
namely that (for most channels) out of several accessible states, only one is `open'
and conducting.  The other states are closed or (and this is different!) inactivated.
If we think about arranging the different states in a kinetic scheme, we might write
\begin{equation}
C_1 \leftrightarrow C_2 \leftrightarrow O,
\end{equation}
which corresponds to two closed states and one open state, with the constraint that
the molecule must pass through the second closed state in order to open.
If the open state also equilibrates with an `inactive' state $I$ that is connected to
$C_1$,
\begin{equation}
C_1 \leftrightarrow C_2 \leftrightarrow O \leftrightarrow I \leftrightarrow C_1\,\,,
\label{bigkinetics}
\end{equation}
then depending on the rate constants for the different transitions the channel can be
forced to pass through the inactive state and then through all of the closed states
before opening again.  This is interesting because the physical processes of `closing'
and `inactivating' are often different, and this means that the transition rates can
differ by orders of magnitude:  there are channels that  can flicker open and
closed in a millisecond, but require minutes to recover from inactivation.  If we
imagine that channels open in response to certain inputs to the cell, this process of
inactivation endows the cell with a memory of how many of these inputs have occurred
over the past minute---the states of individual molecules are keeping count, and the
cell can read this count because the molecular states influence the dynamics of
current flow across the membrane.

Individual amino acids have dipole moments, and this means that when the protein makes
a slight change in structure (say $C_2 \rightarrow O$) there will be a change in the
dipole moment of the protein unless there is an incredible coincidence.  But this has
the important consequence that the energy differences among the different states of
the channel will be modulated by the electric field and hence by the voltage across
the cell membrane.  If the difference in dipole moment were equivalent to moving one
elementary charge across the membrane, then we could shift the equilibrium between
the two states by changing the voltage over $\sim k_B T /q_{\rm e} = 25\,{\rm mV}$,
while if there are order ten charges transferred the channel will switch from one
state to another over just a few mV.  While molecular rearrangements within the
channel protein do not correspond to charge transfer across the whole thickness of the
membrane, the order of magnitude change in dipole moment is in this range.

It is important to understand that one can measure the current flowing through single
channels in a small patch of membrane, and hence one can observe the statistics of
opening and closing transitions in a single molecule.  From such experiments one can
build up kinetic models like that in Eq. (\ref{bigkinetics}), and these provide an
essentially exact description of the dynamics at the single molecule level.  The arrows
in such kinetic schemes are to be interpreted not as macroscopic chemical reaction
rates but rather as probabilities per unit time for transitions among the states,
and from long records of single channel dynamics one can extract these probabilities
and their voltage dependences.  Again, this is not easy, in part because one can
distinguish only the open state---different closed or inactivated states all have
zero conductance and hence are indistinguishable when measuring current---so that
multiple closed states have to be inferred from the distribution of times between
openings of the channel.  This is a very pretty subject, driven by the ability to do
extremely quantitative experiments; it is even possible to detect the shot noise as
ions flow through the open channel,  as well as a small amount of excess noise due to
the `breathing' of the channel molecule while it is in the open state.   The first
single channel experiments were by Neher and Sakmann \cite{neher+sakmann76}, and a modern
summary of what we have learned is given in textbooks \cite{hille,johnston+wu}.

{\small
\medskip
\addtocounter{problem}{1}
\noindent {\bf Problem \theproblem: Closed time distributions.} 
For a channel with just two states, show that the distribution of times between
one closing of the channel and the next opening is exponential in form.  How is this
distribution changed if there are two closed states?  Can you distinguish a
second closed state (``before'' opening) from an inactive state (``after''
opening)?
\bigskip}
 
We would like to pass from a description of single channels to a description of a
macroscopic piece of membrane, perhaps even the whole cell.  If we can assume that
the membrane is homogeneous and isopotential then there is one voltage $V$ across the
whole membrane, and each channel has the same stochastic dynamics.  If the region we
are talking about has enough channels, we can write approximately deterministic
equations for the number of channels in each state.  These equations have coefficients
(the transition probabilities) that are voltage dependent, and of course the voltage
across the membrane has a dynamics driven by the currents that pass through the open
channels.  Let's illustrate this with the simplest case.

Consider a neuron that has one type of
ion channel that is sensitive to voltage, and a
`leak' conductance (some channels that we haven't studied in detail, and which don't
seem to open and close in the interesting range of voltages).  Let the channel have
just two states, open and closed, and a conductance $g$ when it is open. 
Assume that the number of open channels $n(t)$
relaxes to its equilibrium value $n_{\rm eq} (V)$
with a time constant $\tau (V)$.  In addition
assume that the gated channel is (perfectly)
selective for ions that have a chemical potential
difference of $V_{\rm ion}$ across the membrane,
while the leak conductance $G_{\rm leak}$ pulls
the membrane potential back to its resting level
$V_{\rm rest}$. Finally, assume that the cell has
a capacitance $C$, and allow for the possibility
of injecting a current $I_{\rm ext}$ across the membrane. 
Then the equations of motion for the coupled
dynamics of channels and voltage are
\begin{eqnarray}
C{{dV}\over{dt}} &=& - gn (V- V_{\rm ion}) - G_{\rm leak} (V - V_{\rm
rest}) + I_{\rm ext}, 
\label{channeldynamics1}\\
{dn \over dt} &=& -{1 \over {\tau(V)}}[n- n_{\rm eq}(V)] .
\label{channeldynamics2}
\end{eqnarray}
These equations already have a lot in them:
\begin{itemize}
\item If we linearize around a steady state we find that the effect of the channels can
be thought of as adding some new elements to the effective circuit describing the
membrane.  In particular these elements can include an (effective) inductance and a
negative resistance.
\item Inductances of course make for resonances, which actually can be tuned by cells to
build arrays of channel--based electrical filters \cite{fettiplace}.  If the negative
resistance is large enough, however, the filter goes unstable and one gets oscillations.
\item One can also arrange the activation curve $n_{\rm eq}(V)$ relative to $V_{\rm
ion}$ so that the system is bistable, and the switch from one state to the other can be
triggered by a pulse of current.  In an extended structure like the axon of a neuron this
switching would propagate as a front at some fixed velocity.
\item In realistic models there is more than one kind of channel, and the nonlinear
dynamics which selects a propagating front instead selects a propagating pulse, which is
the action potential or spike generated by that neuron.
\end{itemize}

It is worth recalling the history of these ideas, at least briefly.  
In a series of papers, Hodgkin and Huxley \cite{hh1,hh2,hh3,hh4} wrote down
equations similar to Eq's. (\ref{channeldynamics1},\ref{channeldynamics2}) as a
phenomenological description of  ionic current flow across the
cell membrane.  They studied the squid giant axon, which is a
single nerve cell that is a small gift from nature, so large that
one can insert a wire along its length!   This axon, like that in
all neurons, exhibits propagating action potentials, and the task
which Hodgkin and Huxley set themselves was to understand the
mechanism of these spikes. It is important to remember that action
potentials provide the only mechanism for long distance
communication among specific neurons, and so the question of how
action potentials arise is really the question of how information
gets from one place to another in the brain. The  first step taken
by Hodgkin and Huxley was to separate space and time:  suspecting
that current flow along the length of the axon involved only
passive conduction through the fluid, they `shorted' this process
by inserting a wire and thus forcing the entire axon to become
isopotential.  By measuring the dynamics of current flow between
the wire and an electrode placed outside of the cell they were then
measuring the average properties of current flow across a patch of
membrane.

It was already known that the conductance of the
cell membrane changes during an action potential, and Hodgkin and Huxley studied this
systematically by holding the voltage across the membrane at one value and then
stepping to another.  With dynamics of the form in Eq. (\ref{channeldynamics2}), the
fraction of open channels will relax exponentially ... and after some effort one
should be able to pull out the equilibrium fraction of open channels and the
relaxation rates, each as functions of of voltage; again it is important to have the
physical picture that the channels in the membrane are changing state in response to
voltage (or, more naturally, electric field) and hence the dynamics are simple if the
voltage is (piecewise) constant.

There are two glitches in this simple picture.   First, the relaxation of
conductance or current is not exponential.  Hodgkin and Huxley
interpreted this (again, phenomenologically!) by saying that the
equations for elementary `gates' were as in Eq. (\ref{channeldynamics2})
but that conductance of ions trough a pore might require that several
independent gates are open.  
So instead of writing 
\begin{equation}
C{{dV}\over{dt}} = - gn (V- V_{\rm ion}) - G_{\rm leak} (V - V_{\rm
rest}) + I_{\rm ext}, 
\nonumber
\end{equation}
they wrote, for example,
\begin{equation}
C{{dV}\over{dt}} = - gn^4 (V- V_{\rm ion}) - G_{\rm leak} (V - V_{\rm
rest}) + I_{\rm ext}, 
\end{equation}
which is saying that four gates need to open in order for the channel to
conduct (their model for the potassium channel).
To model the inactivation of sodium channels they used equations in
which the number of open channels was proportional to $m^3 h$, where $m$
and $h$ each obey equations like  Eq. (\ref{channeldynamics2}), but the
voltage dependences $m_{\rm eq}(V)$ and $h_{\rm eq}(V)$ have opposite
behaviors---thus a step change in voltage can lead to an increase in
conductance as $m$ relaxes toward its increased equilibrium value, then a
decrease as $h$ starts to relax to its decreased equilibrium value.  In
modern language we would say that the channel molecule has more than two
states, but the phenomenological picture of multiple gates works quite
well; it is interesting that Hodgkin and Huxley themselves were careful
not to take  too seriously any particular molecular interpretation of
their equations. The second problem in the analysis is that there are
several types of channels, although this is easier in the squid axon
because `several' turns out to be just two---one selective for sodium
ions and one selective for potassium ions.   

The great triumph of Hodgkin and Huxley was to show that, having described the
dynamics of current flow across a single patch of membrane, they could predict the
existence, structure, and speed of propagating action potentials.  This was a
milestone, not least because it represents one of the few cases where a fundamental
advance in our understanding of biological systems was marked by a successful
quantitative prediction.  Let me remind you that, in 1952, the idea that nonlinear
partial differential equations like the HH  equations would
generate propagating stereotyped pulses was by no means obvious;
the numerical methods used by Hodgkin and Huxley were not so
different from what we might use today,  while rigorous proofs came only
much later.  

Of course I have inverted the historical order in this presentation, describing the
properties of ion channels (albeit crudely) and then arguing that these can be put
together to construct the macroscopic dynamics of ionic currents.  In fact the path
from Hodgkin and Huxley to the first observation of single channels took nearly twenty
five years.  There were several important steps.  First, the 
HH model makes definite predictions about the
magnitude of ionic currents flowing during an action potential,
and in particular the relative contributions of sodium and
potassium; these predictions were confirmed by measuring the flux
of radioactive ions.\footnote{It also turns out that the different
types of channels can be blocked, more or less independently, by
various molecules.  Some of the most potent channel blockers are
neurotoxins, such as tetrodotoxin from puffer fish, which is a
sodium channel blocker.  These different toxins allow a
pharmacological `dissection' of the molecular contributions to
ionic current flow.}  Second, as mentioned already, the
transitions among different channels states are voltage dependent
only because these different states have different dipole moments.
This means that changes in channel state should be accompanied by
capacitive currents, called `gating currents,' which persist even
if conduction of ions through the channel is blocked, and this is
observed.  The next crucial step is that if we have a patch of
membrane with a finite number of channels, then it should be
possible to observe fluctuations in current flow due to the
fluctuations in the number of open channels---the opening and
closing of each channel is an independent, thermally activated
process.  Kinetic models make unambiguous predictions about the
spectrum of this noise, and again these predictions were confirmed
both qualitatively and quantitatively; noise measurements also led
to the first experimental estimates of the conductance through a
single open channel.  Finally, observing the currents through
single channels required yet better amplifiers and improved
contact between the electrode and the membrane to insure that the
channel currents are not swamped by Johnson noise in stray
conductance paths. 

{\small
\medskip
\addtocounter{problem}{1}
\noindent {\bf Problem \theproblem: Independent opening and closing.} 
The remark that channels open and close independently is a bit glib.
We know that different states have different dipole moments, and you might
expect that these dipoles would interact.
Consider  an area $A$ of membrane  with
$N$ channels that each have two states.   Let the two states
differ by an effective displacement of charge $q_{\rm gate}$
across the membrane, and this charge interacts with the voltage
$V$ across the membrane in the usual way.  In addition, there is an energy associated
with the voltage itself, since the membrane has a capacitance.  
If we represent the two states of each channel by an Ising  spin
$\sigma_{\rm n}$, convince yourself that the energy of the system
can be written as
\begin{equation}
E = {1\over 2} C V^2 + {1\over 2}\sum_{{\rm n} =1}^N (\epsilon + q_{\rm gate} V )
\sigma_{\rm n} .
\end{equation}
Set up the equilibrium statistical mechanics of this system, and average over the
voltage fluctuations.  Show that the resulting model is a mean field interaction among
the channels, and state the condition that this interaction be weak, so that the
channels will gate independently.  Recall that both the capacitance and the number of
channels are proportional to the area. Is this condition met in real neurons?  In
what way does this condition limit the `design' of a cell? 
Specifically, remember that increasing $q_{\rm gate}$ makes the
channels more sensitive to voltage changes, since they make their
transitions over a voltage range $\delta V \sim k_B T/q_{\rm
gate}$; if you want to narrow this range, what do you have to
trade in order to make sure that the channels gate independently? 
And why, by the way, is it desirable to have independent gating?
\bigskip
}

So, in terms of Hodgkin--Huxley style models we would describe a neuron by
equations of the form
\begin{eqnarray}
C {{dV}\over {dt}} &=&- \sum_{\rm i} G_{\rm i}  m_{\rm i} ^{\mu_{\rm i} }
h_{\rm i} ^{\nu_{\rm i} } (V- V_{\rm i}) - G_{\rm leak} (V - V_{\rm
rest}) + I_{\rm ext},\\
{{dm_{\rm i}}\over{dt}} &=& -{1\over {\tau_{\rm act}^{\rm i}}}
[m_{\rm i} - m_{\rm eq}^{\rm i} (V)],\\
{{dh_{\rm i}}\over{dt}} &=& -{1\over{\tau_{\rm inact}^{\rm i}}}
[h_{\rm i} - h_{\rm eq}^{\rm i} (V)],
\end{eqnarray}
where $\rm i$ indexes a class of channels specific for ions with an
equilibrium potential $V_{\rm i}$ and we have separate kinetics for
activation and inactivation.  Of course there have been many studies of
such systems of equations.  What is crucial is that by doing, for
example, careful single channel experiments on patches of membrane from
the cell we want to study, we measure essentially every parameter of
these equations except for the total number of each kind of channel.
This is a level of detail that is not available in any other biological
system as far as I know.  

If we agree that the activation and inactivation variables run from zero
to unity, representing probabilities, then the number of channels is in
the parameters $G_{\rm i}$ which are the conductances we would observe if
all channels of class $\rm i$ were to open.  With good single
channel measurements, these are the only parameters we don't know.

For many years it was a standard exercise to identify the types of
channels in a cell, then try to use these Hodgkin--Huxley style dynamics
to explain what happens when you inject currents etc..  It probably is fair to say
that this program was successful beyond the wildest dreams of Hodgkin and Huxley
themselves---myriad different types of channel from many different types of neuron
have been described effectively by the same general sorts of equations.  On the other
hand (although nobody ever said this) you have to hunt around to find the right
$G_{\rm i}$s to make everything work in any reasonably complex cell.  It was Larry
Abbott, a physicist, who realized that if this is a problem for his graduate student
then it must also be a problem for the cell (which doesn't have graduate students to
whom the task can be assigned).   So, Abbott and his colleagues
realized that there must be regulatory mechanisms that control the channel numbers in
ways that stabilized desirable functions of the cell in the whole neural circuit
\cite{lemasson93}.  This has stimulated a beautiful series of experiments by Turrigiano
and collaborators, first in ``simple'' invertebrate neurons \cite{gina1} and then in
cortical neurons \cite{gina2}, showing that indeed these different cells can change
the number of each different kind of channel in response to changes in
the environment, stabilizing particular patterns of activity or response.
Mechanisms are not yet clear.  I believe that this is an early example of
the robustness problem \cite{elad-robust,goldman01} that was emphasized by Barkai and
Leibler for biochemical networks \cite{barkai}; they took adaptation in bacterial
chemotaxis as an example (cf. the lectures here by Duke) but the question clearly is more
general.  For more on these models of self--organization of channel densities see
\cite{moreLFA1,moreLFA2,moreLFA3}.

The problem posed by Abbott and coworkers was, to some approximation, about
homeostasis:  how does a cell hold on to its function in the network, keeping everything
stable. In the models, the ``correct'' function is defined implicitly.  The fact that we
have seen adaptation processes which serve to optimize information transmission or
coding efficiency makes it natural to ask if we can make models for the dynamics which
might carry out these optimization tasks.  There is relatively little work in this area
\cite{optmech1}, and I think that any effort along these lines will have to come to grips
with some tough problems about how cells ``know'' they are doing the right thing (by any
measure, information theoretic or not).  

Doing the right thing, as we have emphasized repeatedly, involves both the right
deterministic transformations and proper control of noise.  We know a grat deal about
noise in ion channels, as discussed above, but I think the conventional view has been
that most neurons have lots of channels and so this source of noise isn't really crucial
for neural function.  In recent work Schneidman and collaborators have shown that this
dismissal of channel noise may have been a bit too quick \cite{elad-noise}: neurons 
operate in a regime where the number of channels that participate in the ``decision'' to
generate an action potential is vastly smaller than the total number of  channels, so
that fluctuation effects are much more important that expected naively.  In particular,
realistic amounts of channel noise may serve to jitter the timing of spikes on time
scales which are comparable to the degree of reproducibility observed in the
representation of sensory signals (as discussed in Section 4).  In this way the problems
of molecular level noise and the optimization of information transmission may be
intertwined \cite{elad-robust}.

It should be emphasized that the molecular components we have been discussing are
strikingly universal.  Thus we can recognize homologous potassium channels in primate
cortex (the stuff we think with) and in the nerves of an earthworm.  There are vast
numbers of channels coded in the genome, and these can be organized into families of
proteins that probably have common ancestors \cite{hille}.   With such a complete
molecular description of electrical signalling in single cells, one would imagine that we
could answer a deceptively simple question:  what do individual neurons compute?  In
neural network models, for example, neurons are cartooned as summing their inputs and
taking a threshold. We  could make this picture a bit more dynamical by using an
`integrate and fire' model in which input currents are filtered by the RC time constant
of the cell membrane and all the effects of channels are summarized by saying that when
the resulting voltage reaches threshold there is a spike and a resetting to a lower
voltage. We would like to start with a more realistic model and show how one can identify
systematically some computational functions, but really we don't know how to do this.
One attempt is discussed in Refs. \cite{blaise1,blaise2}, where we use the ideas of
dimensionsality reduction \cite{88,features,brenner00a} to pass from the Hogdkin--Huxley
model to a description of the neuron as projecting dynamic input currents onto a low
dimensional space and then performing some nonlinear operations to determine the
probablity of generating a spike.  If a simple summation and threshold picture (or a
generalized `filter and fire' model) were correct, this approach would find it, but it
seems that even with two types of channels neurons can do something richer than this. 
Obviously this is just a start, and understanding will require us to face the deeper
question of how we can indentify the computational function of a general dynamical
system.

In this discussion I have focused on the dynamics of ion channels within one neuron. To
build  a brain we need to make connections or synapses between cells, and of course
these have their own dynamics and molecular mechanisms.  There are also problems of
noise, not least because synapses are very small structures, so that crucial biochemical
events are happening in cubic micron volumes or less.   The emergence of optical
techniques that allow us to look into these small volumes, deep in the living brain,
will quite literally bring into focus a number of questions about noise in biochemical
networks that are of interest both because they relate to how we learn and remember
things and because they are examples of problems that all cells must face as they carry
out essential functions.

\section{Speculative thoughts about the hard problems}

It is perhaps not so surprising that thinking like a physicist helps us to understand how
rod cells count single photons, or helps to elucidate the molecular events that
underlie the electrical activity of single cells.  A little more surprising, perhaps, is
that physical principles are still relevant when we go deeper into a fly's brain and ask
about how that brain extracts interesting features such as motion from a complex array
of data in the retina, or how these dynamic signals are encoded in streams of action
potentials.   As we come to the problems in learning, we have built an interesting
theoretical structure with clear roots in statistical physics, but we
don't yet know how to connect these ideas with experiment.  Behind this uncertainty is a
deeper and more disturbing question:  maybe as we progress from sensory inputs toward the
personal experiences of that world created by our brains we will encounter a real
boundary where physics stops and biology or psychology begins.  My hope, as you might
guess, is that this is not the case, and that we eventually will understand perception,
cognition and learning from the same principled mathematical point of view that we now
understand the inanimate parts of the physical world.  This optimism was shared, of
course, by Helmholtz and others more than a century ago.  In this last lecture I want to
collect some of my reasons for keeping faith despite obvious  problems.

In Shannon's original work on information theory, he separated the problem
of transmitting information from the problem of ascribing meaning to this
information \cite{shannon48}:
\begin{quote}
Frequently the messages have {\em meaning;} that is they
refer to or are correlated according to some system with certain
physical or conceptual entities.  These semantic aspects
of communication are irrelevant to the engineering problem.
\end{quote}
This quote is from the second paragraph of a very long paper; italics appeared in the
original. 
Arguably this is {\em the} major stumbling block in the 
use of information theory or any other ``physical'' approach to analyze cognitive
phenomena: our brains presumably are interested only in information
that has meaning or relevance, and if we are in a framework that
excludes such notions then we can't even get started.

Information theory is a statistical approach, and there is a widespread
belief that there must be ``more than just statistics'' to our
understanding of the world.  The clearest formulation
of this claim was by Chomsky \cite{chomsky56}, in a
rather direct critique of Shannon and his statistical
approach to the description of English.
Shannon had used $N^{\rm th}$ order Markov approximations
to the distribution of letters or words, and other
people used this $N$--gram method in a variety of ways,
including the amusing ``creative writing'' exercises of
Pierce and others.  Chomsky claims that all of this
is hopeless, for several reasons:
\begin{enumerate}
\item The significance of words or phrases is
unrelated to their frequency of occurrence.
\item Utterances can be arbitrarily long, with arbitrarily
long range dependences among words, so that no finite
$N^{\rm th}$ order approximation is adequate.
\item A rank ordering of sentences by their probability
in such models will have grammatical and ungrammatical
utterances mixed, with little if any tendency for
the grammatical sentences to be more probable.
\end{enumerate}
There are several issues here,\footnote{Some time after writing
an early draft of these ideas I learned that Abney \cite{abney} had expressed
similar thoughts about the nature of the Chomsky/Shannon debate;
he is concerned primarily with the first of the issues below.
I enjoyed especially his introduction to the problem:
``In one's introductory linguistics course, one learns that Chomsky
disabused the field once and for all of the notion that there was
anything of interest to statistical models of language.  But one usually
comes away a little fuzzy on the question of what, precisely, he
proved.''} and while I am far from being an expert on language I think if we try to
dissect these issues we'll get a feeling for the general problems of thinking about the
brain more broadly in information theoretic terms.

First we have the distinction between the true probability
distribution of sentences (for example) and any
finite $N^{\rm th}$ order approximation.
There are plenty of cases in physics where analogous
approximations fail, so this shouldn't bother
us, nor is it a special feature of language.
Nonetheless, it {\em is} important to ask how we can go
beyond these limited models.  There is a theoretical
question of how to characterize
statistics beyond $N$--grams,
and there is an experimental issue of how to measure
these long range dependencies in real languages or, more
subtly, in people's knowledge of languages.
I think that we know a big part of the answer to the
first question, as explained above:
The crucial measure of long range correlation is a divergence
in the {\em predictive information} $I_{\rm pred}(N)$,
that is the information
that a sequence of $N$ characters or words provides about the
remainder of the text.  
We can distinguish logarithmic divergence, which means
roughly that the sequence of words allows us to learn
a model with a finite number of parameters (the coefficient
of the log then counts the dimensionality of the parameter space),
from a power law divergence, which is what happens when longer
and longer sequences allow us to learn a more and more
detailed description of the underlying model.  There are hints
that language is in the more complex power law class.

A second question concerns the learnability of
the relevant distributions.  It might be that
the true distribution of words and phrases
contains everything we want to know about the language,
but that we cannot learn this distribution from examples.
Here it is important that ``we'' could be a child
learning a language, or a group of scientists
trying to analyze a body of data on language
as it is used. Although learnability is a crucial issue, 
I think that there is some confusion in the literature.
Thus, even in very recent work, we find comments that confuse the frequency of
occurrence of examples that we have seen with 
the estimate that an observer might make of the underlying
distribution.\footnote{In particular, a widely discussed paper by Marcus et al. 
\cite{marcus99} makes the clear statement that all unseen combinations
of words must be assigned probability zero in
a statistical learning scheme, and this simply
is wrong. The commentary on this paper by Pinker \cite{pinker99}
has some related confusions about what
happens when one learns a distribution from
examples.  He notes that we can be told ``that a whale
is not a fish ... overriding our statistical experience ...'' .
In the same way that reasonable learning algorithms
have to deal with unobserved combinations, they also
have to deal with outliers in distributions;
the existence of outliers, or the evident difficulty
in dealing with them, has nothing to do with
the question of whether our categories of
fish and mammals are built using a probabilistic
approach. The specific example of whales may be 
a red herring: does being told that a whale
is not a fish mean that ``all the fish in the sea''
cannot refer to whales?}  The easiest way to see this is
to think about distributions of continuous variables,
where obviously we have to interpolate or smooth
so that our best estimate of the probability
is not zero at unsampled points nor is it singular
at the sampled points.  There are many ways of doing this, and I think that developments
of the ideas in Ref. \cite{bcs} are leading us toward a view of this problem which at
least seems principled and natural from a physicist's point of view
\cite{periwal1,holy,periwal2,aida,ilya-pre}. 
On the other hand, the question of how one does such
smoothing or regularization in the case of discrete
distributions (as for words and phrases) is much less
clear (see, for example, \cite{nsb}).

Even if  we can access the full
probability distribution of utterances (leaving
aside the issue of learning this distribution from examples), there
is a question of whether this distribution captures
the full structure of the language. 
At one level this is trivial:  if we really have
the full distribution we can generate samples, and there will be no statistical test that
will distinguish these samples from real texts.  Note again that probability
distributions are ``generative'' in the sense that Chomsky
described grammar, and hence that no reasonable
description of the probability distribution is limited
to generating sequences which were observed
in some previous finite sampling or learning
period.  Thus, if we had an accurate model of the
probability distribution for texts, we could pass
some sort of Turing test.  The harder question is
whether this description of the language would contain
any notions of syntax or semantics.  Ultimately
we want to ask about meaning:  is it possible for a 
probabilistic model to encode the meanings of words
and sentences,
or must these be supplied from the outside?
 Again the same question arises in other domains: 
in what sense does knowing the distribution of all possible natural movies
correspond to ``understanding'' what we are seeing?

Recently, Tishby, Pereira and I have worked on the problem
of defining and extracting {\em relevant
information}, trying to fill the gap left by Shannon \cite{tpb}. 
Briefly, the idea is that we observe some signal
$x \in X$ but are interested in another signal $y\in Y$.
Typically a full description of $x$ requires
many more bits than are relevant to the
determination of $y$, and we would like to separate
the relevant bits from the irrelevant ones.
Formally we can do this by asking for a compression
of $x$ into some new space $\tX$ such that
we keep as much information as possible about
$Y$ while throwing away as much as possible about $X$.
That is, we want to find a mapping $x \rightarrow \tx$
that maximizes the information about $Y$  
while holding the information about $X$ 
fixed at some small value.  This problem turns out to be  equivalent
to a set of self--consistent equations for the mapping
$x\rightarrow \tx$, and is very much like a
problem of clustering.  It is important that, unlike
most clustering procedures, there is no
need to specify a notion of similarity or distance
among points in the $X$ or $Y$ spaces---all notions of
similarity emerge directly from the joint statistics of
$X$ and $Y$.

To see a little of how this works, let's start with a somewhat fanciful question:
What is the information content of the morning
newspaper?  Since entropy provides
the only measure of information that is
consistent with certain simple and plausible
constraints (as emphasized above),  it is tempting 
to suggest that the information content of
a news article is related to the entropy
of the distribution from which newspaper
texts are drawn.  This is troublesome---more
random texts have higher entropy and hence would
be more informative---and also  incorrect.
Unlike entropy, information always is  {\em about}
something.  We can ask how much an article
tells us about, for example, current events in France,
or about the political bias of the editors, and in a foreign country we
might even use the newspaper to measure our own comprehension of the
language. In each case, our question of interest has a distribution of
possible answers, and (on average) this distribution shifts to one with a
lower entropy once we read the news; this decrease in
entropy is the information that we gain by reading.
This relevant information typically is much smaller
than the entropy of the original signal:
information about the identity of a person is
much smaller than the entropy of images
that include faces, 
information about words is much smaller than
the entropy of the corresponding speech sounds, and so on.
Our intuitive notion of `understanding' these 
high entropy input signals corresponds
to isolating the relevant information
in some reasonably compact form:  summarizing the
news article, replacing the image with a name,
enumerating the words that have been spoken.

When we sit down to read, we have in mind
some question(s) that we expect to have answered by
the newspaper text.  We can enumerate all of the possible answers
to these questions, and label the answers by $y\in Y$:  this is
the relevant variable, the information of value to us.
On the other hand, we can also imagine the ensemble of
all possible newspaper texts, and  label each possible text by
$x\in X$:  these are the raw data that we have to work with.
Again, there are many more bits in the text $x$ than are relevant
to the answers $y$, and understanding the text means that we must
separate these relevant bits from the majority of irrelevant bits.
In practice this means that we can `summarize' the text, and in the
same way that we  enumerate all possible texts we can
also  enumerate all possible summaries, and labelling them
by $\tx \in \tX$.  If we can generate good or efficient summaries then we
can construct a mapping
of the raw data
$x$ into the summaries
$\tx$  such that we discard most of the information about the text
but preserve as much information as possible about the relevant variable
$y$.

The statement that we want our summaries to be compact
or efficient  means that we want to discard as much information
as possible about the original signal.  Thus, we want to `squeeze'
or minimize the information that the summary provides about the
raw data, $I(\tx ; x)$. On the other hand, if the summary is going
to capture our understanding, then it must preserve information about $y$,
so we want to maximize $I(\tx ; y )$. More precisely, there is going to
be some tradeoff between the level of detail [$I(\tx ; x)$]
that we are willing to tolerate and the amount of relevant information
[$I\tx ; y)$] that we can preserve.   The optimal procedure
would be to find rules for generating summaries which provide the maximum
amount of relevant information given their level of detail.  The way to
do this is to maximize the weighted difference between the two
information measures,
\begin{equation}
-{\cal F} = I(\tx ; y) - T I(\tx ; x),
\label{defF}
\end{equation}
where $T$ is a dimensionless parameter that measures the amount of extra
detail we are willing to accept for a given increase in relevant
information.  We will refer to this parameter as the temperature, for
reasons that become clear below.  So, to find optimal summaries we want
to search all possible rules for mapping $x \rightarrow \tx$ until we
find a maximum of $-{\cal F}$, or equivalently a minimum of the `free
energy' $\cal F$. Note that the structure of the optimal procedure generating
summaries will evolve as we change the temperature $T$; there is no
`correct' value of the temperature, since different values correspond to
different ways of striking a balance between detail and effectiveness in
the summaries.

There are several different interpretations of the principle that we
should minimize $\cal F$.  One view is that we are optimizing the
weighted difference of the two informations, counting one as a benefit
and one as a cost.  Alternatively, we can see minimizing $\cal F$  as
maximizing the relevant information while holding fixed the level of
detail in the summary, and  in this case we interpret $T$ as a Lagrange
multiplier that implements the constraint holding $I(\tx; x)$
fixed.  Similarly, we can divide through by $T$ and
interpret our problem as one of squeezing the summary as much as
possible---minimizing $I(\tx ; x)$---while holding fixed the amount of
relevant information that the summaries convey; in this case $1/T$ serves
as the Lagrange multiplier.

It turns out that the problem of
minimizing the free energy
$\cal F$ can solved, at least formally.  To begin we need to say what we
mean by searching all possible rules for mapping $x \rightarrow \tx$.  We
consider here only the case where the summaries forma discrete set, and
for simplicity we (usually) assume that the data $x$ and the relevant
variable $y$ also are drawn from discrete sets of possibilities.
The general mapping from $x$ to $\tx$ is probabilistic, and the set
of mapping rules is given completely if we specify the set of
probabilities $P(\tx |x)$ that any raw data point $x$ will be assigned to
the summary $\tx$.  These probabilities of course must be normalized, so
we must enforce
\begin{equation}
\sum_{\tx \in \tX} P(\tx | x) = 1
\label{normalize}
\end{equation}
for each $x\in X$.  We can do this by introducing a Lagrange multiplier
$\Lambda (x)$ for each $x$ and then solving the constrained optimization
problem
\begin{equation}
\min_{P(\tx | x)} \left[ {\cal F} - \sum_{x\in X} \Lambda (x)
\sum_{\tx\in \tX} P(\tx | x) \right] ,
\label{variation}
\end{equation}
and at the end we have choose the values of $\Lambda (X)$ to satisfy the
normalization condition in Eq. (\ref{normalize}).

As shown in Ref. \cite{tpb}, the Euler--Lagrange equations for this variational problem
are equivalent to a set of self--consistent equations for the probability distribution
$P(\tx | x)$:
\begin{eqnarray}
P(\tx |x)
&=&
{ {P(\tx )} \over{Z(x,T)}}
\exp
{\Bigg\{}
-{1\over T}
\sum_{y \in Y} P(y|x) \ln \left[ {{P(y|x)}\over {P(y|\tx)}} \right]
{\Bigg\}}
\\
P (y | \tx ) &=& \sum_{x\in X} P(y|x) P(x|\tx )
\nonumber\\
&=& {1\over{P(\tx)}} \sum_{x\in X} P(y|x) P(\tx |x) P(x) .
\label{centerdist}
\end{eqnarray}

Up to this point, the set of summaries
$\tX$ is completely abstract.  If we
choose a fixed number of possible summaries
then the evolution with temperature is continuous, and as we
lower the temperature the summaries become progressively more detailed
[$I(\tx ; x)$ is increasing] and more informative [$I(\tx; y)$ is increasing]; the local
coefficient that relates the extra relevant
information per increment of detail is 
the temperature itself.  

If $X$ is discrete, then the detail in the summary can never exceed the
entropy $S(X)$, and of course the relevant information provided by the
summaries can never exceed the relevant information  in the original
signal.  This means that there is a natural set of normalized coordinates
in the information plane $I(\tx ; y)$ vs. 
$I(\tx ;x)$, and different signals are
characterized by different trajectories in these normalized
coordinates.  If signals are `understandable' in the colloquial
sense, it must be that most of the available relevant information
can be captured by summaries that are very compact, so that
$I(\tx ; y)/I(x;y)$ is near unity even
when $I(\tx; x)/S(X)$ is very small. At the
opposite extreme are signals that have been  encrypted (or texts
which are so convoluted) so that no small piece of the original data
contains any significant fraction of the relevant information.

Throughout most of the information plane the
optimal solution has a probabilistic structure---the assignment rules
$P(\tx|x)$ are not deterministic. This means
that our problem of providing informative but compact summaries is very
different from the usual problems in classification or recognition, where
if we ask for assignment rules that minimize errors we will always find
that the optimal solution is deterministic (recall Problem 2). Thus the information
theoretic approach encompasses automatically a measure of confidence in which the
optimal strategy involves (generically) a bit of true random guessing
when faced with uncertainty.  Returning to our example of the newspaper,
this has an important consequence.  If asked to provide a summary of the
front page news,  the optimal summaries have probabilistic assignments to
the text---if asked several times, even an `optimal reader' will have a
finite probability of giving different answers each time she is asked. 
The fact that assignment rules are probabilistic means also that these
rules can be perturbed quantitatively by finite amounts of additional
data, so that small amounts of additional information about, for example,
the a priori likelihood of different interesting events in the world  can
influence the optimal rules for summarizing the text.  It is
attractive that a purely objective procedure, which provides an optimal
extraction of relevant information, generates these elements of
randomness and subjectivity.

Extracting relevant information has
bene called the ``information bottleneck'' because
we squeeze the signal $X$ through a narrow channel
$\tilde X$ while trying to preserve information about $Y$.
This approach has been used, at least in preliminary
form, in several cases of possible interest for the
analysis of language.  First, we can take $X$ to be one
word and $Y$ to be the next word.   If we
insist that there be very few categories for
$\tilde X$---we squeeze through a very narrow bottleneck---then to a good
approximation the mapping from $X$ into $\tilde X$ constitutes a clustering
of words into sensible syntactic categories (parts of speech, with a
clear separation of proper nouns).  It is interesting that in the
cognitive science literature there is also a discussion
of how one might acquire syntactic categories from such
``distributional information'' (see, for example, \cite{redington}), although
this discussion  seems to use somewhat arbitrary
metrics on the space of conditional distributions.

If we allow for $\tilde X$ to capture more bits about $X$ in the
next word problem, then we start to see the general
part of speech clusters break into groups
that have some semantic relatedness,
as noted also by Redington et al. \cite{redington}.
A more direct demonstration was given by
Pereira,  Tishby and Lee \cite{ptl}, who took $X$ and $Y$ to be the
noun and verb of each sentence.\footnote{This
was done before the development of the bottleneck
ideas so we need to be a bit careful.   Tishby et al.
proposed clustering $X$ according the conditional distributions
$P(y|x)$ and suggested the use of the Kullback--Leibler
divergence ($D_{\rm KL}$)
as a natural measure of distance.  In the bottleneck
approach there is no need to postulate a distance
measure, but what emerges from the analysis
is essentially the soft clustering based on $D_{\rm KL}$ as
suggested by Tishby et al..}
Now one has the clear impression that the clusters of
words (either nouns or verbs) have similar meanings,
although this certainly is only a subjective remark at
this point.  Notice, however, that in this formulation the absolute
frequency of occurrence of individual words or even word
pairs is not essential (connecting to Chomsky's point \#1 above);
instead the clustering of words with apparently similar meanings
arises from the total structure of the set of conditional
distributions $P(y|x)$.

Yet another possible approach to ``meaning''
involves taking $X$ as the identity of a document
$Y$ as a word in the document.  Slonim and Tishby \cite{noam}
did this for documents posted to twenty different
news groups on the web, of course hiding any information
that directly identifies the news group in the
document.  The result is that choosing roughly
twenty different values for $\tilde X$ captures
most of the mutual information between $X$ and $Y$,
and these twenty clusters have a very strong overlap
with the actual newsgroups.   This procedure---which
is `unsupervised' since the clustering algorithm
does not have access to labels on the documents---yields
a categorization that is competitive with state of the art methods for supervised
learning of the newsgroup identities.  While one may object
that some of these tasks are too easy, these results
at least go in the direction of suggesting that
analysis of word statistics alone can identify the
``topic'' of a document as it was assigned by the author.

I think that some reasonably concrete questions emerge from all of this:
\begin{itemize}
\item How clear is the evidence that language---or other (colloquially) complex natural
signals---fall into the power-law class defined  through the analysis of predictive
information?
\item On a purely theoretical matter, can we regularize the problem of learning
distributions over discrete variables in a (principled) way which puts such learning
problems in the power--law class?
\item Can we use the  information bottleneck ideas to find the features of words
and phrases that efficiently represent the large amounts of predictive information that
we find in texts?  
\item Can we test, in psycholinguistic experiments, the hypothesis that this clustering
of words and phrases through the bottleneck collects together items with similar meaning?
\item If we believe that meanings are related to statistical structure, can we shift our
perceptions of meaning by exposure to texts with different statistics? 
\end{itemize}
This last experiment would, I think, be quite compelling (or at least provocative).  When
we set out to test the idea that neural codes are ``designed'' to optimize information
transmission, the clear qualitative prediction was that the code would have to change in
response to the statistics of sensory inputs, and this would have to work in ways beyond
the standard forms of adaptation that had been characterized previously.  This
prediction was confirmed, and it would be quite dramatic if we could design a parallel
experiment on our perception of meaning.  Surely this is the place to stop.

\vfill\newpage

{\small
The text and reference list make clear the enormous debt I owe to my colleagues
and collaborators.  I especially want to thank Rob de Ruyter, since it is literally
because of his experiments that I had the courage to ``think about the brain'' in the
sense which I tried to capture in these lectures.  The adventure which Rob and I have had
in thinking about fly brains in particular has been one we enjoyed sharing with many
great students, postdocs and collaborators over the years:   A. Zee, F. Rieke, D. K.
Warland, M. Potters, G. D. Lewen, S. P. Strong, R. Koberle, N. Brenner, E. Schneidman,
and A. L. Fairhall (in roughly historical order).  While all of this work blended theory
and experiment to the point that there are few papers which are `purely' one or the
other, I've also been interested in some questions which, as noted, have yet to make
contact with data.  These ideas also have developed in a very collaborative way with C.
G. Callan, I. Nemenman, F. Pereira, and N. Tishby (alphabetically).  Particularly for the
more theoretical topics, discussions with my current collaborators J. Miller and S. Still
have been crucial.  I am grateful to the students at Les Houches for making the task of
lecturing so pleasurable, and to the organizers both for creating this opportunity and
for being so patient with me as I struggled to complete these notes.  My thanks also to
S. Still for reading through a (very) late draft and providing many helpful suggestions.}

\end{document}